\documentstyle[12pt,axodraw,epsf]{article} 
\catcode`@=11
\newcount\@tempcntc
\def\@citex[#1]#2{\if@filesw\immediate\write\@auxout{\string\citation{#2}}\fi
  \@tempcnta\z@\@tempcntb\m@ne\def\@citea{}\@cite{\@for\@citeb:=#2\do
    {\@ifundefined
       {b@\@citeb}{\@citeo\@tempcntb\m@ne\@citea\def\@citea{,}{\bf ?}\@warning
       {Citation `\@citeb' on page \thepage \space undefined}}%
    {\setbox\z@\hbox{\global\@tempcntc0\csname b@\@citeb\endcsname\relax}%
     \ifnum\@tempcntc=\z@ \@citeo\@tempcntb\m@ne
       \@citea\def\@citea{,}\hbox{\csname b@\@citeb\endcsname}%
     \else
      \advance\@tempcntb\@ne
      \ifnum\@tempcntb=\@tempcntc
      \else\advance\@tempcntb\m@ne\@citeo
      \@tempcnta\@tempcntc\@tempcntb\@tempcntc\fi\fi}}\@citeo}{#1}}
\def\@citeo{\ifnum\@tempcnta>\@tempcntb\else\@citea\def\@citea{,}%
  \ifnum\@tempcnta=\@tempcntb\the\@tempcnta\else
   {\advance\@tempcnta\@ne\ifnum\@tempcnta=\@tempcntb \else \def\@citea{--}\fi
    \advance\@tempcnta\m@ne\the\@tempcnta\@citea\the\@tempcntb}\fi\fi}
\catcode`@=12
\def\theequation{\arabic{section}.\arabic{equation}}
\def\simlt{\stackrel{<}{{}_\sim}}

\begin{document}

\begin{flushright}
CERN-TH/99-34\\
hep-ph/9902371\\
February 1999
\end{flushright}

\begin{center}
{\Large {\bf Higgs Bosons in the Minimal Supersymmetric}}\\[0.4cm]
{\Large {\bf Standard Model with Explicit CP Violation}}\\[2.4cm]
{\large Apostolos Pilaftsis and Carlos E.M. Wagner}\\[0.4cm]
{\em Theory Division, CERN, CH-1211 Geneva 23, Switzerland}
\end{center}
\vskip1.4cm \centerline{\bf  ABSTRACT} 
We  study the Higgs-boson mass  spectrum of the minimal supersymmetric
standard  model, in  which the tree-level  CP  invariance of the Higgs
potential is broken explicitly  by  loop effects of  soft-CP-violating
Yukawa interactions related to scalar  quarks of the third generation. 
The analysis    is   performed  by   considering   the CP-non-invariant
renormalization-group    improved     effective    potential   through
next-to-leading order that includes leading logarithms due to two-loop
Yukawa and  QCD  corrections.  We find  that the  three neutral  Higgs
particles  predicted by the theory  may strongly mix with one another,
thereby significantly modifying their tree-level couplings to fermions
and to  the $W^\pm$ and $Z$  bosons.  We  analyze the phenomenological
consequences of such a minimal supersymmetric  scenario of explicit CP
violation on the production rates  of the lightest Higgs particle, and
discuss   strategies   for  its   potential discovery   at high-energy
colliders.

\newpage

\setcounter{equation}{0}
\section{Introduction}

Despite the great   phenomenological success of the  minimal  standard
model (SM) at    collider and lower  energies,  its  full experimental
vindication is not yet complete.   The  Higgs boson $H$, the  ultimate
cornerstone of  the SM responsible for  endowing the observed fermions
and the $W^\pm$ and $Z$ bosons with masses, still remains elusive thus
far.  Recent experiments at the  CERN Large Electron Positron Collider
operating at energies of 189 GeV (LEP2) place a  severe lower bound on
$M_H$,    i.e.\ $M_H \ge 95.5$  GeV   at 95$\%$  confidence level (CL)
\cite{Higgs1}.   On the    other hand, global   experimental  data and
theoretical  analyses of  radiative  effects  suggest  that if  nature
indeed realizes the SM Higgs boson, it is  then very unlikely that its
mass be much larger than  about 250 GeV \cite{Higgs2}.   Nevertheless,
there  are many theoretical reasons  to believe that the SM represents
the low-energy  limit of a more  fundamental  theory whose first clear
signals are  expected to be seen in  experiments accessing energies in
the range of 0.1 to  1 TeV.  Especially, supersymmetry (SUSY)  appears
theoretically to  be   a  compelling   ingredient  for  a   successful
unification of gravity with all other fundamental  forces in nature by
means of supergravity and superstrings.   In its minimal  realization,
the minimal supersymmetric standard model  (MSSM), SUSY must be broken
softly, in   agreement with   experimental observations   \cite{SUSY}. 
Unlike  the  SM, the  MSSM offers  an appealing  solution to the gauge
hierarchy problem, which is reflected by the perturbative stability of
such a  theory from  the   electroweak to  the  Planck energy   scale. 
Because of the holomorphicity  of  the superpotential, the MSSM   must
contain  at  least two Higgs doublets, denoted as $\tilde{\Phi}_1$ and
$\Phi_2$, with opposite hypercharges, $Y(\Phi_2 ) = - Y(\tilde{\Phi}_1
) = 1$, so as to give tree-level masses to both  up and down families,
and to cancel the triangle anomalies.

Even  though SUSY requires  the presence of  two Higgs doublets in the
theory,   the so-extended Higgs    sector of  the   MSSM remains  very
predictive.  This is because, at the  tree level, all four-dimensional
quartic  couplings in  the Higgs potential  are  not independent,  but
related to the known electroweak  coupling constants $g_w$ and $g'$ of
the   gauge  groups   SU(2)$_L$    and U(1)$_Y$,  respectively.    The
CP-conserving MSSM  predicts  three Higgs  states:  two  of the  Higgs
bosons,  $h$ and $H$, are even  under CP and   the Higgs boson $A$ has
CP-odd parity.  Beyond  the Born approximation, extensive  theoretical
studies based  on renormalization-group  (RG) methods and diagrammatic
techniques  have shown  \cite{Mh,HH,CEQW}  that  the lightest  CP-even
Higgs boson, $h$, must possess a mass below 130 GeV.  This upper bound
is reached  for large values of the  ratio of Higgs vacuum expectation
values,     $\tan\beta    \equiv   \langle   \Phi_2\rangle  /  \langle
\tilde{\Phi}_1 \rangle  > 15$.  For low  values  of $\tan\beta \approx
2$, the upper bound on the mass  of $h$ decreases substantially, i.e.\ 
$M_h \simlt 110$ GeV.  For such low $\tan\beta$ scenarios, the current
experimental limit on $M_h$ is almost equal to the SM Higgs-mass bound
for large   values of $M_A$,   i.e.\ $M_h \ge  95$  GeV  at 95$\%$ CL,
decreasing slightly for low values of $M_A$, of the  order of the weak
scale.   Therefore,  the present    experimental   bounds put   strong
restrictions on models with  low values of  $\tan\beta$, close  to the
infrared fixed-point     value  \cite{CHE,CCPW}. Therefore, next-round
experiments at  LEP2  turn out   to be   very   crucial, as  they  can
potentially exclude  a significant portion  of  the parameter space of
the CP-conserving MSSM \cite{CCPW}.

Most analyses  of the Higgs-boson mass spectrum  of the MSSM have been
performed,  in  the existing  literature,  only  within the restricted
framework of an effective  CP-invariant Higgs potential.  Recently, it
has been shown \cite{APLB}, however, that the tree-level CP invariance
of the  MSSM Higgs sector can be  broken sizeably by  one-loop effects
that  involve trilinear CP-violating couplings  of the  scalar top and
bottom quarks to the Higgs states.  As  a consequence, the high degree
of   the  tree-level  mass  degeneracy between   $H$   and $A$  may be
considerably  lifted  at one  loop.    Within  the context  of general
two-Higgs-doublet models,  the latter possibility has been extensively
discussed by several authors,  in connection with observable phenomena
of resonant CP violation   through $HA$ mixing \cite{ANPB} at   future
high-energy colliders, such as the Next Linear $e^+e^-$ Collider (NLC)
\cite{PN,Osland}, the CERN Large Hadron Collider (LHC) \cite{ANPB,LHC}
and  the proposed First  Muon  Collider (FMC) \cite{APRL,CD}.  Another
important consequence of Higgs-sector CP violation in the MSSM is that
the loop-induced CP-violating $hA$ mixing can be  of a size comparable
to $M_h$, which  may affect the  predictions obtained for the  mass of
the lightest Higgs  boson.   CP violation  and  a light neutral  Higgs
boson are  essential ingredients to  account for the observed baryonic
asymmetry  in the   Universe,  through  the  mechanism of  electroweak
baryogenesis in the MSSM \cite{EWbau}.

In   this paper, we  shall  systematically study  the mass spectrum of
Higgs  bosons  in the  MSSM with  explicit  CP violation.    In such a
scenario, both  Higgs-boson  masses and their   couplings to fermions,
$W^\pm$  and $Z$ bosons are significantly  affected by the presence of
CP-violating   interactions.  Therefore,   we  shall  pay   particular
attention to the predictions obtained for  the production rates of the
lightest Higgs boson at LEP2  and other high-energy machines, such  as
the upgraded option of the Tevatron at Fermilab.  Our analysis will be
based on the computation of the CP-non-invariant RG-improved effective
potential up to  next-to-leading order.  The dominant contributions to
the  effective potential  come from the  top  ($t$)  and bottom  ($b$)
quarks,  as  well as from  their   supersymmetric partners.  For  this
purpose,  we neglect   chargino  and neutralino quantum   corrections. 
However, we include leading logarithms due to two-loop QCD and $t$ and
$b$ Yukawa corrections \cite{KYS,CEQR}.   As has been explicitly shown
in \cite{CEQR,CEQW},  these corrections  improve  the effective  Higgs
potential by minimizing its scale dependence significantly.

CP-violating low-energy constraints, especially  those coming from the
electric dipole moment (EDM) of the neutron  and the electron, play an
important  role in   our  analysis \cite{EDM,DGH,FOS,KO,IN,AF,CKP}.    
However,  there    have  been several  suggestions      to evade these
constraints, without   suppressing  the  CP-violating  phases of   the
theory.  One  option is to  make  the first two generations  of scalar
fermions  rather  heavy,   e.g.\  above  1   TeV  \cite{KO}.   Another
possibility    is to  arrange  for   partial  cancellations among  the
different  EDM contributions \cite{IN}.   Finally,  it is possible  to
make  the quantum  corrections of  the first  two generations to  EDMs
negligible   by    requiring  a  kind    of   non-universality  in the
soft-trilinear Yukawa    couplings    \cite{AF}.    Nevertheless,  the
aforementioned options do not prevent the supersymmetric analog of the
two-loop Barr--Zee  mechanism for generating  EDMs from becoming large
for high values  of $\tan\beta$ \cite{CKP}.   Therefore, in our study,
we  shall consider these last  `direct' EDM constraints related to the
CP-violating phases of scalar quarks of the third generation.

The organization of the paper is as follows:  in Section 2 we consider
the  CP-violating RG-improved effective  Higgs potential  of the MSSM,
and  derive the minimization  conditions related  to the Higgs  ground
state.  In addition,  we calculate  the  general mass  matrices of the
neutral and charged Higgs bosons.  Technical  details are relegated to
the  Appendix.  In Section 3 we  compute the three mass eigenvalues of
the $(3\times 3)$-dimensional mass matrix of the neutral Higgs bosons,
and the respective mixing angles. In  Section 4 we consider the effect
of EDM   constraints  on the  CP-violating  parameters  related to the
third-generation     scalar   fermions.    Section    5  discusses the
interactions   of the  Higgs particles  with   fermions, and  with the
$W^\pm$ and $Z$ bosons in the presence  of CP violation.  Furthermore,
we analyze the phenomenological implications of these interactions for
Higgs-boson searches  at LEP2,  and discuss  the  prospects of probing
such a minimal  SUSY scenario of explicit CP  violation at an upgraded
Tevatron machine.  Section 6 summarizes our conclusions.

\setcounter{equation}{0}
\section{The CP-violating Higgs potential of the MSSM}

The MSSM introduces   several new parameters  in  the theory  that are
absent from the  SM  and could,  in principle,  possess  CP-odd phases
\cite{EDM}.   Specifically, the new  CP-odd  phases may come from  the
following parameters: (i) the mass parameter $\mu$, which involves the
bilinear   mixing  of  the   two Higgs    chiral  superfields  in  the
superpotential;   (ii)   the  soft-SUSY-breaking   gaugino      masses
$m_\lambda$,  where    $\lambda$  collectively   denotes  $\tilde{g}$,
$\tilde{W}$ and $\tilde{B}$, i.e.\   the gauginos of the  gauge groups
SU(3)$_c$, SU(2)$_L$  and   U(1)$_Y$,  respectively; (iii) the    soft
bilinear  Higgs-mixing mass $m^2_{12}$, which  is sometimes denoted as
$B\mu$ in the literature; and (iv) the soft trilinear Yukawa couplings
$A_f$ of the Higgs particles to  scalar fermions.  If the universality
condition  is imposed on all gaugino  masses at  the unification scale
$M_X$, then $m_\lambda$ has  a common phase.  Likewise, the  different
trilinear couplings $A_f$ are all  equal at $M_X$,  i.e.\  $A_f = A$.  
Here, one  may slightly  deviate  from exact universality  by assuming
that $A$ is a matrix  in the flavour  space \cite{AF}.  In particular,
it  has been argued  recently  \cite{CKP} that many dangerously  large
contributions   to  the electron and  neutron   EDMs may  naturally be
avoided by choosing the trilinear coupling of the  Higgs fields to the
scalar quarks of  the first and second  generation to be  much smaller
than the one of the third generation, i.e.  $A_f \simeq (0,0,1) A$.

It is known  that the conformal-invariant  part of the  supersymmetric
Lagrangian possesses two global Peccei--Quinn-type symmetries:
\begin{itemize}
  
\item[ (i)] The U(1)$_Q$ symmetry,  with $Q$ assignments $Q(\hat{H}_1)
  = 1$,    $Q(\hat{H}_2)   = -2$,  $Q(\hat{Q})   =    Q(\hat{L}) = 0$,
  $Q(\hat{U}) = 2$ and $Q(\hat{D}) = Q(\hat{E}) = -1$, where the caret
  on the  fields symbolizes  superfields.  $\hat{H}_1$ and $\hat{H}_2$
  are   the  Higgs   superfields,   which have  opposite  hypercharges
  $Y(\hat{H}_2)  = -Y(\hat{H}_1)   =  1$,  and $\hat{Q}$  ($\hat{L}$),
  $\hat{U}$ and $\hat{D}$ ($\hat{E}$) are the chiral multiplets of the
  quark (lepton) left-handed  doublet,  the right-handed up quark  and
  the right-handed  down  quark  (lepton), respectively.   The  chiral
  multiplets carry the hypercharges: $Y(\hat{Q}) = 1/3$, $Y(\hat{L}) =
  -1$, $Y(\hat{U}) = -4/3$, $Y(\hat{D}) =  2/3$, and $Y(\hat{E}) = 2$. 
  The  U(1)$_Q$   symmetry is   broken  by  the $\mu$   and $m_{12}^2$
  parameters.
  
\item[(ii)]   The U(1)$_R$  symmetry   acting on  the Grassmann-valued
  coordinates $\theta$  and  $\bar{\theta}$, i.e.\ $e^{i\alpha}\theta$
  and   $e^{-i\alpha}\bar{\theta}$.  So, the   $\theta$ coordinate  of
  superspace carries charge 1.  Moreover, all matter superfields carry
  charge 1  and  all Higgs superfields  carry charge  0.  Under such a
  transformation,  the    gaugino  fields   carry   charge   1.    The
  superpotential carries  charge  2.  Consequently,   this symmetry is
  broken by  the Majorana masses   of the gauginos  as  well as by the
  scalar-fermion--Higgs  trilinear  couplings $A_f$  and the parameter
  $\mu$.

\end{itemize}
As a consequence, not  all phases of  the four complex  parameters $\{
\mu ,\    m^2_{12},\ m_\lambda  ,\  A   \}$ turn   out to be  physical
\cite{DGH},  i.e.\ two phases may  be removed by redefining the fields
accordingly.  Employing the global symmetries (i) and (ii), one of the
Higgs doublets and the  gaugino fields $\lambda$  can be rephased in a
way such that $m_\lambda$  and $m^2_{12}$ become  real numbers.  As we
will see, the fact that $m^2_{12}$ is made real complies also with the
CP-odd  tadpole constraint on  the Higgs  potential  at the tree level
\cite{APLB}.  Thus, arg($\mu$)   and arg($A$)  are  the only  physical
CP-violating phases  in  the MSSM supplemented  by  universal boundary
conditions  at  low  energies.\footnote{Observe that, owing  to the RG
  evolution, even starting  with universal boundary conditions at high
  energies, the low-energy parameters tend to be non-universal.}

Denoting the  scalar  components of the Higgs  superfields $\hat{H}_1$
and $\hat{H}_2$ by $\tilde{\Phi}_1  =  i\tau_2 \Phi^*_1$ ($\tau_2$  is
the usual Pauli  matrix) and $\Phi_2$,  the most general  CP-violating
Higgs  potential  of the MSSM  may  conveniently be   described by the
effective Lagrangian
\begin{eqnarray}
  \label{LV}
{\cal L}_V &=& \mu^2_1 (\Phi_1^\dagger\Phi_1)\, +\, 
\mu^2_2 (\Phi_2^\dagger\Phi_2)\, +\, m^2_{12} (\Phi_1^\dagger \Phi_2)\, 
+\, m^{*2}_{12} (\Phi_2^\dagger \Phi_1)\, +\, 
\lambda_1 (\Phi_1^\dagger \Phi_1)^2\, +\, 
\lambda_2 (\Phi_2^\dagger \Phi_2)^2\,\nonumber\\
&& +\, \lambda_3 (\Phi_1^\dagger \Phi_1)(\Phi_2^\dagger \Phi_2)\, +\, 
\lambda_4 (\Phi_1^\dagger \Phi_2)(\Phi_2^\dagger \Phi_1)\, +\, 
\lambda_5 (\Phi_1^\dagger \Phi_2)^2\, +\, 
\lambda^*_5 (\Phi_2^\dagger \Phi_1)^2\, \\
&&+\, \lambda_6 (\Phi_1^\dagger \Phi_1)(\Phi_1^\dagger \Phi_2)\, +
\lambda^*_6 (\Phi_1^\dagger \Phi_1)(\Phi_2^\dagger \Phi_1)\,
+\, \lambda_7 (\Phi_2^\dagger \Phi_2)(\Phi_1^\dagger \Phi_2)\, +
\lambda^*_7 (\Phi_2^\dagger \Phi_2)(\Phi_2^\dagger \Phi_1)\, .
\nonumber
\end{eqnarray}
At the tree level, the kinematic parameters are given by
\begin{eqnarray}
  \label{LVpar}
\mu^2_1 &=& -m^2_1 - |\mu|^2\, ,\qquad \mu^2_2\ =\ -m^2_2 - |\mu|^2\, ,
                                                            \qquad
\lambda_1\ =\ \lambda_2\ =\ -\, \frac{1}{8}\, (g_w^2 + g'^2)\, ,\nonumber\\
\lambda_3 &=& -\frac{1}{4}\, (g_w^2 -g'^2)\, ,\qquad 
\lambda_4\ =\ \frac{1}{2}\, g_w^2\, , \qquad \lambda_5\ =\ \lambda_6\ 
=\ \lambda_7\ =\ 0\, .
\end{eqnarray}
In Eqs.\ (\ref{LV}) and (\ref{LVpar}), $m^2_1$, $m^2_2$ and $m^2_{12}$
are soft-SUSY-breaking parameters related to the Higgs sector.  Beyond
the Born  approximation, the quartic couplings  $\lambda_5, \lambda_6,
\lambda_7$ receive significant   radiative corrections from  trilinear
Yukawa  couplings of the  Higgs fields to scalar-top and scalar-bottom
quarks.   These parameters    are  in general   complex. The  analytic
expressions of the quartic couplings are given in the appendix.

Our next step  is  to determine the  ground  state of  the MSSM  Higgs
potential.  To this end, we consider  the linear decompositions of the
Higgs fields
\begin{equation}
  \label{Phi12}
\Phi_1\ =\ \left( \begin{array}{c}
\phi^+_1 \\ \frac{1}{\sqrt{2}}\, ( v_1\, +\, \phi_1\, +\, ia_1)
\end{array} \right)\, ,\qquad
\Phi_2\ =\ e^{i\xi}\, \left( \begin{array}{c}
\phi^+_2 \\  \frac{1}{\sqrt{2}}\, ( v_2 \, +\, \phi_2\, +\, ia_2 )
 \end{array} \right)\, ,
\end{equation}
where $v_1$ and $v_2$ are the moduli  of the vacuum expectation values
(VEVs)  of the  Higgs  doublets and $\xi$    is their relative phase.  
Without any  loss of generality, in the  parameterization of the Higgs
doublets in Eq.\ (\ref{Phi12}), we have adopted  a weak basis in which
the VEV $v_1$ ($v_2$) and  the quantum fluctuation $\phi_1$ ($\phi_2$)
have the same phase. Furthermore, we assume the  absence of any CP-odd
component  due   to spontaneous  CP  violation  \cite{BRLW}.  Although
radiative  corrections   can,  in  principle,   lead to  a spontaneous
breakdown of  CP invariance in the MSSM   \cite{NM}, such a particular
scenario, however, predicts an unacceptably  small mass for the CP-odd
Higgs scalar $A$, i.e.\ $M_A < 40$ GeV, and  it is therefore ruled out
experimentally \cite{APNH,KL}.

The VEVs $v_1$ and $v_2$ and the phase  $\xi$ can now be determined by
the minimization conditions  on   ${\cal L}_V$.  This  is  achieved by
requiring that the following tadpole parameters vanish:
\begin{eqnarray}
  \label{Tphi1}
T_{\phi_1} &\equiv& \langle\,\frac{\partial {\cal L}_V}{\partial
  \phi_1}\, \rangle\ =\ 
v\, c_\beta\, \Big\{\, \mu^2_1\ +\ {\rm Re} (m^2_{12}e^{i\xi})\,
  \tan\beta\, +\, v^2\, \Big[\, \lambda_1 c^2_\beta\, +\,
  \frac{1}{2}\, (\lambda_3 + \lambda_4) s^2_\beta\, \nonumber\\
&&+\, {\rm Re}(\lambda_5 e^{2i\xi})s^2_\beta\, +\, 
\frac{3}{2}\, {\rm Re}(\lambda_6 e^{i\xi})s_\beta c_\beta\, +\,
\frac{1}{2}\, {\rm Re}(\lambda_7 e^{i\xi})s^2_\beta\tan\beta\, \Big]\,
  \Big\}\, ,\\
  \label{Tphi2}
T_{\phi_2} &\equiv& \langle\,\frac{\partial {\cal L}_V}{\partial
  \phi_2}\, \rangle\ =\ 
v\, s_\beta\, \Big\{\, \mu^2_2\ +\ {\rm Re} (m^2_{12}e^{i\xi})\,
  \cot\beta\, +\, v^2\, \Big[\, \lambda_2 s^2_\beta\, +\,
  \frac{1}{2}\, (\lambda_3 + \lambda_4) c^2_\beta\, \nonumber\\
&&+\, {\rm Re}(\lambda_5 e^{2i\xi})c^2_\beta\, +\, 
\frac{1}{2}\, {\rm Re}(\lambda_6 e^{i\xi})c^2_\beta\cot\beta\, +\,
\frac{3}{2}\, {\rm Re}(\lambda_7 e^{i\xi})s_\beta c_\beta\, \Big]\,
  \Big\}\, ,\\
  \label{Ta1}
T_{a_1} &\equiv& \langle\,\frac{\partial {\cal L}_V}{\partial a_1}\,\rangle
\ =\ v\, s_\beta\, \Big[\, {\rm Im} (m^2_{12}e^{i\xi})\, +\,
{\rm Im}(\lambda_5 e^{2i\xi})\, v^2 s_\beta c_\beta\, 
+\, \frac{1}{2}\, {\rm Im}(\lambda_6 e^{i\xi}) v^2 c^2_\beta\, \nonumber\\
&&+\, \frac{1}{2}\, {\rm Im}(\lambda_7 e^{i\xi}) v^2 s^2_\beta \, \Big]\, ,\\
  \label{Ta2}
T_{a_2} &\equiv& \langle\,\frac{\partial {\cal L}_V}{\partial a_2}\,\rangle
\ =\ -\, v\, c_\beta\, \Big[\, {\rm Im} (m^2_{12}e^{i\xi})\, +\,
{\rm Im}(\lambda_5 e^{2i\xi})\, v^2 s_\beta c_\beta\, +\, 
\frac{1}{2}\, {\rm Im}(\lambda_6 e^{i\xi}) v^2 c^2_\beta\, \nonumber\\
&& +\, \frac{1}{2}\, {\rm Im}(\lambda_7 e^{i\xi}) v^2 s^2_\beta \, \Big]\, ,
\end{eqnarray}
where $s_x\equiv \sin x$,  $c_x\equiv  \cos x$, $\tan\beta =  v_2/v_1$
and $v^2  =  v^2_1 + v^2_2$.  Furthermore,  we  assume the absence  of
charge-breaking minima, i.e.\  variations of ${\cal L}_V$ with respect
to $\phi^+_1$ and $\phi^+_2$ vanish identically. It is now easy to see
that the orthogonal rotation of the CP-odd fields,
\begin{equation}
  \label{G0A}
\left( \begin{array}{c} a_1 \\ a_2\end{array}\right)\ =\
\left( \begin{array}{cc} \cos\beta & -\sin\beta \\ 
\sin\beta & \cos\beta\end{array}\right)\,
\left( \begin{array}{c} G^0 \\ a\end{array}\right)\, ,
\end{equation}
gives rise to a flat direction of the  Higgs potential with respect to
the  $G^0$ field, i.e.\    $\langle \partial {\cal  L}_V/\partial  G^0
\rangle = 0$. In the newly defined weak basis, the $G^0$ field becomes
the would-be  Goldstone boson, which  is  absorbed by the longitudinal
component   of  the $Z$   boson.  Moreover,   the  orthogonal rotation
(\ref{G0A}) of  the CP-odd fields yields  a non-trivial CP-odd tadpole
parameter:
\begin{eqnarray}  
  \label{Ta}
T_a &\equiv& \langle\,\frac{\partial {\cal L}_V}{\partial a}\,\rangle
\ =\ -\, v\, \Big[\, {\rm Im} (m^2_{12}e^{i\xi})\, +\,
{\rm Im}(\lambda_5 e^{2i\xi})\, v^2 s_\beta c_\beta\, +\, 
\frac{1}{2}\, {\rm Im}(\lambda_6 e^{i\xi}) v^2 c^2_\beta\, \nonumber\\
&& +\, \frac{1}{2}\, {\rm Im}(\lambda_7 e^{i\xi}) v^2 s^2_\beta \, \Big]\, .
\end{eqnarray}
In the CP-invariant  limit of  the  theory, both  $T_a$ and the  phase
$\xi$ vanish. Since $m^2_{12}$ is taken to be  real at the tree level,
a non-zero value of the phase $\xi$ is first generated at the one-loop
level \cite{APLB}.  Nevertheless, in a general two-Higgs-doublet model
with Higgs-sector CP violation, the  phase $\xi$ already occurs in the
Born approximation.   For the sake   of generality, we  shall keep the
full $\xi$ dependence in the analytic results.

It is now interesting to discuss  the conditions under which the Higgs
sector of the  MSSM respects  the CP  symmetry.  We find  that the  CP
invariance of the Higgs potential is assured only if
\begin{equation}
  \label{CPconds}
{\rm Im}( m^4_{12} \lambda^*_5 )\ =\ {\rm Im}( m^2_{12} \lambda^*_6 )\ 
=\ {\rm Im}( m^2_{12} \lambda^*_7 )\ =\ 0\, .
\end{equation}
If  we assume  a kind  of  universality  between  the trilinear Yukawa
couplings at low  energies, $A  = A_t  = A_b$,  and neglect the  small
chargino and neutralino    contributions, the phases of  the   quartic
couplings $\lambda_5$, $\lambda_6$ and $\lambda_7$ are then related to
one another.  Employing the analytic  results of the quartic couplings
given in the Appendix, it is easy to show that
\begin{equation}
  \label{CPlambda}
{\rm Im}(\lambda^*_5 \lambda^2_6 )\ =\  {\rm Im}(\lambda^*_5
\lambda^2_7 )\ = \ 0\, .
\end{equation}
However, even in  this case, the phase  of the  complex soft parameter
$m^2_{12}$  is not  restricted  by  any  universal boundary  condition
imposed  by minimal supergravity models at  the unification point.  In
other words, the  CP  invariance of  the Higgs potential  of the  MSSM
holds true only if the condition
\begin{equation}
  \label{CPinv}
{\rm Im} ( m^{*2}_{12}\, \mu\, A )\ =\ 0\, 
\end{equation}
is satisfied.    Within the most  general  framework of  the MSSM, the
equality  (\ref{CPinv}) [or    equivalently (\ref{CPconds})]   can  be
violated, thus giving rise to observable CP violation.

CP violation in the Higgs  potential of the  MSSM leads to mixing mass
terms  between the CP-even and CP-odd  Higgs fields \cite{APLB}. Thus,
one has to consider  a $(4\times 4)$-dimensional  mass matrix for  the
neutral Higgs  bosons. In the  weak basis $(G^0,  a, \phi_1, \phi_2)$,
the neutral  Higgs-boson mass matrix ${\cal M}^2_0$  may  be cast into
the form
\begin{equation}
  \label{NHiggs}
{\cal M}^2_0 \ =\ 
\left(\begin{array}{cc} \widehat{\cal M}^2_P  & {\cal M}^2_{PS} \\
                {\cal M}^2_{SP} &  {\cal M}^2_S \end{array} \right)\, ,
\end{equation}
where  $\widehat{\cal   M}^2_P$  and   ${\cal  M}^2_S$  describe   the
CP-conserving transitions  $(G^0,a)\to (G^0,a)$ and   $(\phi_1,\phi_2)
\to  (\phi_1,\phi_2)$,  respectively,  and ${\cal   M}^2_{PS} = ({\cal
  M}^{2}_{SP})^T$   contains   the   CP-violating   mixings   $(G^0,a)
\leftrightarrow (\phi_1,\phi_2)$. The analytic form of the submatrices
is given by
\begin{eqnarray}
  \label{M2Phat}
\widehat{\cal M}^2_P & =&  \left( \begin{array}{cc} 
-\, \frac{\displaystyle c_\beta T_{\phi_1} + s_\beta
  T_{\phi_2}}{\displaystyle  v} 
& \frac{\displaystyle s_\beta T_{\phi_1} - c_\beta 
T_{\phi_2}}{\displaystyle v}\\
\frac{\displaystyle s_\beta T_{\phi_1} - c_\beta T_{\phi_2}}{\displaystyle v}&
\quad M^2_a\, -\, 
\frac{\displaystyle s_\beta\tan\beta\, T_{\phi_1} +
c_\beta\cot\beta\, T_{\phi_2}}{\displaystyle v} 
\end{array} \right)\, ,\\[0.35cm]
  \label{M2SP}
{\cal M}^2_{SP} &=& v^2\! 
\left( \begin{array}{cc} 0 & {\rm Im}(\lambda_5 e^{2i\xi}) s_\beta\,
    +\, {\rm Im}(\lambda_6 e^{i\xi}) c_\beta \\
    0 & {\rm Im}(\lambda_5 e^{2i\xi}) c_\beta\,
    +\, {\rm Im}(\lambda_7 e^{i\xi}) s_\beta \end{array}\right)\, 
-\, \frac{T_a}{v} \left(\begin{array}{cc} 
s_\beta & c_\beta \\ - c_\beta & s_\beta \end{array} \right)\, ,\\[0.35cm]
  \label{M2S}
{\cal M}^2_S &=& M^2_a\, \left( \begin{array}{cc} 
 s^2_\beta & -s_\beta c_\beta \\ 
 -s_\beta c_\beta & c^2_\beta \end{array} \right)\, -\, 
\left( \begin{array}{cc} 
 \frac{\displaystyle T_{\phi_1}}{\displaystyle v\,c_\beta} & 0  \\ 
 0 & \frac{\displaystyle T_{\phi_2}}{\displaystyle v\,s_\beta} 
\end{array} \right)\ \\
 &&\hspace{-1.95cm} -v^2\! \left(\! \begin{array}{cc} 
 2\lambda_1 c^2_\beta + 2{\rm Re}(\lambda_5e^{2i\xi}) s^2_\beta
+ 2 {\rm Re}(\lambda_6 e^{i\xi}) s_\beta c_\beta  & 
\lambda_{34} s_\beta c_\beta + 
{\rm Re}(\lambda_6 e^{i\xi}) c^2_\beta + 
{\rm Re}(\lambda_7 e^{i\xi}) s^2_\beta \\
\lambda_{34} s_\beta c_\beta + 
{\rm Re}(\lambda_6 e^{i\xi}) c^2_\beta + 
{\rm Re}(\lambda_7 e^{i\xi}) s^2_\beta &
 2\lambda_2 s^2_\beta + 2{\rm Re}(\lambda_5 e^{2i\xi}) c^2_\beta
+ 2 {\rm Re}(\lambda_7 e^{i\xi}) s_\beta c_\beta
 \end{array}\!\right)\!.\nonumber
\end{eqnarray}
In    Eqs.\  (\ref{M2Phat})  and (\ref{M2S}),    we    have used   the
abbreviations $\lambda_{34} = \lambda_3 + \lambda_4$ and
\begin{equation}
\label{SMa}
M^2_a\ =\ \frac{1}{s_\beta c_\beta}\, 
\Big\{ {\rm Re}( m^2_{12} e^{i\xi}) + v^2 \Big[ 
2{\rm Re}(\lambda_5 e^{2i\xi}) s_\beta c_\beta +
\frac{1}{2}\, {\rm Re}(\lambda_6 e^{i\xi}) c^2_\beta
+ \frac{1}{2}\, {\rm Re}(\lambda_7 e^{i\xi}) s^2_\beta \Big]\, \Big\}\,.
\end{equation}
If  the MSSM is  invariant under the  CP symmetry,  $M_a$  is then the
physical mass of the CP-odd Higgs scalar \cite{HH}.

Correspondingly, the  charged Higgs-boson  mass matrix  $\widehat{\cal
  M}^2_\pm$, spanned in  the   mass  basis $(G^\pm,  H^\pm)$,  may  be
obtained by the Lagrangian
\begin{equation}
  \label{LHpm}
{\cal L}^\pm_{\rm mass}\ =\  -\, (G^+,\ H^+)\ \widehat{\cal M}^2_\pm\, 
\left( \begin{array}{c} G^- \\ H^- \end{array} \right)\, ,
\end{equation}
with
\begin{equation}
  \label{MHpm}
\widehat{\cal M}^2_\pm\ =\ 
\left( \begin{array}{cc} 
-\, \frac{\displaystyle c_\beta T_{\phi_1} + s_\beta
  T_{\phi_2}}{\displaystyle  v} 
& \frac{\displaystyle s_\beta T_{\phi_1} - c_\beta 
T_{\phi_2}}{\displaystyle v} 
- i\, \frac{\displaystyle T_a}{\displaystyle v} \\
\frac{\displaystyle s_\beta T_{\phi_1} - c_\beta
  T_{\phi_2}}{\displaystyle v} + i\,\frac{\displaystyle T_a}{\displaystyle v}&
\quad M^2_{H^\pm}\, -\, \frac{\displaystyle s_\beta\tan\beta\, 
T_{\phi_1} + c_\beta\cot\beta\, T_{\phi_2}}{\displaystyle v} 
\end{array} \right)\, 
\end{equation}
and 
\begin{eqnarray}
  \label{MHplus}
M^2_{H^\pm} &=& \frac{1}{s_\beta c_\beta}\, 
\Big\{ {\rm Re}( m^2_{12} e^{i\xi}) + v^2 \Big[\,
\frac{1}{2}\,\lambda_4 s_\beta c_\beta +
{\rm Re}(\lambda_5 e^{2i\xi}) s_\beta c_\beta +
\frac{1}{2}\, {\rm Re}(\lambda_6 e^{i\xi}) c^2_\beta \nonumber\\
&&+ \frac{1}{2}\, {\rm Re}(\lambda_7 e^{i\xi}) s^2_\beta \Big]\,
\Big\}\, .
\end{eqnarray}
{}From  Eqs.\  (\ref{SMa})    and (\ref{MHplus}),   we  observe   that
$M^2_{H^\pm}$  is  related to  the  mass of the would-be  CP-odd Higgs
scalar $M^2_a$ through
\begin{equation}
  \label{MaMH}
M^2_a\ =\ M^2_{H^\pm}\ -\ \frac{1}{2}\, \lambda_4 v^2\ +\
{\rm Re}(\lambda_5 e^{2i\xi}) v^2\, .
\end{equation}
Taking  this very  last  relation into  account,   we may express  the
neutral Higgs-boson masses  as functions  of $M_{H^+}$, $\mu$,  $A_t$,
$A_b$, the common SUSY   scale  $M_{\rm SUSY}$, $\tan\beta$  and   the
physical  phase $\xi$.    Since we  neglect  chargino and   neutralino
contributions, we can absorb the radiatively  induced phase $\xi$ into
the definition of the $\mu$ parameter. 

It is  worth stressing that, even though  CP violation  decouples from
the  sector   of  the  lightest  Higgs  boson    for large  values  of
$M_{H^+}\approx M_a$, this decoupling  property of CP non-conservation
does  not generally persist for  the system of  the two heaviest Higgs
bosons. This point  may  formally be  seen as  follows:  in the  large
$M^2_a$ limit, assuming that the quartic couplings are kept fixed, the
submatrix ${\cal  M}^2_S$   given in  Eq.\  (\ref{M2S})  has  one mass
eigenvalue which approaches  $M^2_a$, and corresponds  to the heaviest
CP-even Higgs  boson, while the other one  is small at the electroweak
scale, related to the lightest CP-even  Higgs boson.  Furthermore, the
submatrix $\widehat{\cal  M}^2_P$ in Eq.\  (\ref{M2Phat}) has only one
non-zero  mass eigenvalue equal  to $M^2_a$, corresponding to the mass
of the would-be CP-odd Higgs scalar.  Thus,  for large $M^2_a$ values,
it is  easy  to   see that the   effective  $(2\times  2)$-dimensional
submatrix, which may be formed by the would-be CP-odd and the heaviest
CP-even  Higgs  bosons, also  contains off-diagonal CP-violating terms
coming from  $M^2_{SP}$   in Eq.\  (\ref{M2SP}).    These off-diagonal
CP-violating  matrix elements are   generically  of the  order  of the
difference of the diagonal entries of the effective submatrix, thereby
giving rise to a strong mixed system of CP  violation.  As we will see
in Section 5, numerical  estimates of CP  violation in the heavy Higgs
sector offer firm support of this observation.

\setcounter{equation}{0}
\section{Higgs-boson masses and mixing angles}

In this section, we shall evaluate the  physical masses of the neutral
Higgs bosons and  the mixing angles related  to the diagonalization of
the general $4\times 4$ matrix ${\cal M}^2_0$  in Eq.\ (\ref{NHiggs}). 
Even though  our primary  interest is  in the  CP-violating  MSSM, the
validity of the analytic expressions that we shall derive here extends
to the  most general class  of CP-violating two-Higgs-doublet  models. 
After setting all tadpole parameters to zero, we easily see that $G^0$
does  not  mix with  the  other  neutral  fields   and so becomes   an
independent  massless  field, as   it should  be    on account  of the
Goldstone  theorem  \cite{APLB}.  Then,   ${\cal  M}^2_0$  effectively
reduces to a $(3 \times  3)$-dimensional matrix, ${\cal M}^2_N$, which
is spanned in the weak basis $(a,\phi_1,\phi_2)$.

The  mass eigenvalues  of ${\cal M}^2_N$   are obtained by solving the
characteristic equation of cubic order
\begin{equation}
  \label{lcubic}
x^3\ +\ r x^2 \ +\ s x\ +\ t\ =\ 0\, ,
\end{equation}
with 
\begin{eqnarray}
  \label{trace}
r & = & -\, {\rm Tr} ({\cal M}^2_N )\, ,\nonumber\\
s & = & \frac{1}{2}\, \Big[\, {\rm Tr}^2 ({\cal M}^2_N )\ -\
{\rm Tr} ({\cal M}^4_N )\, \Big]\, ,\nonumber\\
t &=& -\, {\rm det} ({\cal M}^2_N )\, . 
\end{eqnarray}
To  this   end, it proves  useful to    define the following auxiliary
parameters:
\begin{eqnarray}
  \label{p}
p &=& \frac{3s\, -\, r^2}{3}\ ,\nonumber\\
  \label{q}
q &=& \frac{2r^3}{27}\ -\ \frac{rs}{3}\, +\, t\, ,\nonumber\\
  \label{D}  
D &=& \frac{p^3}{27}\ +\ \frac{q^2}{4}\, .
\end{eqnarray}
To ensure that the three eigenvalues are positive, it is necessary and
sufficient to require that
\begin{equation}
  \label{condition}
D\ <\ 0\, ,\qquad r\ <\ 0\, ,\qquad s\ >\ 0\, ,\qquad t\ <\ 0\, .   
\end{equation}
Imposing these inequalities on the kinematic parameters of the theory,
we may express the three mass eigenvalues of ${\cal M}^2_N$ as
\begin{eqnarray}
  \label{e1}
\rho^2_1 & =& -\frac{1}{3}\, r\ +\ 2\, \sqrt{-p/3}\, 
\cos\Big(\, \frac{\varphi}{3}\, \Big)\, ,\nonumber\\
\rho^2_2 & =& -\frac{1}{3}\, r\ +\ 2\, \sqrt{-p/3}\, 
\cos\Big(\, \frac{\varphi}{3}\, +\, \frac{2\pi}{3}\, \Big)\, ,\nonumber\\
\rho^2_3 & =& -\frac{1}{3}\, r\ +\ 2\, \sqrt{-p/3}\, 
\cos\Big(\, \frac{\varphi}{3}\, -\, \frac{2\pi}{3}\,\Big)\, ,
\end{eqnarray}
with
\begin{equation}
  \label{phi}
\varphi \ =\  {\rm arccos}\, \Big( - \frac{q}{2\sqrt{-p^3/27}} \Big)\, .
\end{equation}

Since the Higgs-boson mass matrix ${\cal  M}^2_N$ is symmetric, we can
diagonalize it by means of an orthogonal rotation $O$ as follows:
\begin{equation}
  \label{Odiag}
O^T\, {\cal M}^2_N\, O\ =\ {\rm diag}\, ( M^2_{H_3},\ M^2_{H_2},\ 
M^2_{H_1})\ .   
\end{equation}
Note that some arbitrariness  exists in assigning the mass eigenvalues
$\rho_1$, $\rho_2$ and $\rho_3$ in Eq.\ (\ref{e1}) to those related to
the mass eigenfields $H_1$, $H_2$ and $H_3$ in Eq.\ (\ref{Odiag}).  For
clarity of the presentation, we define these fields such that
\begin{equation}
M_{H_1}\ \leq M_{H_2}\ \leq M_{H_3}\, .
\end{equation}
Alternatively, these  fields  could be defined in  such  a way that we
have,   in  the CP-invariant   limit  of the   theory, $H_1\equiv  h,\ 
H_2\equiv H,\ H_3\equiv A$, where $h$ and $H$  denote the lightest and
heaviest  CP-even Higgs bosons, respectively,  and  $A$ is the  CP-odd
Higgs scalar.   However, we should  use the  former definition, as the
latter often  leads to discontinuities in the  $H_i$ mass values, when
plotted as a function of the MSSM parameters.

In general, the  orthogonal matrix $O$   in Eq.\ (\ref{Odiag})  can be
described  in  terms of the  three physical  Euler-type angles $\chi$,
$\psi$ and $\theta$.  We  parameterize $O$, assuming $\chi,\psi \to 0$
or $\pi/2$ in the CP-conserving limit of the theory, as follows:
\begin{equation}
  \label{Orth}
O\ =\ \left( \begin{array}{ccc}
c_\chi c_\psi & - s_\chi c_\theta - s_\psi c_\chi s_\theta &
s_\chi s_\theta - s_\psi c_\chi c_\theta \\
s_\chi c_\psi & c_\chi c_\theta - s_\psi s_\chi s_\theta &
-c_\chi s_\theta - s_\psi s_\chi c_\theta \\
s_\psi & s_\theta c_\psi & c_\theta c_\psi \end{array} \right)\, .
\end{equation}
It is  convenient  to  find first  the    entries $O_{ij}$,  and  then
determine the three rotational angles by the obvious relations:
\begin{equation}
  \label{angles}
\psi\ =\ {\rm arcsin}\, (O_{31})\, ,\qquad
\chi\ =\ {\rm arcsin}\, \Big(\,\frac{O_{21}}{\cos\psi}\,\Big)\, ,\qquad
\theta\ =\ {\rm arcsin}\, \Big(\,\frac{O_{32}}{\cos\psi}\,\Big)\, ,  
\end{equation}
where  the mixing  angles take  on  values  in the  interval $(-\pi/2,
\pi/2]$.

If $M^2_{ij}$,  with   $i,j=1, 2,  3$, denote the   matrix elements of
${\cal   M}^2_N$,   the elements $O_{ij}$  can     then be obtained by
appropriately solving the underdetermined coupled system of equations,
$\sum_k M^2_{ik} O_{kj} = M^2_{H_{(4-j)}} O_{ij}$:
\begin{eqnarray}
  \label{system}
(M^2_{11} - M^2_{H_{(4-i)}} ) O_{1i}\, +\, M^2_{12} O_{2i}\, +\, 
M^2_{13} O_{3i} & =& 0\,, \nonumber\\ 
M^2_{21} O_{1i}\, +\, ( M^2_{22} - M^2_{H_{(4-i)}} ) O_{2i}\, +\, 
M^2_{23} O_{3i} & =& 0\,, \nonumber\\ 
M^2_{31} O_{1i}\, +\, M^2_{32}  O_{2i}\, +\, 
( M^2_{33} - M^2_{H_{(4-i)}} ) O_{3i} & =& 0\, . 
\end{eqnarray}
More explicitly, we have
\begin{equation}
  \label{Oij}
O \ =\   \left( \begin{array}{ccc}
|x_1|/\Delta_1 & x_2/\Delta_2 & x_3/\Delta_3 \\
y_1/\Delta_1 & |y_2|/\Delta_2 & y_3/\Delta_3 \\
z_1/\Delta_1 & z_2/\Delta_2 & |z_3|/\Delta_3 \end{array} \right)\, ,
\end{equation}
where
\begin{equation}
  \label{Deltas}
\Delta_i = \sqrt{ x^2_i\, +\, y^2_i\, +\, z^2_i} 
\end{equation}
and 
{\small \begin{eqnarray}
  \label{xyz}
|x_1| \!\!\!&=&\!\!\! \left|\!\left| \begin{array}{lr}
M^2_{22}-M^2_{H_3} & M^2_{23} \\
M^2_{32} & \hspace{-1.2cm} M^2_{33} - M^2_{H_3} \end{array}\right|\!\right|
,\
y_1 = {\rm s}_{x_1}\! \left| \begin{array}{cc}
M^2_{23} & M^2_{21} \\
M^2_{33} -M^2_{H_3} & M^2_{31} \end{array} \right|,\
z_1 = {\rm s}_{x_1}\! \left| \begin{array}{cc}
M^2_{21} & M^2_{22} - M^2_{H_3} \\
M^2_{31} & M^2_{32} \end{array} \right|,\nonumber\\
x_2 \!\!\!&=&\!\!\! {\rm s}_{y_2}\! \left| \begin{array}{cc}
M^2_{13} & M^2_{12} \\
M^2_{33} - M^2_{H_2} & M^2_{32} \end{array} \right|,\
|y_2| = \left|\!\left| \begin{array}{lr}
M^2_{11} - M^2_{H_2} & M^2_{13} \\
M^2_{31}  & \hspace{-1.2cm} M^2_{33} - M^2_{H_2} \end{array} 
\right|\!\right|,\ 
z_2 = {\rm s}_{y_2}\! \left| \begin{array}{cc}
M^2_{12} & M^2_{11} - M^2_{H_2} \\
M^2_{32} & M^2_{31} \end{array} \right|,\nonumber\\
x_3 \!\!\!&=&\!\!\! {\rm s}_{z_3}\! \left| \begin{array}{cc}
M^2_{12} & M^2_{13} \\
M^2_{22} - M^2_{H_1} & M^2_{23} \end{array} \right|,\
y_3 = {\rm s}_{z_3}\! \left| \begin{array}{cc}
M^2_{13} & M^2_{11} - M^2_{H_1} \\
M^2_{23} & M^2_{21} \end{array} \right|,\ 
|z_3| = \left|\!\left| \begin{array}{lr}
M^2_{11} - M^2_{H_1} & M^2_{12} \\
M^2_{21} & \hspace{-1.2cm} M^2_{22} - M^2_{H_1} \end{array}
\right|\!\right|.\nonumber\\ 
&&
\end{eqnarray}  } 
In  Eq.\ (\ref{xyz}), we have  used the abbreviation ${\rm s}_x \equiv
{\rm sign}\, (x)$, which is an operation that simply gives the sign of
a real expression  $x$.  Notice  that  the parameterization of  $O$ in
terms  of   $x_i,\ y_i,\  z_i$  may  be chosen,  in  a  way  such that
indefinite expressions do not occur in the  CP-conserving limit of the
theory.

\setcounter{equation}{0}
\section{EDM constraints}

As we mentioned  in the introduction, the  EDM of the electron and the
neutron provide the   most stringent constraints  on the  CP-violating
parameters  of the  MSSM \cite{EDM,DGH,FOS,KO,IN,AF,CKP}.  There  have
been several suggestions to suppress the EDM contributions coming from
the   first two   families    of scalar  quarks without    making  the
CP-violating phases of the theory very  small \cite{KO,IN,AF,CKP}.  To
be specific, the following  three possibilities may be considered: one
can make the  first two  families  of  scalar fermions rather   heavy,
having a  mass  of order  few  TeV \cite{KO}.    Another option is  to
arrange for    partial   cancellations  among   the   different    EDM
contributions \cite{IN}.  In this case, the CP-violating phases of the
theory  turn out  to  be rather  correlated.  Finally, an  interesting
alternative  is  to adopt a slightly    non-universal scenario for the
trilinear  couplings   $A_f$ \cite{AF,CKP}.   In  particular, one  may
require that ${\rm  arg}  (\mu ) < 10^{-2}$   and  $A_f =   (0,0,1) A$
\cite{CKP}.  In the latter scheme,  $A_\tau = A_t  = A_b$ are the only
large  trilinear couplings in the theory   with CP-violating phases of
order unity. Furthermore, assuming that gluinos are heavier than about
400 GeV \cite{IN},  one may significantly reduce the  size of the  EDM
effect due to Weinberg's three-gluon operator \cite{SW} well below the
present experimental bound.

%******************************************************************
%%% The EDM Figure 
%******************************************************************
\begin{figure}

\begin{center}
\begin{picture}(400,200)(0,0)
\SetWidth{0.8}
 
\ArrowLine(10,30)(45,60)\Text(30,35)[lb]{$f$}
\DashLine(45,60)(68,85){5}\Text(55,75)[rb]{$a$}
\Photon(92,85)(115,60){3}{3}\Text(105,80)[l]{$\gamma,g$}
\ArrowLine(45,60)(115,60)\Text(80,51)[l]{$f$}
\ArrowLine(115,60)(150,30)\Text(118,35)[lb]{$f$}
\Photon(80,120)(80,150){3}{3}\Text(86,140)[l]{$\gamma,g$}
\DashArrowArc(80,100)(20,0,360){3}
\Text(55,100)[r]{$\tilde{t},\tilde{b},(\tilde{\tau})$} 

\Text(75,20)[]{\bf (a)}

\ArrowLine(210,30)(245,60)\Text(230,35)[lb]{$f$}
\DashLine(245,60)(268,85){5}\Text(255,75)[rb]{$a$}
\Photon(301,100)(315,60){3}{4}\Text(315,80)[l]{$\gamma,g$}
\Photon(301,100)(315,130){3}{3}\Text(292,130)[l]{$\gamma,g$}
\ArrowLine(245,60)(315,60)\Text(280,51)[l]{$f$}
\ArrowLine(315,60)(350,30)\Text(318,35)[lb]{$f$}
\DashArrowArc(280,100)(20,0,360){3}
\Text(255,100)[r]{$\tilde{t},\tilde{b},(\tilde{\tau})$} 

\Text(275,20)[]{\bf (b)}

\end{picture}
\end{center}
\vspace{-1.cm}
\caption{Two-loop contribution to EDM and CEDM of a light fermion $f$
  in  the CP-violating MSSM (mirror-symmetric  graphs are  not shown). 
  Note that $\tilde{\tau}$ does not contribute to the CEDM of a coloured
  fermion $f$.}\label{f1}
\end{figure}
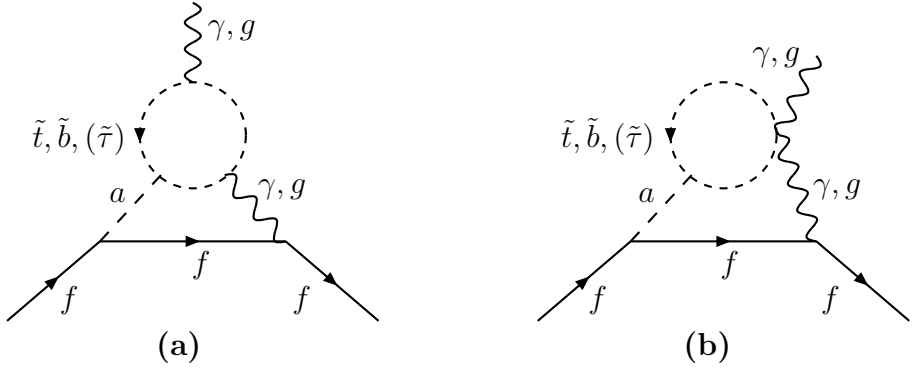

In   the aforementioned    SUSY   scenarios,  however,   the  two-loop
Barr--Zee-type \cite{BZ}  contribution  to the  electron  and  neutron
EDMs,   shown  in  Fig.\ \ref{f1}, can     still be potentially  large
\cite{CKP}.  It is important to notice that the very same kind of loop
graphs generated by scalar top and bottom  quarks are also responsible
for CP violation in the  Higgs sector of  the MSSM.  Therefore, we are
compelled to take these  two-loop EDM constraints into  consideration. 
The graphs displayed  in  Fig.\  \ref{f1}  give  rise to both  an  EDM
($d_f/e$) and a chromo-EDM (CEDM) ($d^C_f/g_s$)  of a light (coloured)
fermion $f$, i.e.\ 
\begin{eqnarray} 
  \label{EDMf}   
\frac{d_f}{e} &=& Q_f\,   
\frac{\alpha_{\rm em}}{64\pi^3}\, \frac{R_f\, m_f}{M^2_a}
 \sum_{\tilde{f} = \tilde{t},\tilde{b},\tilde{\tau}}\ 
N^c_{\tilde{f}}\, \xi_{\tilde{f}}\, Q^2_{\tilde{f}} \,\Big[\, 
F\Big(\, \frac{M^2_{\tilde{f}_1}}{M^2_a}\, \Big)\
 -\
F\Big(\, \frac{M^2_{\tilde{f}_2}}{M^2_a}\, \Big)\, \Big]\, ,\\
  \label{CEDMf}   
\frac{d^C_f}{g_s} &=&
\frac{\alpha_s}{128\pi^3}\, \frac{R_f\, m_f}{M^2_a}\ 
\sum_{\tilde{q} = \tilde{t},\tilde{b}}\ 
\xi_{\tilde{q}}\, \Big[\, 
F\Big(\, \frac{M^2_{\tilde{q}_1}}{M^2_a} \,\Big)\ -\ 
F\Big(\, \frac{M^2_{\tilde{q}_2}}{M^2_a}\, \Big)\, 
       \Big]\, ,
\end{eqnarray}
where  $Q_f$ ($Q_{\tilde{f}}$) stands for   the  electric charge of  a
(scalar) fermion given in $|e|$ units, $N^c_{\tilde{f}}$ is the colour
factor of the scalar  fermion $\tilde{f}$ ($N^c_{\tilde{\tau}} =1$ and
$N^c_{\tilde{t}} = N^c_{\tilde{b}} =   3$), $R_f = \cot\beta$ for  the
up-family  fermions, $R_f = \tan\beta$ for  the  down-family ones.  In
addition, we have defined
\begin{eqnarray} 
  \label{xif}
\xi_{\tilde{f}} &=& -\, R_f\, \frac{\sin 2\theta_f  m_f {\rm Im} ( \mu
  e^{i\delta_f})}{\sin\beta \cos\beta\, v^2}\ ,\\
  \label{Fz}
F(z) &=& \int_0^{1} dx\ \frac{x(1-x)}{z - x(1-x)}\ 
\ln \Big[\,\frac{x(1-x)}{z}\,\Big]\, .
\end{eqnarray}
In Eq.\  (\ref{xif}), $\delta_f  =  {\rm  arg} (A_f   -  R_{\tilde{f}}
\mu^*)$ and $\theta_f$ indicates  the  mixing angle between weak   and
mass eigenstates of $\tilde{f}$.   Since the off-diagonal elements  of
the scalar-quark and lepton mass  matrices of the third generation may
be  larger  than   the  difference of  the   diagonal  entries, angles
$\theta_f$ close to $45^\circ$ are obtained in a natural way.  Further
discussion and   more details of  the   calculation may  be  found  in
\cite{CKP}.

The present status of measurements of the electron and neutron EDMs is
as follows \cite{PDG}:
\begin{eqnarray}
  \label{EDMe1}
\Big(\, \frac{d_e}{e}\, \Big)_{\rm exp} &=& (-0.27 \pm 0.83)
\times 10^{-26}\ {\rm cm}\ \mbox{\cite{abdullah}},\\ 
  \label{EDMe2}
\Big(\, \frac{d_e}{e}\, \Big)_{\rm exp} &=& (0.18 \pm 0.12 \pm 0.10)
\times 10^{-26}\ {\rm cm}\ \mbox{\cite{commins}},\\ 
  \label{EDMn}
\Big(\, \frac{d_n}{e}\, \Big)_{\rm exp} &=& (0.26 \pm 0.40 \pm 0.16)
\times 10^{-25}\ {\rm cm}\ \mbox{\cite{altarev}}.
\end{eqnarray}
The experimental numbers listed here  contain an amount of theoretical
uncertainty originating from the model used  for the heavy atoms, such
as  $^{105}{\rm   Tl}$,  or from  the   description of  neutron's wave
function  in  the heavy   nucleus   \cite{FPT}.  Notwithstanding   the
possible uncertainties in the determination of $d_e$, we should regard
the 1$\sigma$ upper  bound, $|d_e/e| <  1.1\times 10^{-26}$ cm, stated
in   Eq.\ (\ref{EDMe1}) as  a conservative  one,  when compared to the
improved bound  in Eq.\  (\ref{EDMe2}).  In particular,  the 2$\sigma$
upper bound on the  electron EDM coming  from the latter  experimental
analysis is $|d_e/e| < 0.5\times 10^{-26}$ cm.  Finally, the 1$\sigma$
and  2$\sigma$ upper   bounds  on the    neutron EDM  are  $|d_n/e|  <
0.69\times   10^{-25}$ cm  and  $|d_n/e|   < 1.12\times 10^{-25}$  cm,
respectively.

%%%TABLE 1
\begin{table}[t]

\begin{center}

\begin{tabular}{|cc||ccc|| ccc|}
\hline
$\tan\beta$ & $M_{\rm SUSY}$ & & $|d_e/e|$ & $[\,10^{-27}\ {\rm cm}\,]$ &
& $|d_n/e|$ & $[\,10^{-26}\ {\rm cm}\,]$ \\
& $[\,{\rm TeV}\,]$ & $\mu =$ 0.5, & ~~~~~1,~~~~~ & 2~TeV & $\mu =$ 0.5, 
& ~~~~~1,~~~~~ & 2~TeV\\ 
\hline\hline
   & 0.5 & 0.7 & 1.6 & 5.9 & 0.7 & 1.5 & 5.6 \\
2  & 0.6 & 0.3 & 0.7 & 1.6 & 0.3 & 0.7 & 1.5 \\
   & 0.7 & 0.2 & 0.4 & 0.8 & 0.2 & 0.4 & 0.8 \\
\hline
   & 0.5 & 1.2 & 2.5 & 5.5 & 1.1 & 2.2 & 4.9 \\
4  & 0.6 & 0.6 & 1.1 & 2.3 & 0.5 & 1.0 & 2.1 \\
   & 0.7 & 0.3 & 0.6 & 1.3 & 0.3 & 0.6 & 1.2 \\
\hline
   & 0.5 & 2.9 & 5.8 & 12. & 2.6 & 5.2 & 11. \\
10 & 0.6 & 1.3 & 2.7 & 5.4 & 1.2 & 2.4 & 4.8 \\
   & 0.7 & 0.7 & 1.5 & 3.0 & 0.7 & 1.3 & 2.7 \\
\hline
   & 0.5 & 6.0 & 12. & 24. & 5.5 & 11. & 22. \\
20 & 0.6 & 2.8 & 5.5 & 11. & 2.6 & 5.1 & 10. \\
   & 0.7 & 1.5 & 3.1 & 6.2 & 1.4 & 2.9 & 5.7 \\
\hline
\end{tabular}
\end{center}

\caption{Numerical predictions for the electron and neutron EDMs in the 
MSSM, using $|A| = |A_t| = |A_b| = |A_\tau| = 1$ TeV, ${\rm arg} (A) = 
90^\circ$  and   $M_a = 150$  GeV ($m_b (M_Z) = 3$ GeV).}\label{Tab1}
\end{table}

It is interesting to confront the theoretical predictions obtained for
$d_e/e$ and $d_n/e$ in the  MSSM  with the corresponding  experimental
bounds mentioned above.  In  Table  \ref{Tab1}, we present   numerical
estimates for the case $|A| = |A_t| = |A_b| = |A_\tau| = 1$ TeV, ${\rm
  arg}   (A)  = 90^\circ$  and $M_a  =  150$  GeV, while the kinematic
parameters $\tan\beta$,  $\mu$,   $M_{\rm  SUSY}$  have    been varied
discretely.  We  should note  that   the largest  contribution to  the
neutron EDM comes    from the CEDM   of the  $d$  quark.  Furthermore,
scalar-top quarks  have the biggest quantum effect  on the $e$ and $n$
EDMs   for  $2  \simlt   \tan\beta \simlt  30$.  Taking the  numerical
estimates  in  Table \ref{Tab1} into  account,   we shall consider the
following three representative scenarios:
\begin{eqnarray}
  \label{scenarios}
{\rm I.} && M_{\rm SUSY} = 0.5\ {\rm TeV}\,,\quad  |A| = 1\ {\rm TeV}\,,\quad 
\mu = 2\ {\rm TeV}\,,\quad \tan\beta = 2\,, \nonumber\\   
{\rm II.} && M_{\rm SUSY} = 0.5\ {\rm TeV}\,,\quad  |A| = 1\ {\rm TeV}\,,\quad 
\mu = 2\ {\rm TeV}\,,\quad \tan\beta = 4\,, \nonumber\\   
{\rm III.} && M_{\rm SUSY} = 0.5\ {\rm TeV}\,,\quad  
|A| = 1\ {\rm TeV}\,,\quad \mu = 1\ {\rm TeV}\,,\quad \tan\beta = 20\, .
\end{eqnarray}
In   the  next section,   we     shall analyze  the   phenomenological
consequences of  the scenarios given  by Eq.\ (\ref{scenarios}) on the
currently operating and future high-energy colliders.

\begin{figure}
   \leavevmode
 \begin{center}
   \epsfxsize=16.0cm
    \epsffile[0 0 539 652]{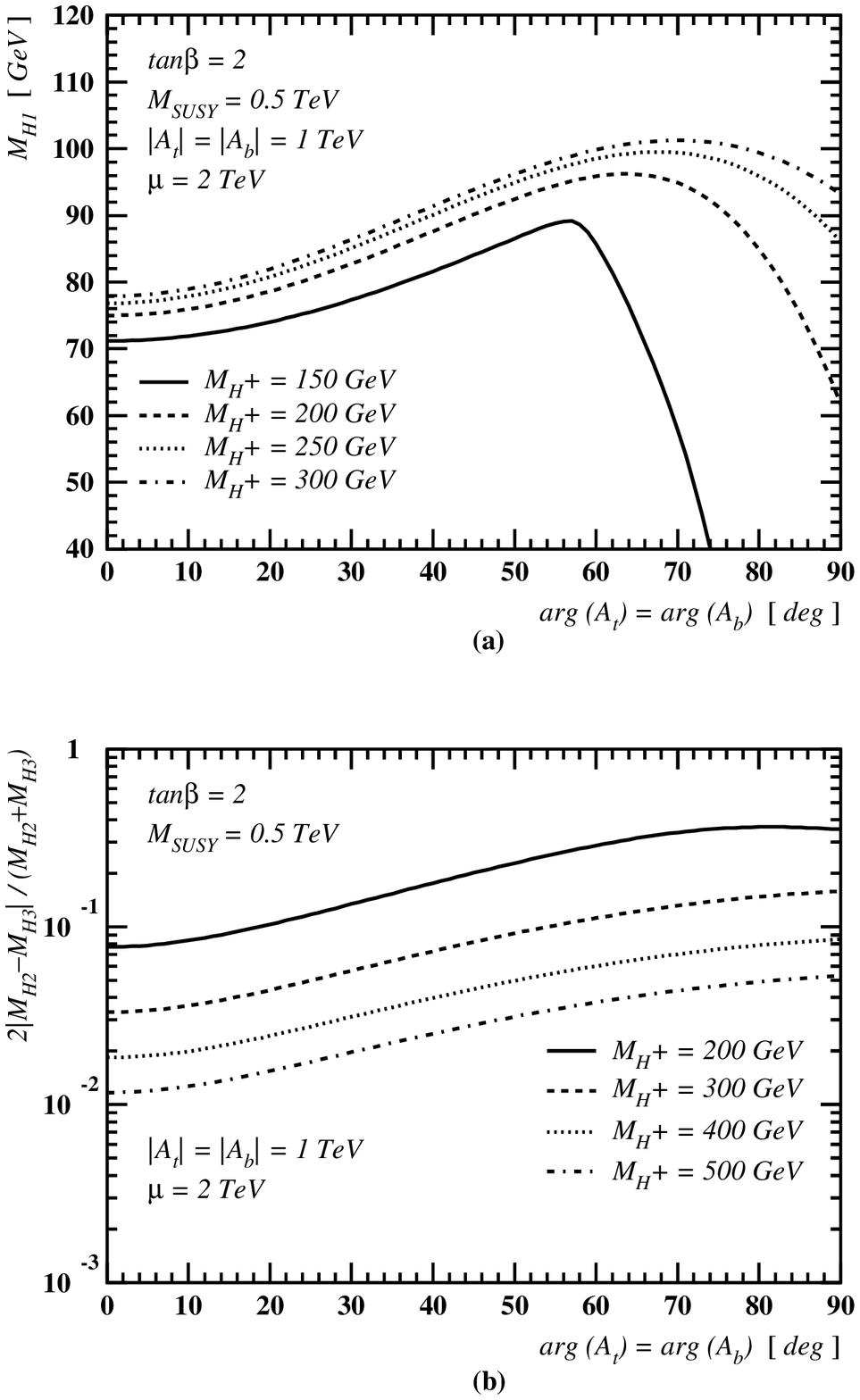}
 \end{center}
 \vspace{-0.5cm} 
\caption{Numerical estimates of (a) $M_{H_1}$ and 
  (b) $2|M_{H_2}-M_{H_3}|/(M_{H_2} + M_{H_3})$ as a function of the
  CP-violating phase arg($A_t$).}\label{fig:scp1}
\end{figure}
\begin{figure}
   \leavevmode
 \begin{center}
   \epsfxsize=16.0cm
    \epsffile[0 0 539 652]{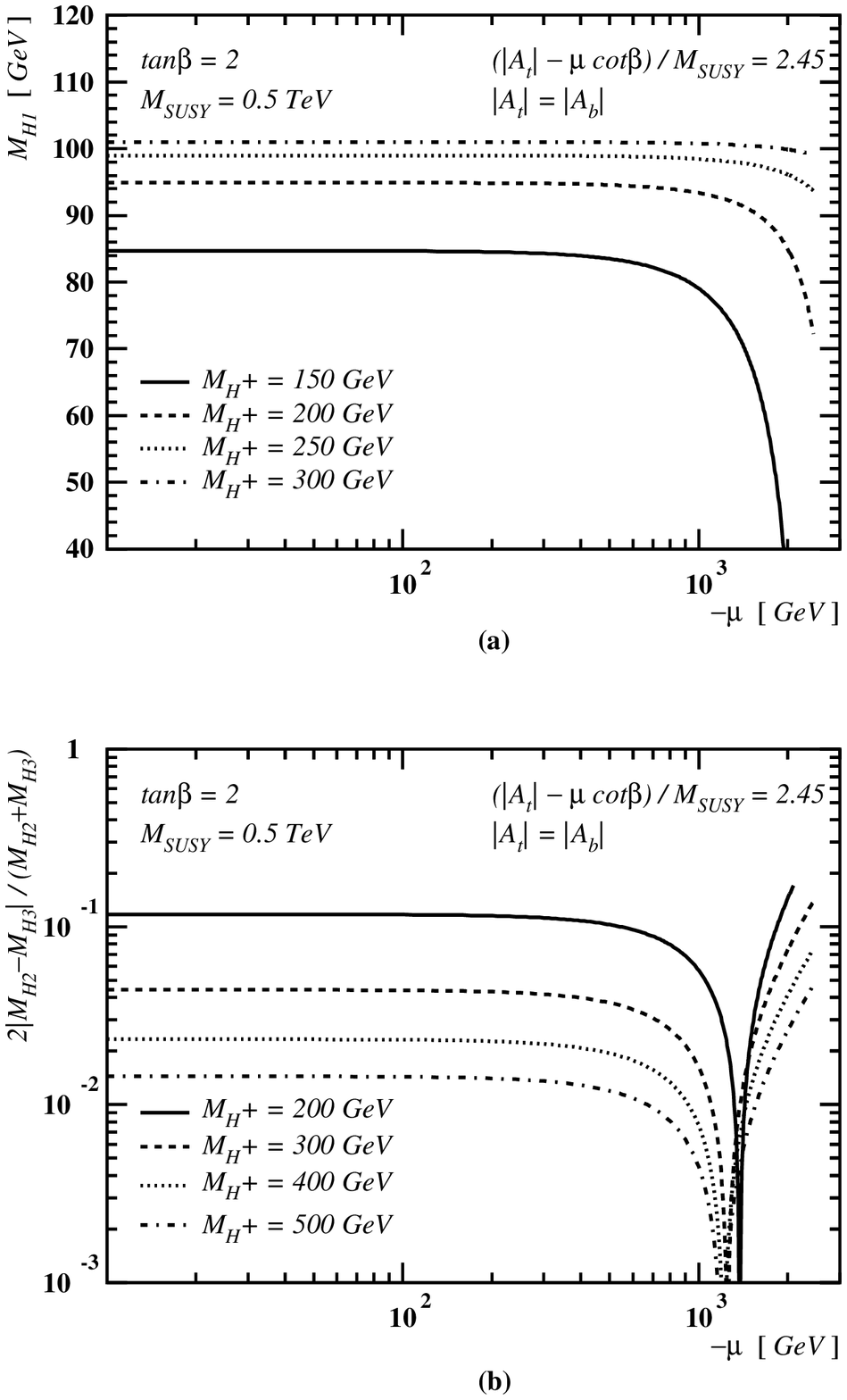}
 \end{center}
 \vspace{-0.5cm} 
\caption{Numerical estimates of (a) $M_{H_1}$ and 
  (b) $2|M_{H_2}-M_{H_3}|/(M_{H_2} +  M_{H_3})$ as a function of $\mu$
  in the CP-conserving MSSM.}
\label{fig:scp2}
\end{figure}
\begin{figure}
   \leavevmode
 \begin{center}
   \epsfxsize=16.0cm
    \epsffile[0 0 539 652]{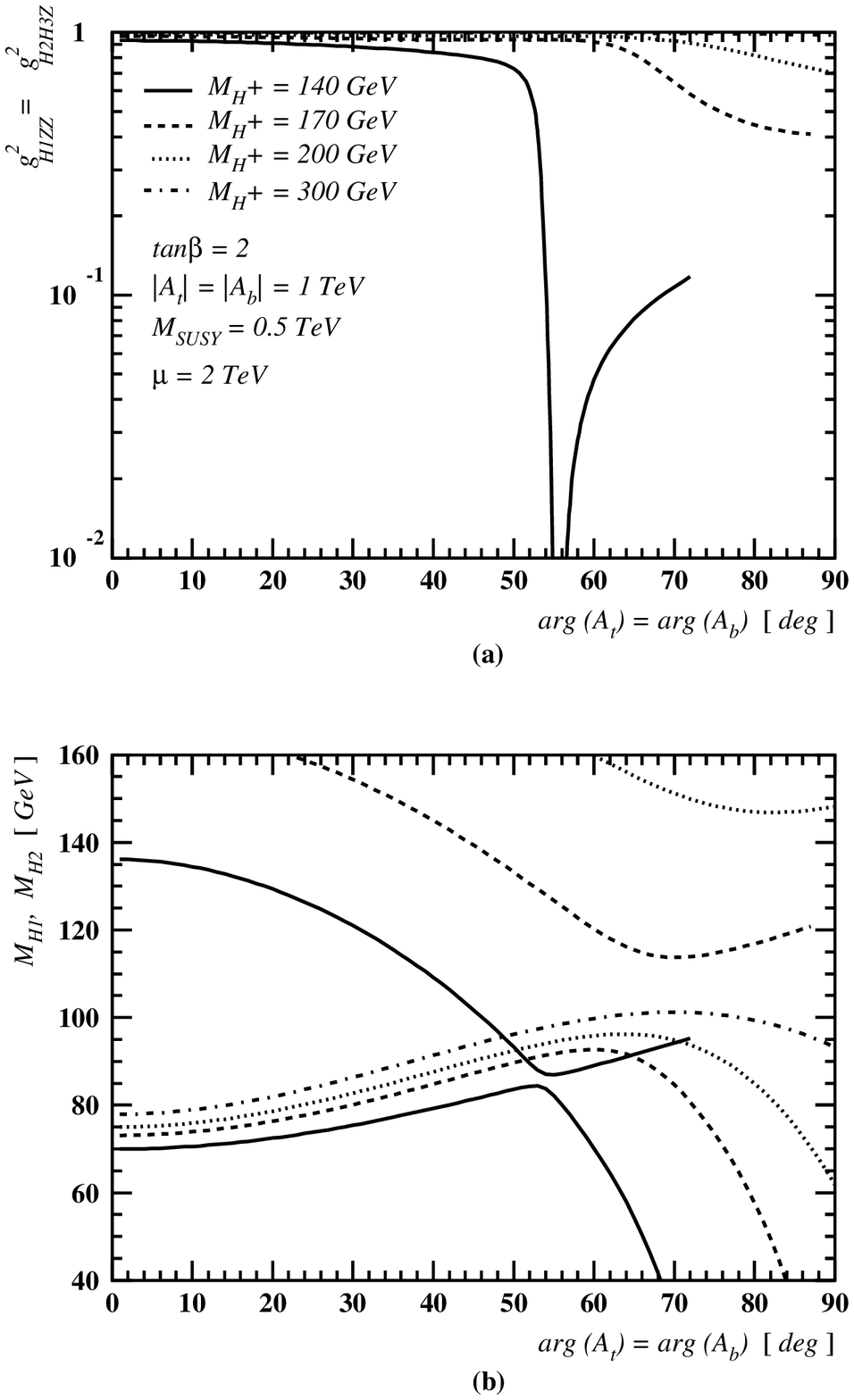}
 \end{center}
 \vspace{-0.5cm} 
\caption{Numerical predictions for (a) $g^2_{H_1ZZ} = g^2_{H_2H_3Z}$
  and  (b)   $M_{H_1} \le M_{H_2}$  as a   function of arg($A_t$). The
  definitions of  $g_{H_1ZZ}$  and $g_{H_2H_3Z}$  are given  in  Eqs.\ 
  (\ref{gHZZ}) and (\ref{gHHZ}), respectively.}
\label{fig:scp3}
\end{figure}
\begin{figure}
   \leavevmode
 \begin{center}
   \epsfxsize=16.0cm
    \epsffile[0 0 539 652]{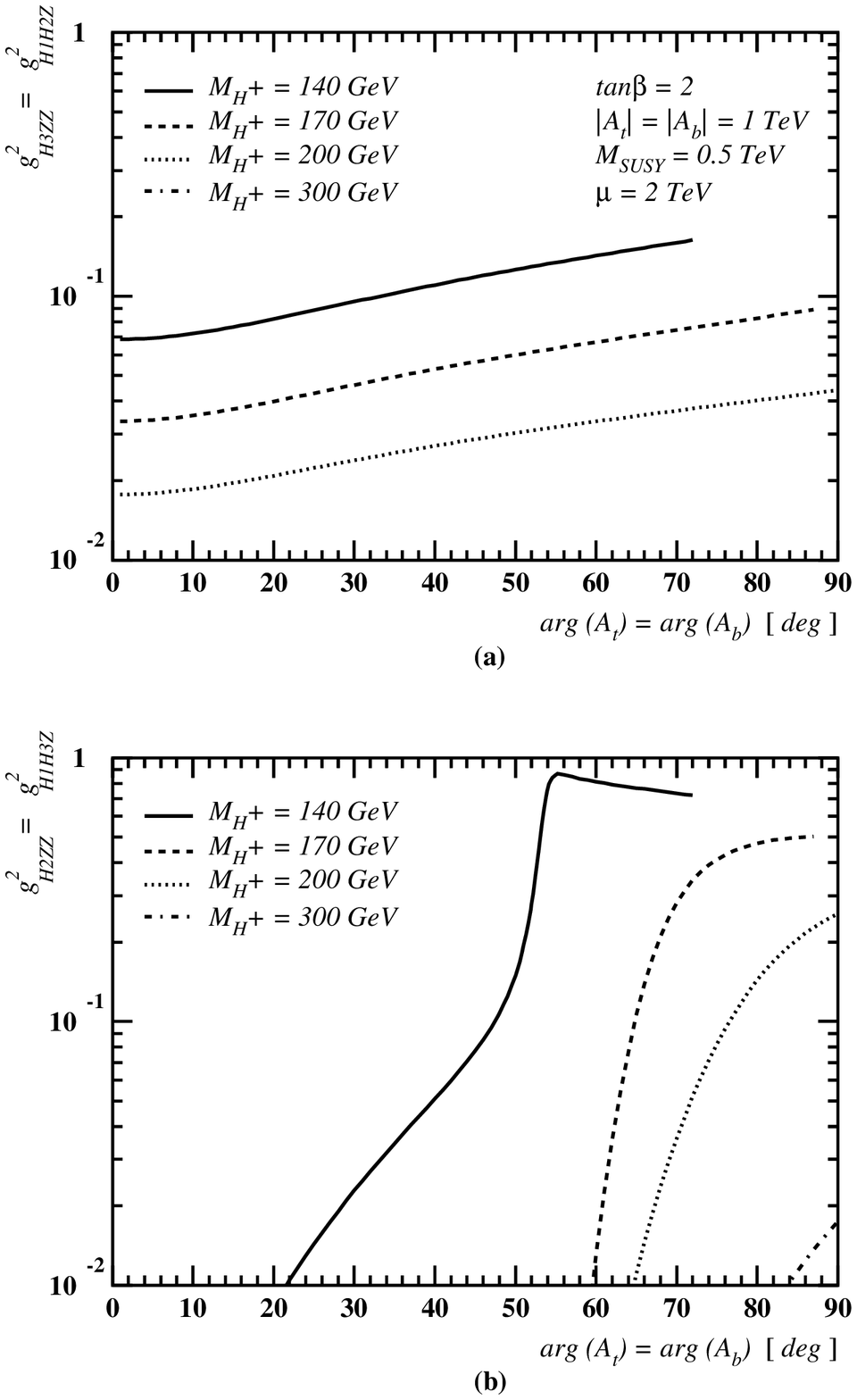}
 \end{center}
 \vspace{-0.5cm} 
\caption{Numerical predictions for (a) $g^2_{H_3ZZ} = g^2_{H_1H_2Z}$
  and (b) $g^2_{H_2ZZ} = g^2_{H_1H_3Z}$ as a function of arg($A_t$).
  The definition of $g_{H_1ZZ}$ and $g_{H_2H_3Z}$ are given in Eqs.\ 
  (\ref{gHZZ}) and (\ref{gHHZ}), respectively.}
\label{fig:scp4}
\end{figure}

\setcounter{equation}{0}
\section{Phenomenological implications for Higgs searches}

We   shall discuss   the  main phenomenological    consequences  of CP
violation  in  the Higgs sector  of  the MSSM  on the Higgs-boson mass
spectrum and  on the production cross sections   of the lightest Higgs
boson $H_1$.  In general,  CP violation modifies  the couplings of the
Higgs particles to fermions and to the $W$ and $Z$  bosons, as well as
their  self-interactions.  Furthermore,   quadrilinear    interactions
involving Higgs  bosons change as  well, when CP-violating effects due
to a   mixing of Higgs   states  are  included.   However, the  latter
interactions  as  well  as  the  trilinear   Higgs  self-couplings are
generally sub-dominant in  production  processes, and hence  we should
not consider these here.

At the high-energy machines LEP2 and Tevatron, the dominant production
mechanism  of  the lightest Higgs   boson is  the  Bjorken process  in
association  with $W$ and $Z$  bosons \cite{LEP2,Workshop}, e.g.\ $e^+
e^- \to H_1  Z$ or the partonic process  $u\bar{d} \to W^+ H_1$.  Such
reactions involve the couplings of Higgs bosons to $W$ and $Z$ bosons.
In the  presence of CP  violation, these couplings  may be read off by
the Lagrangians
\begin{eqnarray}
  \label{HVV}
{\cal L}_{HVV} &=&  g_wM_W\, ( c_\beta O_{2i}\, +\, s_\beta
O_{3i} )\, \Big(\, H_{(4-i)} W^+_\mu W^{-,\mu}\ +\
\frac{1}{2c^2_w}\, H_{(4-i)} Z_\mu Z^\mu\, \Big)\, ,\\
  \label{HpHW}
{\cal L}_{H H^\pm W^\mp} &=& \frac{g_w}{2}\, 
( c_\beta O_{3i}\, -\, s_\beta
O_{2i}\, +\, iO_{1i} )\,
W^{+,\mu}\, ( H_{(4-i)}\, i\!\!
\stackrel{\leftrightarrow}{\vspace{2pt}\partial}_{\!\mu} H^- )\ +\ {\rm
  H.c.},\\
  \label{HHZ}
{\cal L}_{HHZ} &=& \frac{g_w}{4c_w}\, 
\Big[\, O_{1i}\, ( c_\beta O_{3j}\, -\, s_\beta
O_{2j} )\, -\, O_{1j}\, ( c_\beta O_{3i}\, -\, s_\beta
O_{2i} )\, \Big]\, \nonumber\\
&&\times\, Z^\mu\, ( H_{(4-i)}\, \!\!
\stackrel{\leftrightarrow}{\vspace{2pt}\partial}_{\!\mu} H_{(4-j)} )\, ,
\end{eqnarray}
where  $c_w = M_W/M_Z$ and    $\stackrel{\leftrightarrow}{\vspace{2pt}
  \partial}_{\!   \mu}\  \equiv\   \stackrel{\rightarrow}{\vspace{2pt}
  \partial}_{\!   \mu}     -        \stackrel{\leftarrow}{\vspace{2pt}
  \partial}_{\! \mu}$.  Note that the $Z$ boson can only couple to two
different   Higgs particles as  stated in  the Lagrangian (\ref{HHZ}). 
The reason is that Bose symmetry  forbids any antisymmetric derivative
coupling of a vector particle to two identical real scalar fields.

Making  now use  of the following  identity,  which governs the matrix
elements of $O$ (assuming det$\,O=1$):
\begin{equation}
  \label{Oid}
O_{lk}\ =\ \frac{1}{2}  \sum\limits_{n,m,i,j=1}^3\! 
\varepsilon_{nml}\, \varepsilon_{ijk}\, O_{ni} O_{mj}\, ,
\end{equation}
it   is not  difficult  to  derive  an important  relation between the
couplings of the neutral Higgs bosons to the gauge bosons, namely
\begin{equation}
  \label{relation}
g_{H_k V V}\ =\ \varepsilon_{ijk}\, g_{H_i H_j Z}\, ,
\end{equation}
where $g_{H_i VV}$  ($V=W^\pm,Z$) are the Higgs--gauge-boson couplings
normalized to the SM value, i.e.\ 
\begin{equation}
  \label{gHZZ}
g_{H_iVV}\   =\ c_\beta O_{2i}\, +\, s_\beta O_{3i}\,,
\end{equation}
while  $g_{H_i H_j   Z}$  is defined  by  the   expression between the
brackets in Eq.  (\ref{HHZ}), 
\begin{equation}
  \label{gHHZ}
g_{H_i H_j Z}\ =\ O_{1i} ( c_\beta O_{3j}\, -\, s_\beta O_{2j}  )\ -\
 O_{1j} (  c_\beta O_{3i}\,  -\,  s_\beta O_{2i}  )\,.
\end{equation} 
{}From the relation (\ref{relation}) and the unitarity constraint
\begin{equation}
\sum\limits_{i=1}^3 g_{H_i ZZ}^2\ =\ 1\, ,
\end{equation}
the immediate  result is that  the knowledge of two  $g_{H_i Z  Z}$ is
sufficient to  determine the whole  set of  couplings of  the  neutral
Higgs to the gauge bosons \cite{Alex}.

\begin{figure}
   \leavevmode
 \begin{center}
   \epsfxsize=16.0cm
    \epsffile[0 0 539 652]{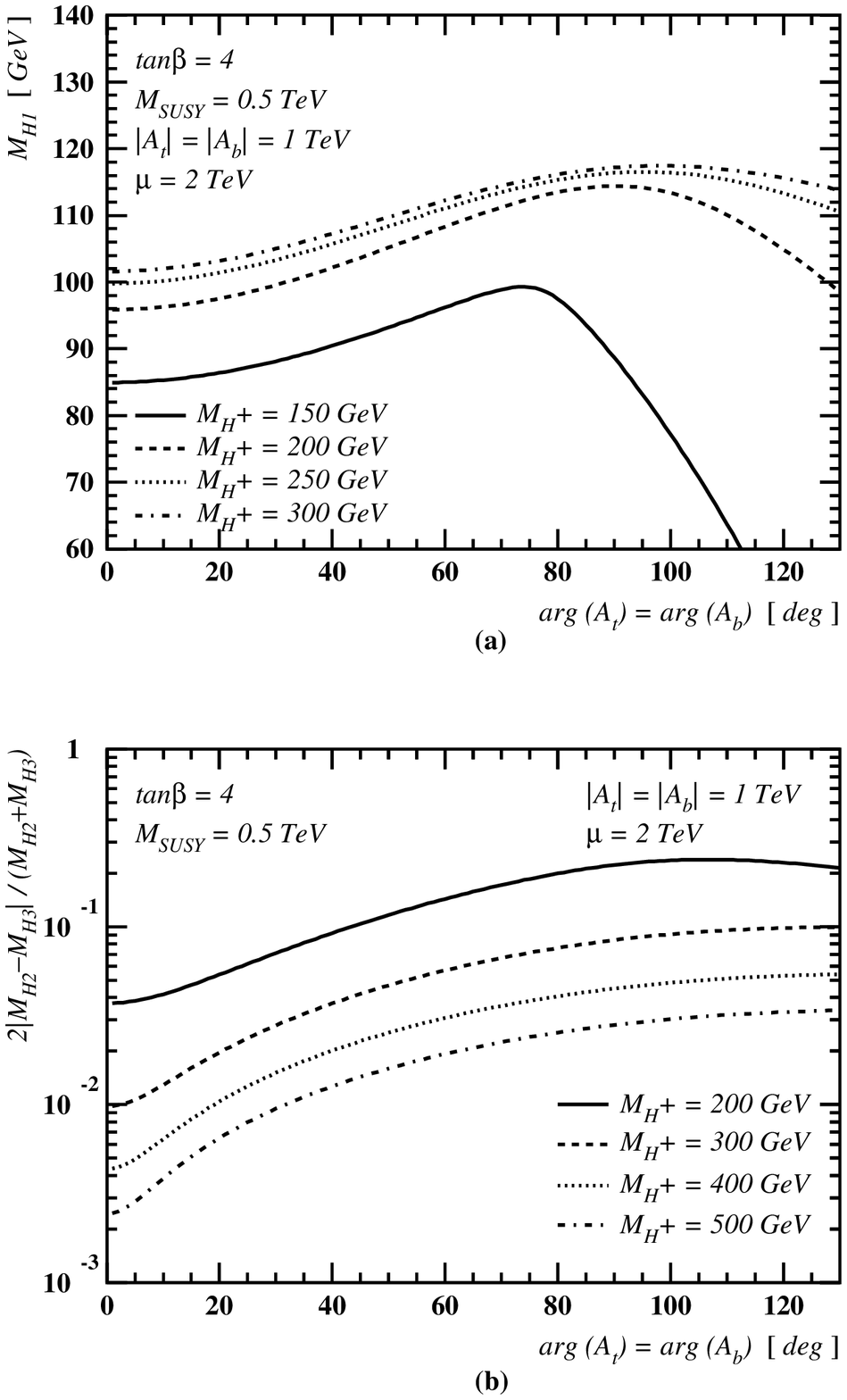}
 \end{center}
 \vspace{-0.5cm} 
\caption{The same as in Fig.\ \ref{fig:scp1}, but with $\tan\beta = 4$.}
\label{fig:scp5}
\end{figure}
\begin{figure}
   \leavevmode
 \begin{center}
   \epsfxsize=16.0cm
    \epsffile[0 0 539 652]{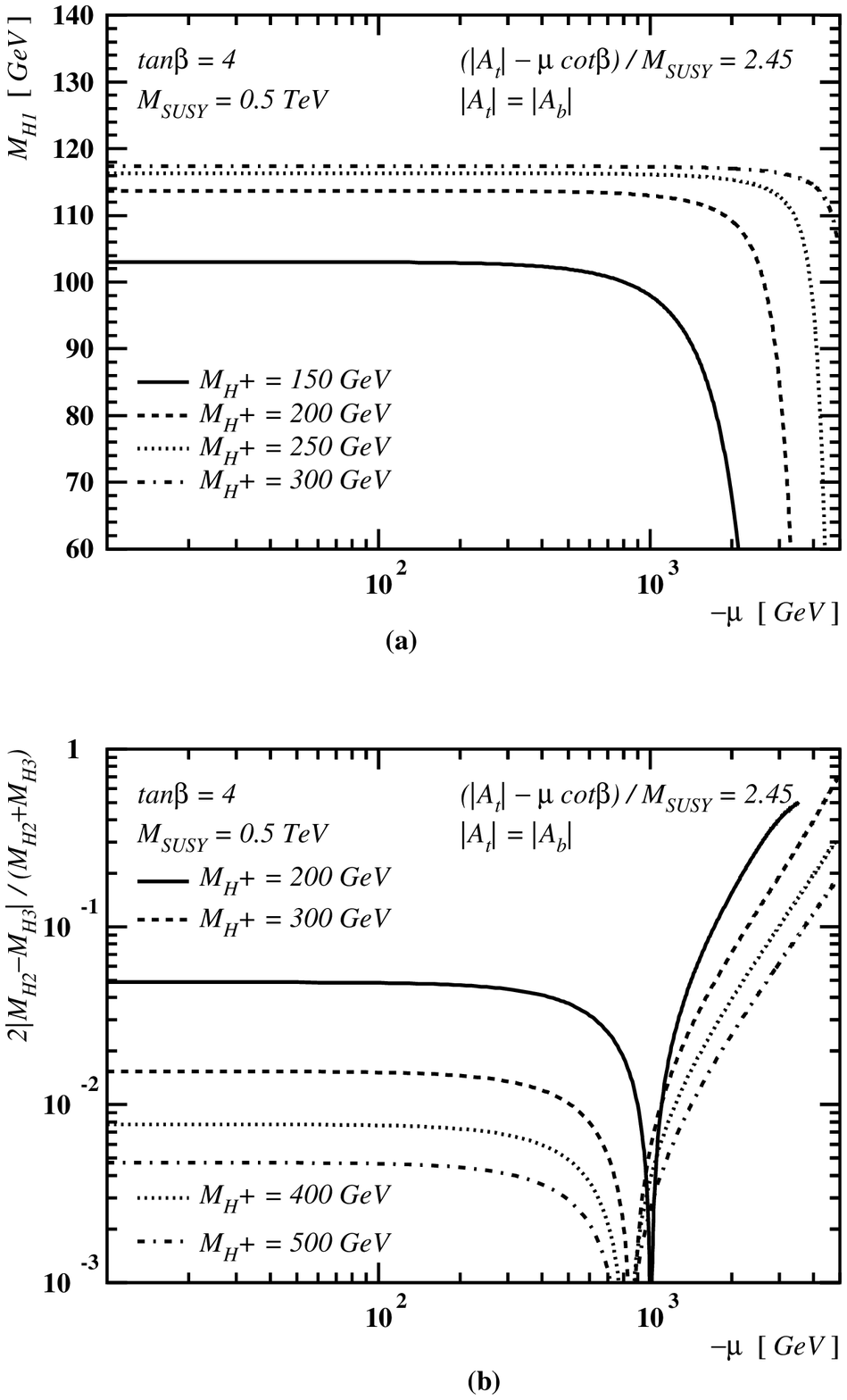}
 \end{center}
 \vspace{-0.5cm} 
\caption{The same as in Fig.\ \ref{fig:scp2}, but with $\tan\beta = 4$.}
\label{fig:scp6}
\end{figure}
\begin{figure}
   \leavevmode
 \begin{center}
   \epsfxsize=16.0cm
    \epsffile[0 0 539 652]{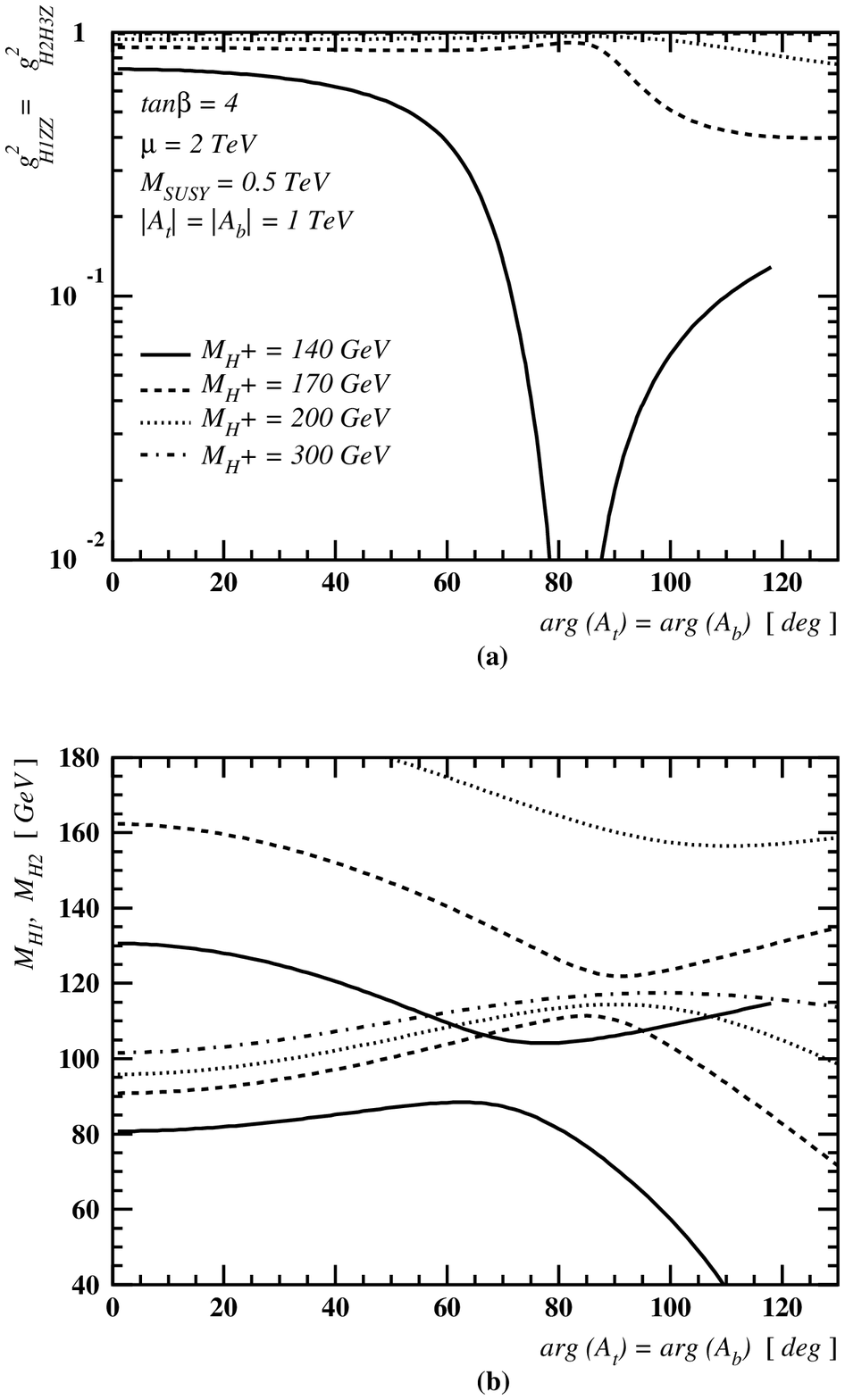}
 \end{center}
 \vspace{-0.5cm} 
\caption{The same as in Fig.\ \ref{fig:scp3}, but with $\tan\beta = 4$.}
\label{fig:scp7}
\end{figure}
\begin{figure}
   \leavevmode
 \begin{center}
   \epsfxsize=16.0cm
    \epsffile[0 0 539 652]{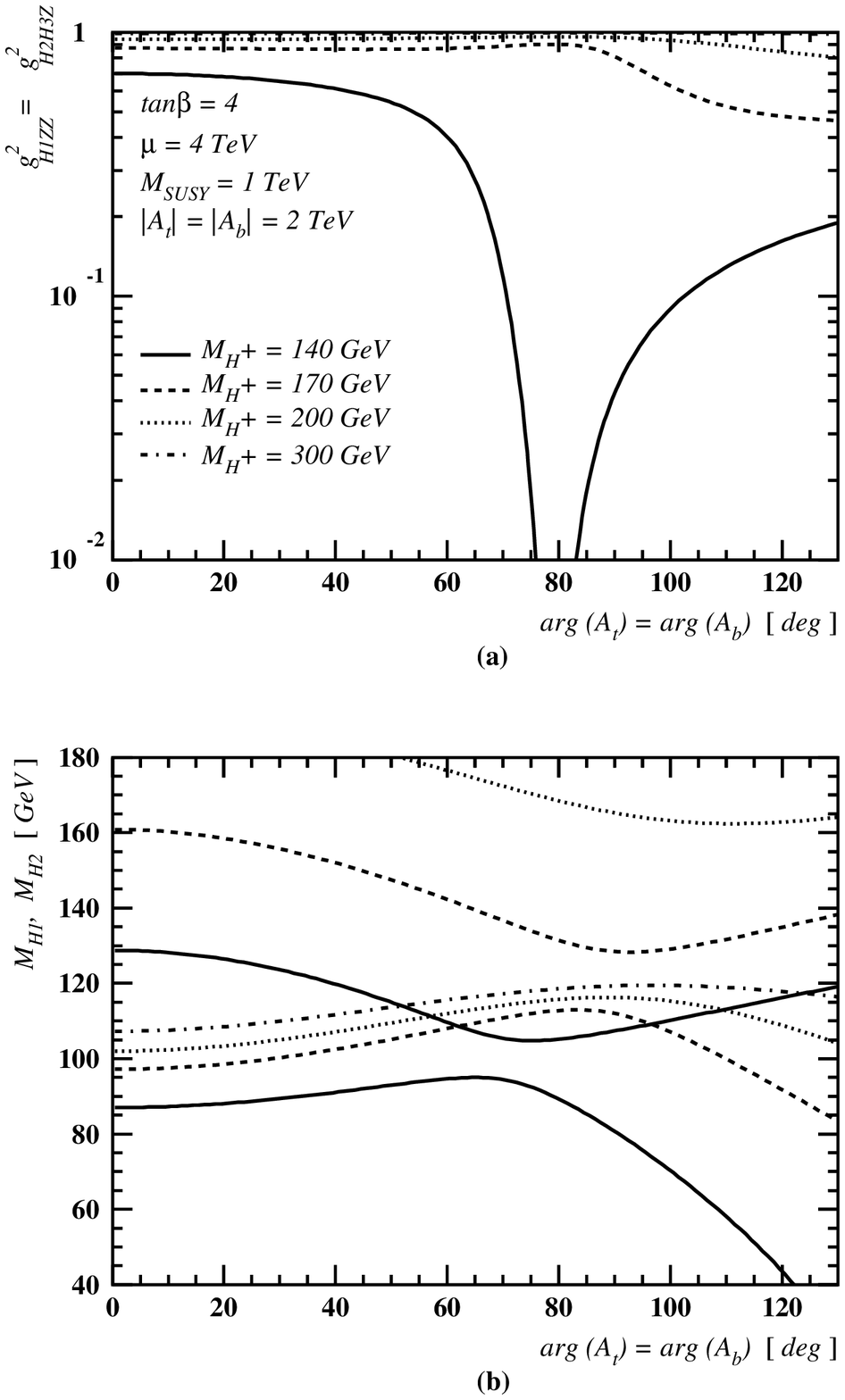}
 \end{center}
 \vspace{-0.5cm} 
\caption{The same as in Fig.\ \ref{fig:scp7}, but setting $M_{\rm SUSY}
  = 1$ TeV, $|A_t| = |A_b| = 2$ TeV and $\mu =4$ TeV.}
\label{fig:scp7new}
\end{figure}

The essential difference between a CP-conserving MSSM Higgs sector and
a CP-violating one is that mixing effects  between the would-be CP-odd
and CP-even  Higgs  bosons  are present  in  the latter   case.   Such
scalar--pseudoscalar    mixing  effects   are   induced   by radiative
corrections to the   Higgs potential \cite{APLB}.  The  characteristic
size of the CP-violating   off-diagonal terms in the  Higgs-boson mass
matrix may be estimated by
\begin{eqnarray}
M^2_{SP} & \simeq & {\cal O} \left( \frac{ m_t^4}{v^2} 
\frac{|\mu| |A_t|}{32 \pi^2M_{\rm SUSY}^2} \right) \sin \phi_{\rm CP} 
\nonumber\\
&&\times\, \left(6,\ \frac{|A_t|^2}{M_{\rm SUSY}^2}\, ,\ \frac{|\mu|^2}
{\tan\beta\, M_{\rm SUSY}^2}\,,\ \frac{\sin 2\phi_{\rm CP}}{\sin
  \phi_{\rm CP}}\, \frac{|\mu||A_t|}{M_{\rm SUSY}^2} \right),
\end{eqnarray}  
where  the last bracket summarizes  the relative size of the different
contributions, and
\begin{equation}
\phi_{\rm CP}\ =\ {\rm arg}(A_t \mu)\, +\, \xi\, .
\end{equation}
For $|\mu|$  and $|A_t|$ values  larger than the arithmetic average of
the  scalar-top-quark masses squared,  denoted as $M_{\rm SUSY}$ (cf.\ 
Eq.\ (\ref{Msusy})), the CP-violating effects can be significant.  For
instance,  if   $|\mu|  \simeq |A_t|   \simeq  2  M_{\rm   SUSY}$, and
$\phi_{\rm CP} \simeq 90^\circ$, the off-diagonal terms of the neutral
Higgs-boson mass matrix  may be of the  order of $(100\ {\rm GeV})^2$. 
These  potentially large mixing   effects have important consequences,
since they lead to drastic variations in the definition of the neutral
Higgs-boson masses and  in the couplings  of the  Higgs states to  the
gauge bosons. Because of the quantum nature of the CP-violating mixing
effects and the known decoupling property  of heavy states in the loop
in SUSY theories,  the phenomenology of the  lightest Higgs boson will
only be important for  low values of $M_{H^+}$.   For the same reason,
$M_{H^+}$   values much larger   than the  electroweak  scale  lead to
predictions for the mass of the lightest Higgs boson $H_1$ and for the
couplings of the $H_1$ scalar to the gauge bosons which are equivalent
to those obtained in the CP-invariant theory.   The only difference in
the   CP-violating case   is    that the  relevant   scalar-top mixing
parameter entering the definition of $M_{H_1}$ is now given by
\begin{equation}
  \label{tildeat}
|\tilde{A}_t|\ =\ |A_t - \mu^* /\tan\beta|\, .
\end{equation}

In addition  to  the  effects induced  by  scalar-top   quarks, Yukawa
interactions due to scalar-bottom  quarks can also be  significant for
large values of $\tan\beta$,  so as to  lead to sizeable contributions
to  the elements of the  Higgs-boson mass-matrix.  As was discussed in
Section 4, however, the contributions  of the scalar-bottom sector are
limited  by constraints that originate  from  the electron and neutron
EDMs.   In  general, unless    cancellations occur  between  different
contributions to the   EDMs, the CP-violating   quantum effects coming
from the scalar-bottom sector are expected to be small.

In  Fig.\ \ref{fig:scp1},   we   present numerical   estimates of  the
lightest Higgs-boson mass $M_{H_1}$ and of the relative mass splitting
of the two  heaviest Higgs bosons, $H_2$  and $H_3$, as a  function of
$\phi_{\rm CP} \equiv  {\rm  arg}(A_t)$, for different  values  of the
charged Higgs-boson mass and for $\tan\beta = 2$, $M_{\rm SUSY} = 0.5$
TeV,  $|A_t| = 1$   TeV  and $\mu  =  2$  TeV.   We  must remark  that
relatively large values of  the  $|\mu|$ and $|A_t|$  parameters  have
been chosen, in  order to make  the CP-violating effects relevant.  In
fact, for these values of the trilinear couplings, we  are at the edge
of  the limit of  validity  of the expansion we   used to compute  the
expression of the quartic couplings in the RG-improved Higgs potential
(see appendix).  However, we may verify that the neglected terms would
lead  to  a  slight increase  of   the CP-violation   effects  we have
computed. For the region of parameters we are considering, the omitted
terms are of the order of finite two-loop corrections to the effective
potential, which  are not included  in the  analytic expression of the
one-loop RG-improved effective potential.   Moreover, if all SUSY-mass
parameters are  rescaled by a factor  2, the validity of the expansion
improves dramatically, while the physical   results are only  slightly
modified.   Later  on, we shall explicitly  demonstrate  this point by
giving   a specific example  (cf.\ Fig.\  \ref{fig:scp7new}).  In this
context,   we  believe  that  our   results   should be  regarded   as
conservative  estimates  of the   possible size  of   the CP-violating
effects in the Higgs sector.

In Fig.\ \ref{fig:scp1}, the dependence of $M_{H_1}$ on arg$(A_t)$ may
be understood as  follows: if the moduli  of the  parameters $A_t$ and
$\mu$ are  kept fixed,  the mass of  the lightest  neutral Higgs boson
$H_1$ starts increasing, as a function of arg$(A_t)$, up to a maximum;
it then  decreases    rapidly.  This   kinematic  dependence  may   be
attributed to  the  fact that the  modulus  of  the  scalar-top mixing
parameter $\tilde{A}_t$ increases monotonically for  the range $0  \le
{\rm arg}(A_t) \le 180^\circ$. We find that the behaviour of $M_{H_1}$
as a function of arg$(A_t)$  is very analogous to  the one that  would
have been  obtained   for the lightest  CP-even  Higgs  boson  in  the
CP-invariant theory   if we had  varied  $|\tilde{A}_t|$ given in Eq.\ 
(\ref{tildeat}).  However, unlike  the CP-conserving case, the effects
of  the Higgs-sector  CP violation can  give rise  to much larger mass
splittings between the two heaviest Higgs bosons $H_2$  and $H_3$ as a
function of arg$(A_t)$  \cite{APLB}. This fact  is difficult to  infer
from   the  kinematic  dependence  of    $M_{H_1}$  alone.  In   Fig.\ 
\ref{fig:scp1}(b), we see  that  even in the  case of  $M_{H^+} = 200$
GeV, the CP-violating effects may lead to a relative mass splitting of
the two heaviest neutral Higgs bosons of the order of 30$\%$.
  
In order  to get an idea of  how much of  the above effects are due to
the presence of   CP violation, we plot  in   Fig.\ \ref{fig:scp2} the
behaviour of the same neutral Higgs-boson  masses as a function of the
parameter $\mu$,  where  $\phi_{\rm   CP}=0$  is considered  and   the
scalar-top mixing parameter $|\tilde{A}_t|$ is fixed to the value that
yields the maximum for the lightest CP-even Higgs-boson mass $M_h$ for
large  values of     $M_{H^+}$.  Figure  \ref{fig:scp2} exhibits   the
dependence of $M_h  \equiv M_{H_1}$ as a  function  of $\mu$. We  find
that large  values  of $\mu$ tend  to  reduce  $M_h$ and  increase the
degree of the mass splitting between  the heaviest CP-even Higgs boson
and the CP-odd Higgs scalar, $2|M_{H_2}-M_{H_3}|/(M_{H_2} + M_{H_3})$.
In  particular, we  observe  that  for the   same value of  $M_{H_1}$,
$2|M_{H_2}-M_{H_3}|/(M_{H_2}   + M_{H_3})$ is   much  smaller  in  the
CP-invariant theory than in the CP-violating one.

It is worth investigating whether the CP-violating Higgs effects could
lead to not-yet-explored  open windows in  the parameter space of  the
theory that cannot be accessed by  the running experiments and are not
present in the CP-conserving  case.  To this  end,  we plot in  Figs.\ 
\ref{fig:scp3} and \ref{fig:scp4}  the Higgs--gauge-boson couplings as
a function  of arg$(A_t)$, for  the same choice  of SUSY parameters as
those in Fig.\ \ref{fig:scp1}.  For low values of $M_{H^+}$, the Higgs
boson that couples predominantly  to the $Z$  boson is $H_1$ ($H_2$).  
For $M_{H^+} = 140$ GeV, there  is an interesting region of parameters
for which the $H_1ZZ$ coupling becomes  small, rendering the detection
of  the $H_1$   boson via  the  $H_1ZZ$ channel  at    LEP2 difficult. 
However,  in   the  same region    of parameters,   the  mass  of  the
next-to-lightest  neutral Higgs state, $H_2$,  is  smaller than 95 GeV
and the $H_2ZZ$ coupling becomes large.  Therefore, in this particular
kinematic range  of parameters,  the   $H_2ZZ$  channel  is the   only
relevant one that helps  to exclude these  small values of $\tan\beta$
and charged  Higgs-boson  masses  at   LEP2.  For  larger   values  of
$M_{H^+}$, the situation resembles more the CP-invariant theory.  Even
for $M_{H^+}  = 170$ GeV, the suppression  of  the $H_1ZZ$ coupling is
not sufficiently large to produce a  significant change in the present
LEP2 bound.   Therefore,  very analogously  to the CP-conserving  case
\cite{CCPW}, only a small region of parameters, for which the lightest
CP-even Higgs boson is heavy  enough, is still experimentally allowed,
i.e.\   for large values  of   the  scalar-top  mixing mass  parameter
$|\tilde{A}_t|$ (${\rm  arg}  (A_t)  \simeq 60^\circ$--$90^\circ$) and
for relatively large values of the charged Higgs-boson mass.

\begin{figure}
   \leavevmode
 \begin{center}
   \epsfxsize=16.0cm
    \epsffile[0 0 539 652]{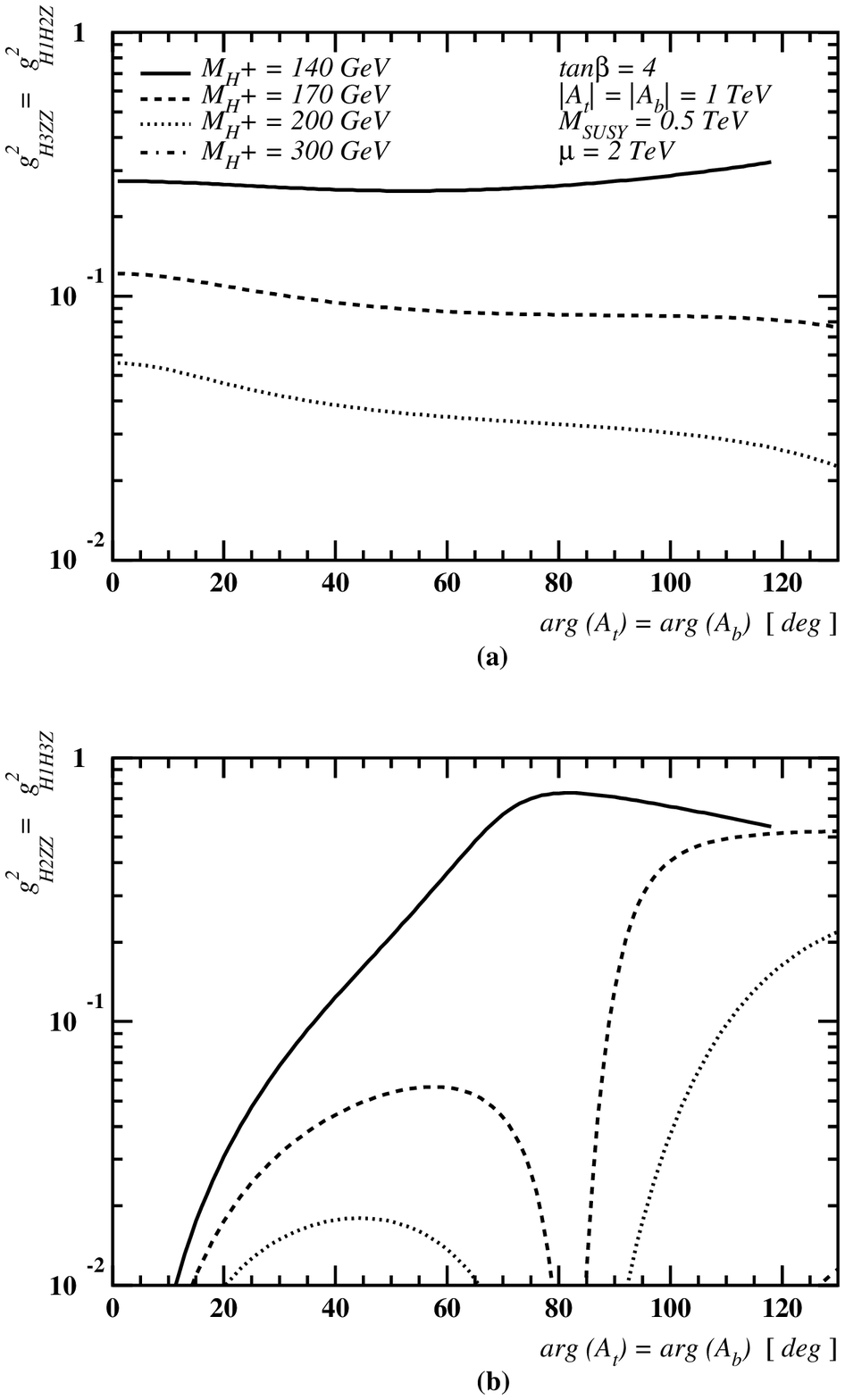}
 \end{center}
 \vspace{-0.5cm} 
\caption{The same as in Fig.\ \ref{fig:scp4}, but with $\tan\beta = 4$.}
\label{fig:scp8}
\end{figure}

More   interesting  is   the   situation  at  intermediate   values of
$\tan\beta$.   Even though, as displayed  in Figs.\ \ref{fig:scp5} and
\ref{fig:scp6}, the dependence of the Higgs masses on the CP-violating
phases and   the scalar-top mixing parameters  is  similar  to the one
obtained for $\tan\beta = 2$, the region for which the coupling of the
lightest  neutral  Higgs boson to  the  gauge bosons becomes small now
displays larger values  of the $H_1$  and $H_2$ masses.  In this range
of  kinematic parameters, the experimental  detection of the $H_1$ and
$H_2$ particles becomes more difficult at  LEP2.  In Table \ref{Tab2},
we present  the current reach  of LEP2  for the  $H_1$-boson mass as a
function of $g_{H_1ZZ}$,  and independently, for $(M_{H_1},  M_{H_2})$
versus $g_{H_1H_2Z}$, assuming $M_{H_1} \approx M_{H_2}$.  As can then
be  seen from Fig.\ \ref{fig:scp7} for  the case of a relatively light
charged Higgs scalar $M_{H^+} = 140$ GeV, we get regions for which the
lightest Higgs-boson mass $M_{H_1}$ is as small as  60--70 GeV and the
$H_1ZZ$  coupling  is small   enough  for the  $H_1$  boson to  escape
detection   at the latest    LEP2  run, with   $\sqrt{s}  = 189$  GeV. 
Moreover,  the $H_2$  boson  is too heavy  to  be detected through the
$H_2ZZ$  channel.  In addition,  either the  coupling $H_1H_2Z$ is too
small  or $H_2$ is   too heavy to allow  the   Higgs detection in  the
$H_1H_2Z$ channel (see also Fig.\ \ref{fig:scp8}). To better gauge the
validity of the expansion that has been used for the quartic couplings
in the  effective  Higgs potential  (see  also the  discussion in  the
appendix),    we have plotted    in Fig.\  \ref{fig:scp7new}  the same
$\tan\beta$ scenario but rescaling $M_{\rm SUSY}$, $|A_t| = |A_b|$ and
$\mu$ by a factor of 2, i.e.\ $M_{\rm SUSY} = 1$ TeV, $|A_t| = |A_b| =
2$ TeV   and $\mu = 4$  TeV.   We then find  that $M_{H_1}$   has only
slightly increased and    the  kinematic dependence of the     $H_1ZZ$
coupling   on  the  CP-violating   phase  arg$(A_t)$  remained  almost
unchanged when compared to Fig.\ \ref{fig:scp7}.

%%%TABLE 2
\begin{table}[t]

\begin{center}

\begin{tabular}{|r|c||r|c|}
\hline
$g^2_{H_1ZZ}$ & $M_{H_1}$ & $g^2_{H_1H_2Z}$ & $M_{H_1}\approx M_{H_2}$\\
 & ~~~$[\,{\rm GeV}\,]$~~~ &   & $[\,{\rm GeV}\,]$ \\ 
\hline\hline
1    & 97 & 1    & 86 \\ 
0.6  & 96 & 0.9  & 85 \\
0.5  & 95 & 0.8  & 84 \\ 
0.4  & 93 & 0.7  & 83 \\
0.3  & 92 & 0.6  & 82 \\
0.2  & 88 & 0.5  & 80 \\ 
0.1  & 77 & 0.4  & 77 \\
0.08 & 70 & 0.3  & 72 \\
\hline
\end{tabular}
\end{center}

\caption{Present experimental sensitivity of the couplings
  $g^2_{H_1ZZ}$ and $g^2_{H_1H_2Z}$ as a function of $M_{H_1}$, 
assuming $M_{H_1} \approx M_{H_2}$. The results are obtained 
from four experiments combined at LEP2 with 
$\sqrt{s} = 189$ GeV.}\label{Tab2}
\end{table}

\begin{figure}
   \leavevmode
 \begin{center}
   \epsfxsize=16.0cm
    \epsffile[0 0 539 652]{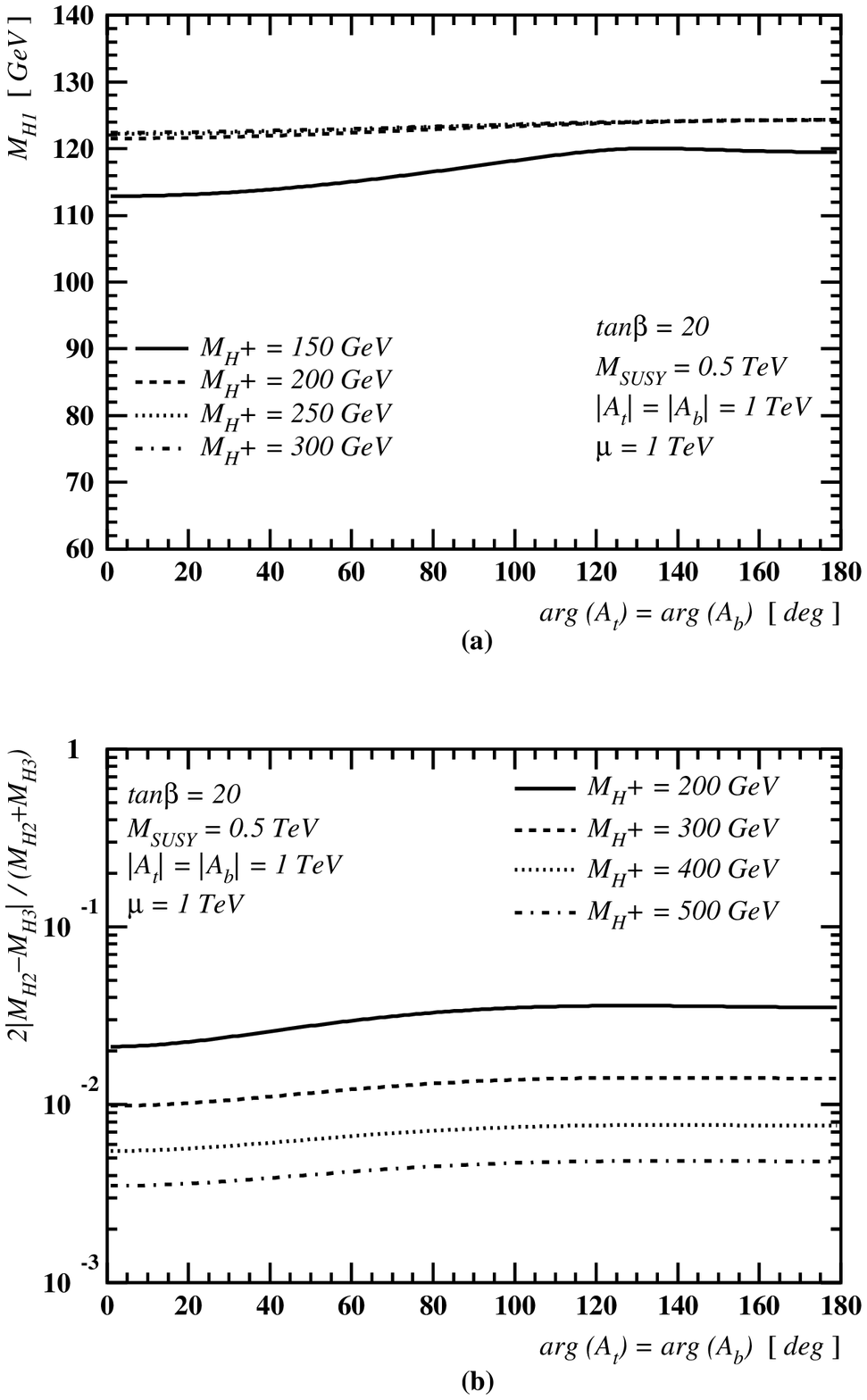}
 \end{center}
 \vspace{-0.5cm} 
\caption{The same as in Fig.\ \ref{fig:scp1}, but with $\tan\beta =
  20$ and $\mu = 1$ TeV.}
\label{fig:scp9}
\end{figure}

\begin{figure}
   \leavevmode
 \begin{center}
   \epsfxsize=16.0cm
    \epsffile[0 0 539 652]{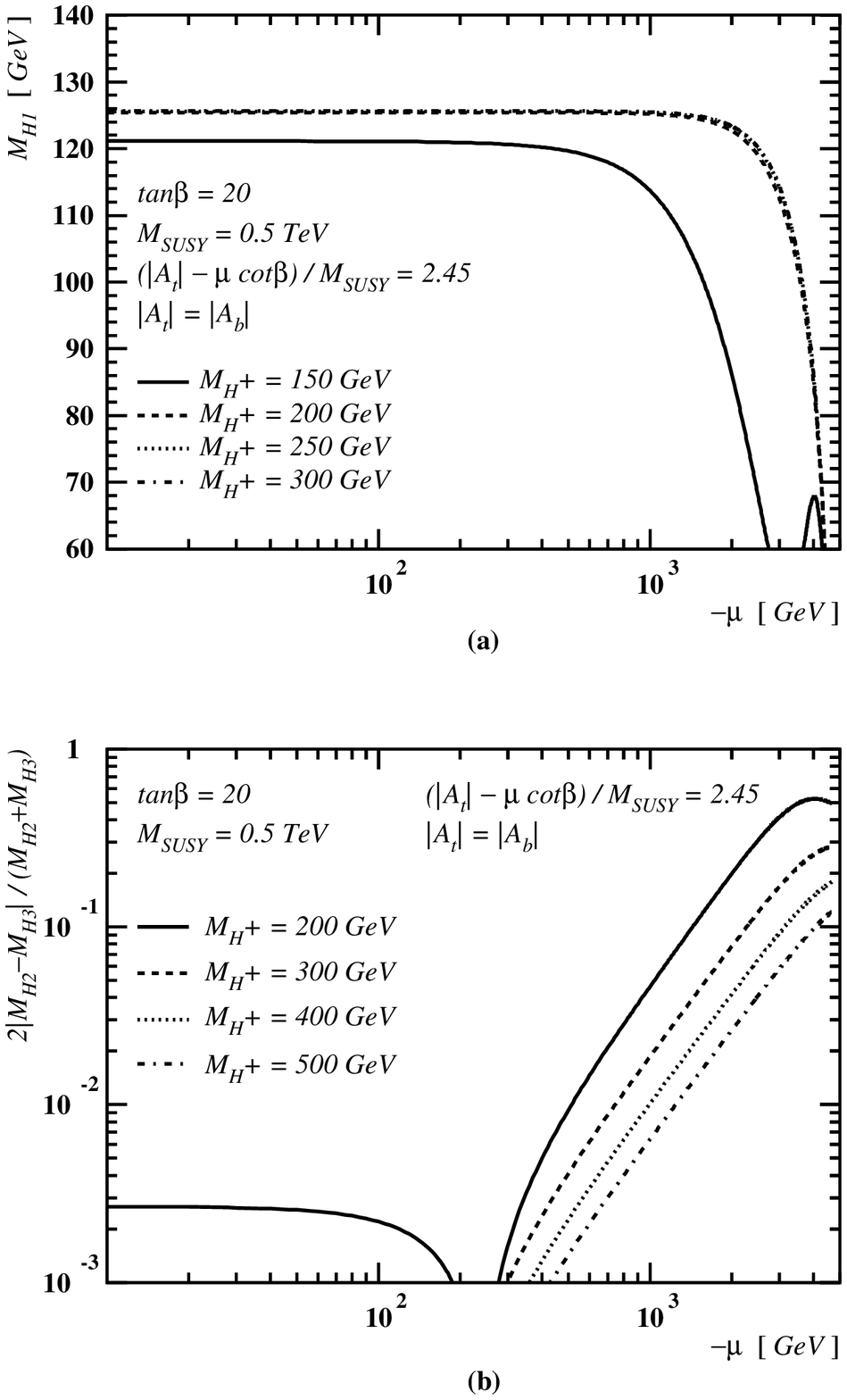}
 \end{center}
 \vspace{-0.5cm} 
\caption{The same as in Fig.\ \ref{fig:scp2}, but with $\tan\beta =
  20$ and $\mu = 1$ TeV.}
\label{fig:scp10}
\end{figure}

\begin{figure}
   \leavevmode
 \begin{center}
   \epsfxsize=16.0cm
    \epsffile[0 0 539 652]{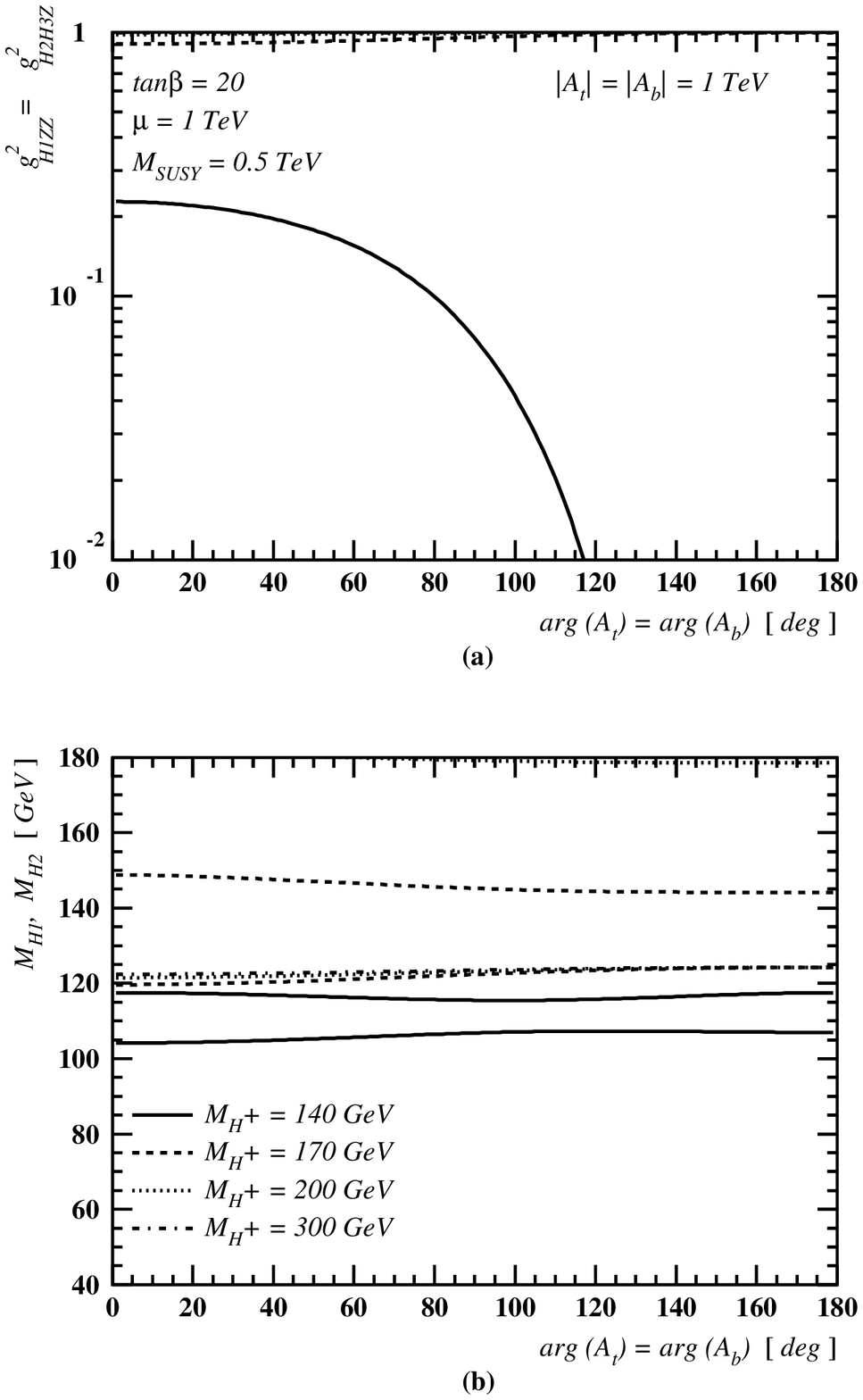}
 \end{center}
 \vspace{-0.5cm} 
\caption{The same as in Fig.\ \ref{fig:scp3}, but with $\tan\beta =
  20$ and $\mu = 1$ TeV.}
\label{fig:scp11}
\end{figure}

\begin{figure}
   \leavevmode
 \begin{center}
   \epsfxsize=16.0cm
    \epsffile[0 0 539 652]{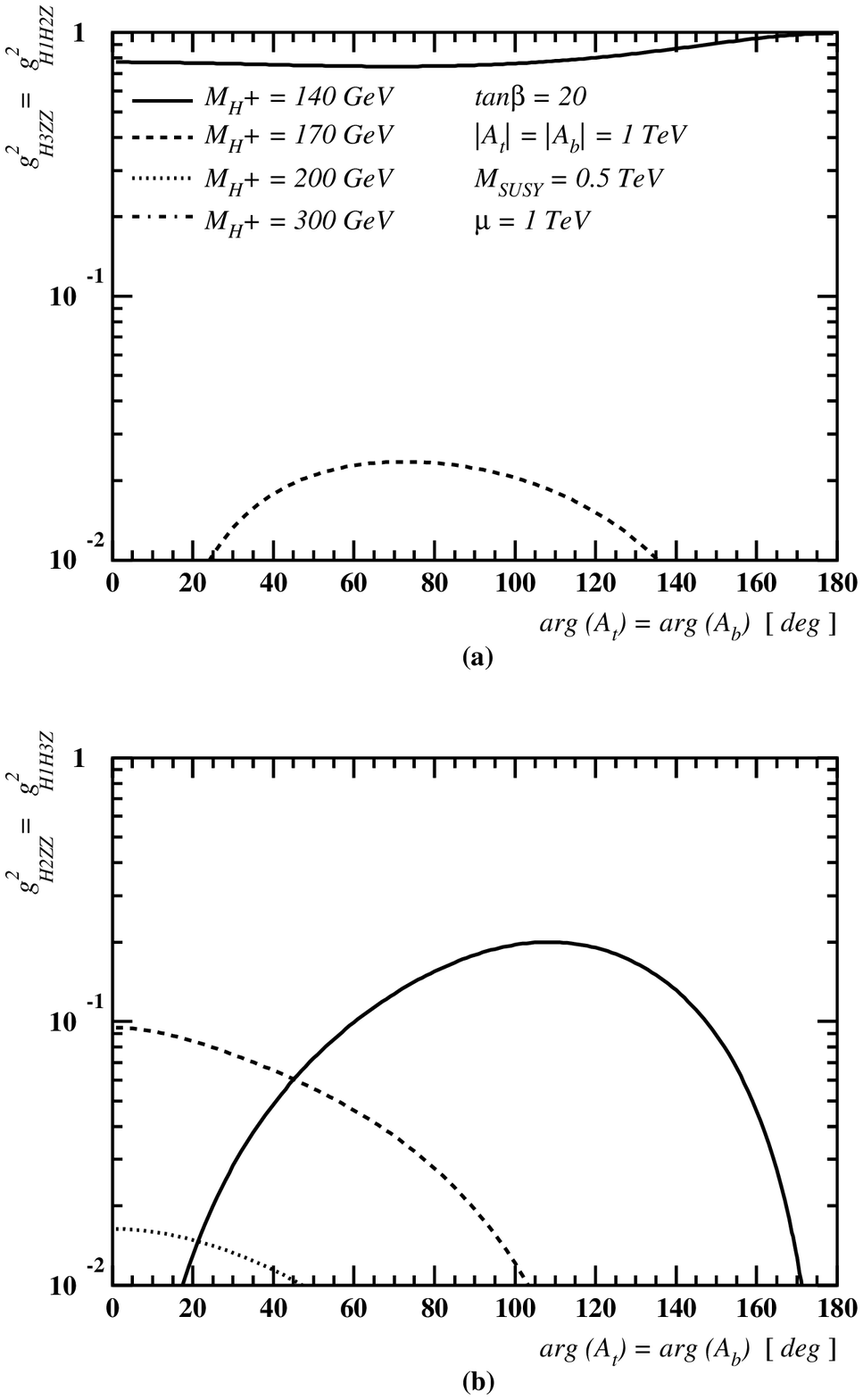}
 \end{center}
 \vspace{-0.5cm} 
\caption{The same as in Fig.\ \ref{fig:scp4}, but with $\tan\beta =
  20$ and $\mu = 1$ TeV.}
\label{fig:scp12}
\end{figure}

For very large values of  $\tan\beta$, the CP-violating effects in the
Higgs sector are constrained by the bounds  on the EDM of the electron
and neutron discussed in  Section 4.  If  no cancellation mechanism is
assumed to  occur   between   the different EDM   contributions,   the
CP-violating effects  in the Higgs sector   are then relatively small,
and  the phenomenological properties of  the  Higgs bosons become very
similar  to those obtained in the  CP-invariant case.  This feature is
shown     in Figs.\   \ref{fig:scp9}--\ref{fig:scp12},    for the same
quantities             as     those     considered     in       Figs.\ 
\ref{fig:scp1}--\ref{fig:scp4}.   Notice   that  the  actual bound  on
$\tan\beta$ coming from EDM constraints strongly  depends on the exact
value of the CP-odd phase, which we choose to  be equal to $90^\circ$. 
In the high-$\tan\beta$ regime, the running $b$-quark mass $m_b$ plays
a central role.  As  we will  see later on,  $m_b$  turns out to be  a
model-dependent  quantity  in  the   MSSM.  So as   to  make  definite
predictions,  however,  we have set  $m_b  (\overline{m}_t)  = 3$ GeV,
where  $\overline{m}_t$ is   the    top-quark pole  mass (cf.\    Eq.\ 
(\ref{hthb})).


It is  now  interesting to   discuss   whether the Tevatron   collider
together with   its future   upgraded  facilities have   the potential
capabilities of exploring the open windows  that are not accessible at
LEP2 \cite{Howie,CMW}.   The Tevatron reach  depends  very strongly on
the final run II  luminosity.   Furthermore, the  reach is very   much
affected  by a suppression of  the Higgs-boson production  rate in the
$H_iVV$ channel.  Hence, it is unlikely that the Tevatron will be able
to observe the lightest Higgs boson in the regions where the effective
$H_1VV$ coupling is suppressed.  The search  for the $H_2$ boson looks
more  promising.  Indeed, for high  luminosities available  at a later
stage of  the  collider, e.g.\    30  fb$^{-1}$ per experiment,    the
Tevatron-discovery reach of the Higgs-boson mass can be as high as 125
GeV for a SM production rate and 115 GeV if there is a 0.7 suppression
factor.\footnote{These values rely on a combination of the $H_1WW$ and
  $H_1ZZ$ channel for  both experiments \cite{Workshop}, assuming that
  the total branching ratio into  $b$ quarks remains almost  unchanged
  with  respect  to the   SM   value.}   From  the results  of  Figs.\ 
\ref{fig:scp5}--\ref{fig:scp8}, we  can  see that, as  happens  in the
CP-conserving case \cite{CMW}, a high-luminosity Tevatron collider may
be capable of covering most of the windows left open at LEP2.

\begin{figure}
   \leavevmode
 \begin{center}
   \epsfxsize=16.0cm
    \epsffile[0 0 539 652]{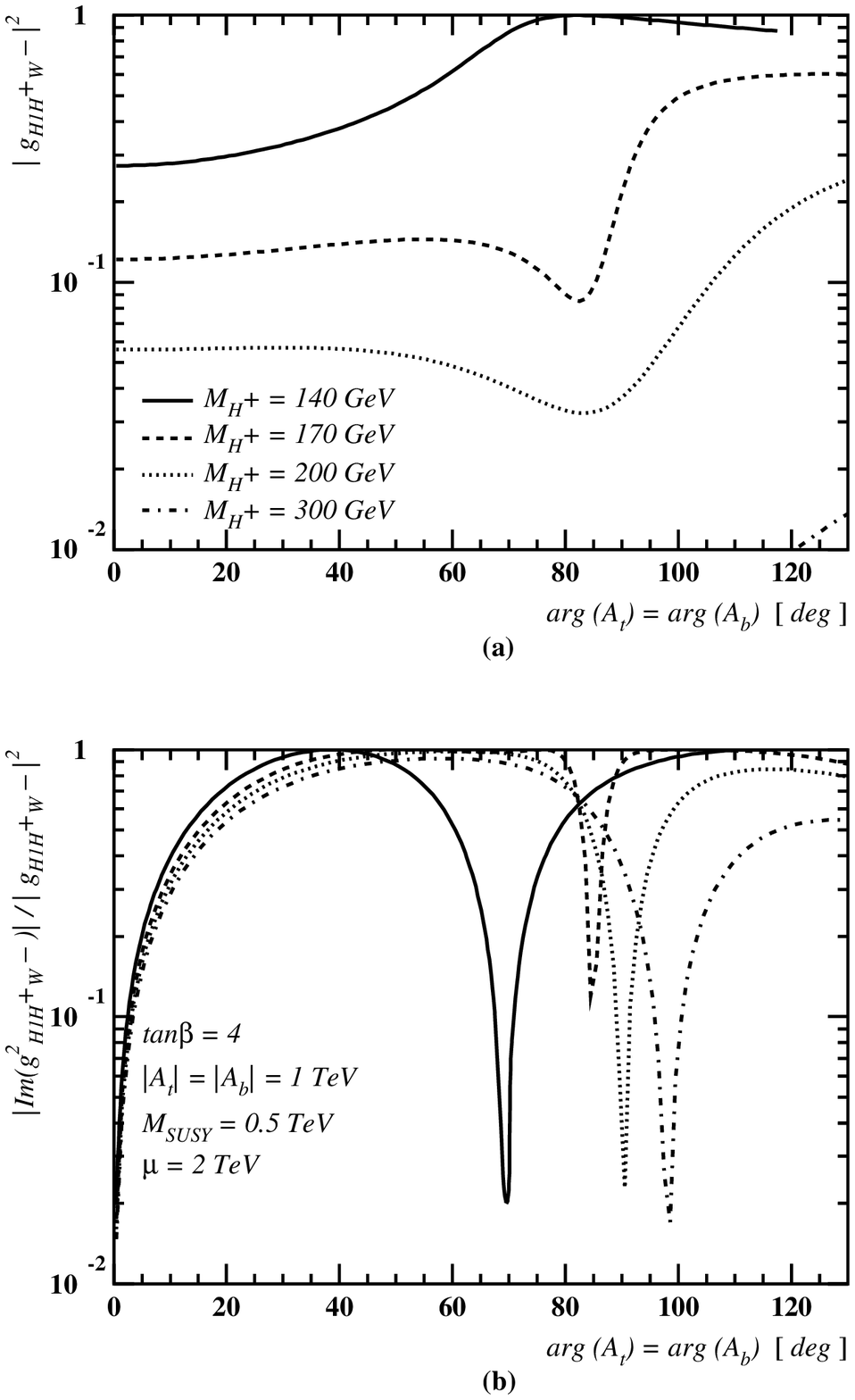}
 \end{center}
 \vspace{-0.5cm} 
\caption{Numerical estimates of (a) $|g_{H_1H^+W^-}|^2$ and (b) 
  $|{\rm Im}( g^2_{H_1H^+W^-} )|/|g_{H_1H^+W^-}|^2$  as a  function of
  ${\rm arg} (A_t )$. The  definition of  $g_{H_1H^+W^-}$ is given  in
  Eq.\ (\ref{gHplus}).}
\label{fig:scp13}
\end{figure}

In Fig.\ \ref{fig:scp13}, we analyze  the variation of the coupling of
the charged  Higgs bosons $H^\pm$ to the  lightest neutral Higgs boson
$H_1$ and the $W^{\mp}$ gauge bosons, for the same parameters as those
considered in Fig.\ \ref{fig:scp5}.   The $H_1H^+W^-$ coupling may  be
defined    from    Eq.\ (\ref{HpHW})  without    including   the  weak
gauge-coupling factor $g_w/2$, i.e.\ 
\begin{equation}
  \label{gHplus}
g_{H_1H^+W^-}\ =\ c_\beta O_{33}\, -\, s_\beta O_{23}\, +\, iO_{13}\, .
\end{equation}
The $H_1H^+W^-$ coupling is relatively small when the coupling $H_1ZZ$
is   large, while it    is  enhanced  when  the  coupling  $H_1ZZ$  is
suppressed.  On  the  other hand, a  measure  of  CP  violation in the
$H_1H^+W^-$  vertex may be obtained  by  analyzing the CP-odd quantity
$|{\rm  Im}(g^2_{H_1H^+W^-})|/|g_{H_1H^+W^-}|^2$  (see    also   Fig.\ 
\ref{fig:scp13}(b)).  This  CP-odd quantity  shows a  very interesting
kinematic behaviour, as it  can become of  order 1 in large regions of
the parameter space  with $M_{H^+} \simlt  300$ GeV.  A consequence of
large CP violation in the $H_1H^+W^-$ coupling is that the decay rates
for $H^+ \to H_1 W^+$ and  $H^-\to H_1 W^-$  may become very different
\cite{LL}.

\begin{figure}
   \leavevmode
 \begin{center}
   \epsfxsize=16.0cm
    \epsffile[0 0 539 652]{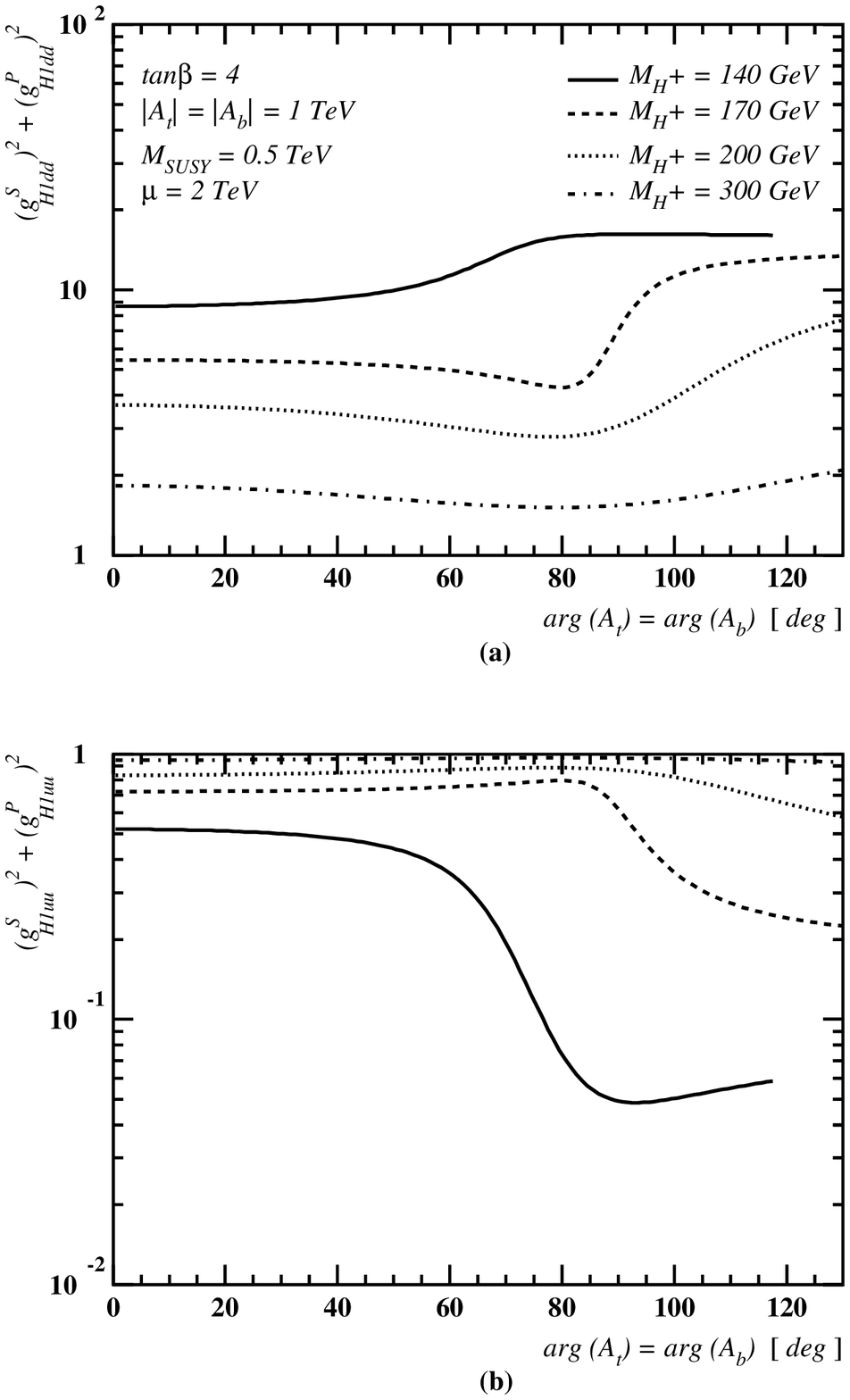}
 \end{center}
 \vspace{-0.5cm} 
\caption{Numerical estimates of (a) $(g^S_{H_1dd})^2 +
  (g^P_{H_1dd})^2$ and (b) $(g^S_{H_1uu})^2 + (g^P_{H_1uu})^2$ versus
  ${\rm arg} (A_t )$.  The definition of the couplings
  $g^{S,P}_{H_idd}$ and $g^{S,P}_{H_iuu}$, with $i=1,2$, is given in
  Eq.\ (\ref{gHff}).}
\label{fig:scp14}
\end{figure}
\begin{figure}
   \leavevmode
 \begin{center}
   \epsfxsize=16.0cm
    \epsffile[0 0 539 652]{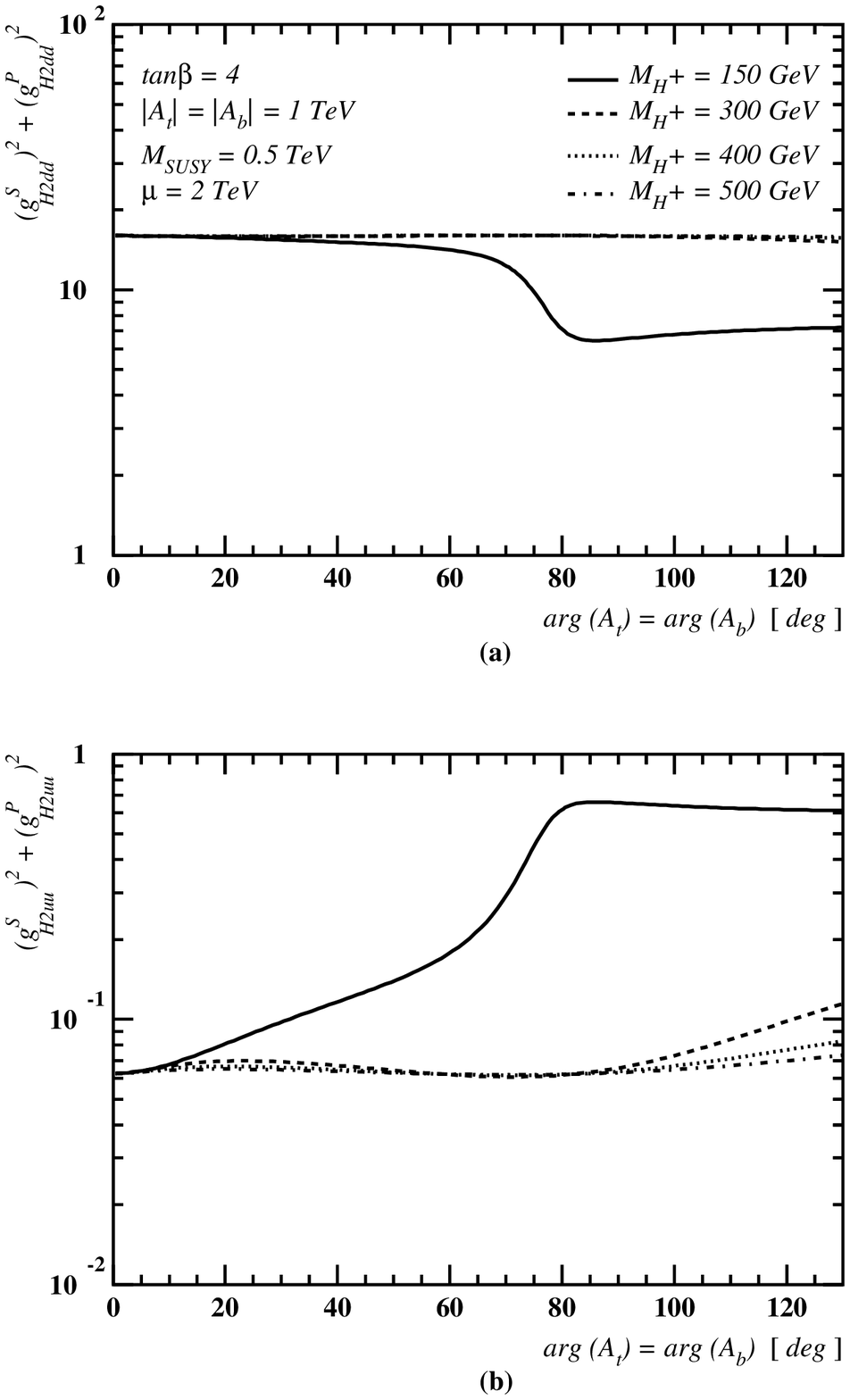}
 \end{center}
 \vspace{-0.5cm} 
\caption{Numerical estimates of (a) $(g^S_{H_2dd})^2 +
  (g^P_{H_2dd})^2$ and (b) $(g^S_{H_2uu})^2 + (g^P_{H_2uu})^2$ as a
  function of ${\rm arg} (A_t )$.  The definition of the couplings
  $g^{S,P}_{H_idd}$ and $g^{S,P}_{H_iuu}$, with $i=1,2$, is given in
  Eq.\ (\ref{gHff}).}
\label{fig:scp16}
\end{figure}
\begin{figure}
   \leavevmode
 \begin{center}
   \epsfxsize=16.0cm
    \epsffile[0 0 539 652]{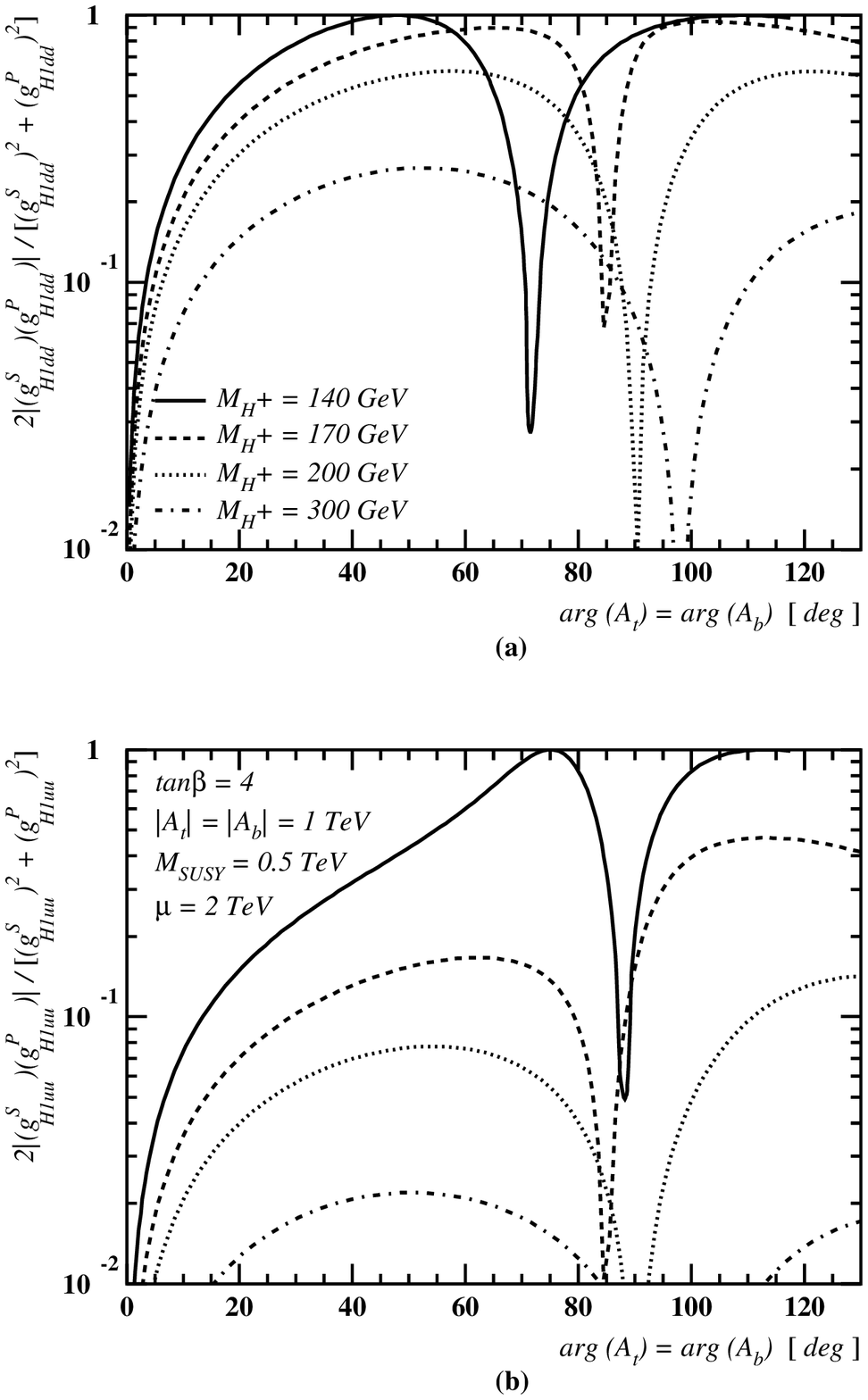}
 \end{center}
 \vspace{-0.5cm} 
\caption{Numerical estimates of (a) $2|(g^S_{H_1dd})\, (g^P_{H_1dd})|/
  [(g^S_{H_1dd})^2 + (g^P_{H_1dd})^2]$ and (b) $2|(g^S_{H_1uu})\,
  (g^P_{H_1uu})|/[(g^S_{H_1uu})^2 + (g^P_{H_1uu})^2]$ as a function of
  ${\rm arg} (A_t )$.  The definition of the couplings
  $g^{S,P}_{H_idd}$ and $g^{S,P}_{H_iuu}$, with $i=1,2$, is given in
  Eq.\ (\ref{gHff})}
\label{fig:scp15}
\end{figure}
\begin{figure}
   \leavevmode
 \begin{center}
   \epsfxsize=16.0cm
    \epsffile[0 0 539 652]{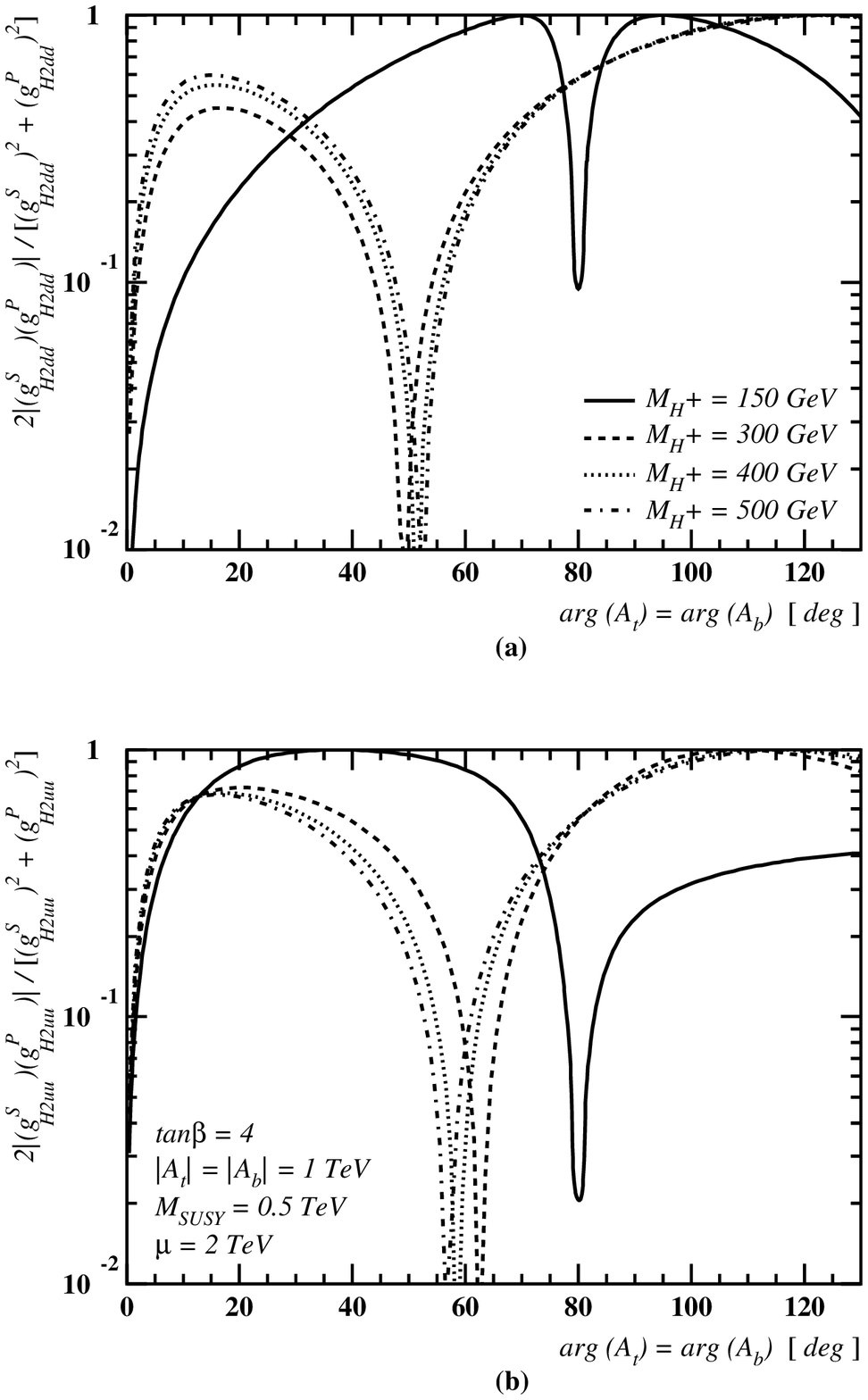}
 \end{center}
 \vspace{-0.5cm} 
\caption{Numerical estimates of (a) $2|(g^S_{H_2dd})\, (g^P_{H_2dd})|/
  [(g^S_{H_2dd})^2 + (g^P_{H_2dd})^2]$ and (b)
  $2|(g^S_{H_2uu})\, (g^P_{H_2uu})|/[(g^S_{H_2uu})^2 + (g^P_{H_2uu})^2]$
  as a function of ${\rm arg} (A_t )$. The definition of the couplings
  $g^{S,P}_{H_idd}$ and $g^{S,P}_{H_iuu}$, with $i=1,2$, is given in
  Eq.\ (\ref{gHff}).}
\label{fig:scp17}
\end{figure}

In  the following, we consider  the  interactions of the neutral Higgs
fields with the  fermions.  These interactions  may be obtained by the
Lagrangian
\begin{eqnarray}
  \label{Hiff}
{\cal L}_{H\bar{f}f} &=& -\, \sum_{i=1}^3 H_{(4-i)}\,
\Big[\, \frac{g_w m_d}{2M_W c_\beta}\, \bar{d}\,( O_{2i}\, -\,
is_\beta O_{1i}\gamma_5 )\, d\nonumber\\
&& +\, \frac{g_w m_u}{2M_W s_\beta}\, \bar{u}\,( O_{3i}\, -\,
ic_\beta O_{1i}\gamma_5 )\, u\, \Big]\, .
\end{eqnarray}
Obviously,  the  Higgs--fermion--fermion couplings are significant for
the third-generation quarks, $t$ and $b$. {}From Eq.\ (\ref{Hiff}), we
readily see that the effect of CP-violating Higgs  mixing is to induce
a simultaneous   coupling of $H_i$,  with $i  =1,2,3$,  to CP-even and
CP-odd   fermionic bilinears  \cite{DM},  e.g.\  to  $\bar{u}   u$ and
$\bar{u}  i\gamma_5 u$.  This can  lead  to sizeable  phenomena of  CP
violation in high-energy processes that involve decays of Higgs bosons
into longitudinally polarized top-quark pairs \cite{darwin/yee}.

For the discussion  that follows, it  proves convenient to  define the
following parameters:
\begin{eqnarray}
  \label{gHff}
g^S_{H_1uu} \!\!\!&=&\!\!\! O_{33}/s_\beta\,,\quad 
g^P_{H_1uu} = O_{13}\, \cot\beta\,,\quad
g^S_{H_2uu} = O_{32}/s_\beta\,,\quad 
g^P_{H_2uu} = O_{12}\, \cot\beta\,,\nonumber\\
g^S_{H_1dd} \!\!\!&=&\!\!\! O_{23}/c_\beta\,,\quad 
g^P_{H_1dd} = O_{13}\, \tan\beta\,,\quad
g^S_{H_2dd} = O_{22}/c_\beta\,,\quad 
g^P_{H_2dd} = O_{12}\, \tan\beta\, .\qquad
\end{eqnarray}
These parameters represent  the  scalar and pseudoscalar couplings  of
the  Higgs bosons $H_1$  and $H_2$ to  the up- and down-type fermions,
normalized to the SM values.  Then, the  partial decay widths of $H_1$
and   $H_2$ in the  MSSM   may  be obtained    by the  SM ones,  after
multiplying the    latter     by the  effective    coupling    factors
$[(g^S_{H_1ff})^2     + (g^P_{H_1ff})^2]$   and   $[(g^S_{H_2ff})^2  +
(g^P_{H_2ff})^2]$   (with    $f=u,d$),     respectively.   In   Figs.\ 
\ref{fig:scp14}  and \ref{fig:scp16},  we plot the  effective coupling
factors   as  a function of arg$(A_t)$.     The behaviour  observed is
similar to  the  CP-conserving case.   To be  specific,  the effective
coupling factor related to  the $H_i\bar{f}f$ coupling (with  $i=1,2$)
is  close to the SM   value, whenever the  respective $H_iVV$ coupling
approaches  1, while  in  the  regions where  the  $H_iVV$ coupling is
suppressed,   the $H_i$ coupling   to down  (up) fermions  is enhanced
(suppressed) by a factor $\tan\beta$ ($1/\tan\beta$).

One may now construct quantities that can  provide a realistic measure
of CP violation  in  the  $H_1\bar{f}f$ and  $H_2\bar{f}f$  couplings,
i.e.\  $2|(g^S_{H_iff})\,   (g^P_{H_iff})|/[(g^S_{H_iff})^2          +
(g^P_{H_iff})^2]$.   Figures    \ref{fig:scp15}   and  \ref{fig:scp17}
exhibit the dependence of these CP-violating quantities related to the
$H_1$ and $H_2$  bosons, respectively,  as a  function of  arg$(A_t)$. 
Figure \ref{fig:scp15} reveals  that the CP-violating component of the
$H_1\bar{f}f$  coupling is large   only for  relatively light  charged
Higgs-boson  masses, $M_{H^+} \simlt 180$  GeV.  On the other hand, as
can be seen from Fig.\ \ref{fig:scp17},  the CP-violating component of
the $H_2\bar{f}f$ coupling may become  of order 1 for  a wide range of
the parameter space, even   for  heavier charged  Higgs-boson  masses,
e.g.\ $M_{H^+} \approx 500$ GeV.

%******************************************************************
%%% The Radiative H2bb Figure 
%******************************************************************
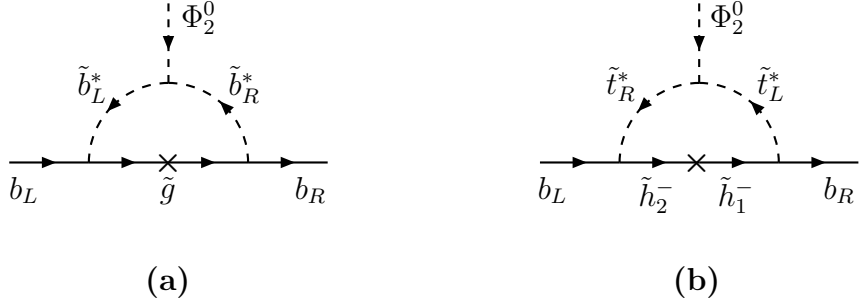
\begin{figure}

\begin{center}
\begin{picture}(320,150)(0,0)
\SetWidth{0.8}
 
\ArrowLine(0,70)(30,70)\ArrowLine(30,70)(60,70)
\ArrowLine(60,70)(90,70)\ArrowLine(90,70)(120,70)
\DashArrowArc(60,70)(30,90,180){3}\DashArrowArc(60,70)(30,0,90){3}
\DashArrowLine(60,130)(60,100){3}
\Text(60,70)[]{\boldmath $\times$}
\Text(0,65)[lt]{$b_L$}\Text(120,65)[rt]{$b_R$}
\Text(60,65)[t]{$\tilde{g}$}\Text(65,125)[l]{$\Phi^0_2$}
\Text(37,100)[r]{$\tilde{b}^*_L$}\Text(83,100)[l]{$\tilde{b}^*_R$}

\Text(60,25)[]{\bf (a)}

\ArrowLine(200,70)(230,70)\ArrowLine(230,70)(260,70)
\ArrowLine(260,70)(290,70)\ArrowLine(290,70)(320,70)
\DashArrowArc(260,70)(30,90,180){3}\DashArrowArc(260,70)(30,0,90){3}
\DashArrowLine(260,130)(260,100){3}
\Text(260,70)[]{\boldmath $\times$}
\Text(200,65)[lt]{$b_L$}\Text(320,65)[rt]{$b_R$}
\Text(245,65)[t]{$\tilde{h}^-_2$}\Text(275,65)[t]{$\tilde{h}^-_1$}
\Text(265,125)[l]{$\Phi^0_2$}\Text(237,100)[r]{$\tilde{t}^*_R$}
\Text(283,100)[l]{$\tilde{t}^*_L$}

\Text(260,25)[]{\bf (b)}

\end{picture}
\end{center}
\vspace{-1.cm}
\caption{Feynman graphs mediated by the exchange of (a) gluinos
  $\tilde{g}$ and (b) Higgsinos $\tilde{h}^-_{1,2}$ that give rise
  to an effective one-loop $\Phi^0_2\bar{b}b$ coupling.}\label{f2}
\end{figure}

Apart from the CP-violating effects  generated by the radiative mixing
of  the Higgs fields  which we consider here in  detail,  there may be
important CP-violating effects induced by one-loop vertex corrections.
In  the   leptonic sector, these     corrections  are generally  small
\cite{BKW}.  Because of the large    Yukawa and colour-enhanced    QCD
interactions, however,  the radiative  corrections  to the  tree-level
$b$-quark  couplings to the  Higgs bosons \cite{EMa}  may be important
when the relevant Higgs-mass eigenstate has dominant components in the
Higgs   doublet $\Phi_2$  \cite{dmb,sola}   or, equivalently, when the
matrix  elements  $O_{2i}$ and $s_\beta  O_{1i}$  related to the Higgs
fields  $H_{(4-i)}$ are small.  In  the  CP-conserving case, the Higgs
boson  that couples predominantly  to the gauge  bosons always fulfils 
these properties \cite{CMW}, i.e.\ $O_{12}$  and $O_{13}$ are zero for
the CP-even Higgs states in this case.

The     general    analytic    expression       for    the   effective
Higgs-boson--bottom-quark  coupling \cite{CMW}  may  be  obtained   by
considering the vertex  graphs shown in  Fig.\ \ref{f2}.  However, the
magnitude of these corrections and  the  phases involved in these  are
strongly model-dependent.  Furthermore, the gluino-exchange diagram in
Fig.\ \ref{f2}(a) usually  represents the dominant contribution, which
has   a  counterpart in the  effective  Higgs  potential   only at the
two-loop level.  To be precise, the effective one-loop Yukawa coupling
of the  $b$  quark to the  neutral   field component of  the  $\Phi_2$
doublet, $\Phi^0_2 = \phi_2 + i a_2$, is given by
\begin{equation}
{\cal L}_{\Phi^0_2 \bar{b}b}\ =\ 
\Delta h_b\, \Phi^0_2\, \bar{b}_L b_R\ +\  {\rm  H.c.}\, ,
\end{equation}
with 
\begin{equation}
  \label{Dhb}
\frac{\Delta h_b}{h_b}\ =\ 
\frac{2 \alpha_s}{3\pi} m_{\tilde{g}} \mu I(m_{\tilde{b}_1}^2,
m_{\tilde{b}_2}^2,|m_{\tilde{g}}|^2)\ +\ \frac{h^2_t}{16\pi^2} A_t \mu
I(m_{\tilde{t}_1}^2,m_{\tilde{t}_2}^2,|\mu|^2)\, ,
\end{equation}
where  $\alpha_s = g^2_s/(4\pi)$   is  the  SU(3)$_c$  fine  structure
constant,  $h_t$   is the    top Yukawa    coupling, $m_{\tilde{b}_1}$
($m_{\tilde{t}_1}$)  and $m_{\tilde{b}_2}$ ($m_{\tilde{b}_2}$) are the
mass eigenvalues of the scalar-bottom  (top) quarks, and $I(a,b,c)$ is
the one-loop function
\begin{equation}
I(a,b,c)\ =\ \frac{ a b \ln (a/b) + b c \ln (b/c) + a c \ln (c/a)}
{(a-b)(b-c)(a-c)}\ .
\end{equation}
Note  that the total bottom-mass  corrections  depend on the  relative
size of   the  gluino-mediated graphs  to  the Higgsino-mediated ones,
shown in Figs.\ \ref{f2}(a) and \ref{f2}(b), respectively.  Therefore,
as was mentioned above, the two different quantum corrections strongly
depend on the  model under study.  In fact,  it is  straightforward to
introduce these quantum   effects    in the  analysis of  the    Higgs
couplings.  Taking both  CP-violating vertex and  Higgs-mixing effects
into  consideration,  the  effective  Lagrangian for the $H_i\bar{b}b$
couplings reads
\begin{eqnarray}
  \label{newHiff}
{\cal L}^{\rm eff}_{H\bar{b}b} &=& -\, \sum_{i=1}^3 H_{(4-i)}\ \bar{b}\,
\Big\{ \Big[\, h_b\, O_{2i}\, +\, {\rm Re}(\Delta h_b) O_{3i}\, -\,
{\rm Im}(\Delta h_b) c_\beta
O_{1i} \,\Big]
\nonumber\\
&& -\, i \Big[\, h_b\, s_\beta O_{1i}\, -\, 
{\rm Re}(\Delta h_b)\,c_\beta O_{1i}\, -\, {\rm Im}(\Delta h_b)
O_{3i} \,\Big]
\gamma_5  \Big\}\, b \, ,
\end{eqnarray}
where 
\begin{equation}
h_b\ =\ \frac{g_w m_b}{2 M_W c_\beta\, | 1 + (\Delta h_b/h_b)
 \tan\beta e^{i\xi}|}
\end{equation}
and $\Delta h_b/h_b$ is given in Eq. (\ref{Dhb}).

As we have detailed above, the actual size  of the CP-violating vertex
effects depends both on $\phi_{\rm CP}$ and arg$(m_{\tilde{g}} \mu )$.
The latter phase   does not  directly enter   the  calculation of  the
one-loop effective Higgs potential.   To avoid  excessive complication
in the analysis, we have decided  to present the results assuming that
the vertex effects are very small, which is  typically true for values
of $\tan\beta \simlt 4$, like  the ones  considered here.  For  larger
values  of $\tan\beta$,  instead,   these  effects can   no  longer be
ignored.  However, a specific model is then needed in order to be able
to  determine  their significance. For    example, for an  appropriate
choice of SUSY-mass parameters and phases, the vertex effects can even
be  tuned  to   zero independently  of  the CP-violating  Higgs-mixing
effects.

\section{Conclusions}

We have performed a  systematic study of the  Higgs sector of the MSSM
with explicit CP  violation,  and analyzed the   main phenomenological
implications  of such a theory for  direct searches of Higgs bosons at
LEP2 and the upgraded  Tevatron collider.  In this general theoretical
framework, the tree-level CP invariance of the MSSM Higgs potential is
considered to be sizeably  broken  by loop graphs involving  trilinear
CP-violating  couplings of the scalar  top and  bottom quarks to Higgs
bosons  \cite{APLB}.  These loop   effects  are taken into account  by
calculating the CP-violating RG-improved effective potential up to the
next-to-leading order, in  which two-loop leading  logarithms of $t$-,
$b$-Yukawa and QCD corrections have been included \cite{CEQW}.

The analysis shows that the   upper bound on the lightest  Higgs-boson
mass $M_{H_1}$  obtained in the CP-violating  MSSM is almost identical
to that already  derived in the  CP-invariant theory, for both low and
high   $\tan\beta$   values.    Nevertheless,    the   Higgs-sector CP
non-conservation  may  drastically modify the   couplings of the $H_1$
scalar to the $Z$ and $W$ bosons.  In fact, the production rate of the
$H_1$ boson is  considerably  affected at  LEP2, for relatively  light
charged Higgs-boson masses,  i.e.\ for $120  < M_{H^+} < 200$ GeV, and
for small   and intermediate values  of   $\tan\beta$, i.e.\ $2 \simlt
\tan\beta \simlt  5$.     For  larger  values  of  $\tan\beta$,    the
soft-CP-violating   couplings of  the  third  generation  are severely
constrained by  the  two-loop SUSY Barr--Zee-type contribution  to the
electron  EDM \cite{CKP}, leading  to   much  weaker CP-odd  effects.  
Because of the  drastic   modification of the $H_1ZZ$   and  $H_1H_2Z$
couplings  for low- and  intermediate-$\tan\beta$  scenarios, we  find
that the  current experimental lower bound  on  the mass of  the $H_1$
particle may  be dramatically  relaxed up  to the 60-GeV  level in the
presence of large  CP violation in  the Higgs sector  of the MSSM (see
Fig.\ \ref{fig:scp7}(b)).  Therefore, a combined experimental analysis
is required by considering all possible reactions that contain the two
lightest neutral Higgs bosons, $H_1$  and $H_2$, and the charged Higgs
bosons $H^\pm$ in the final state, which  are produced either singly or
in pairs.  In  this respect, the upgraded Tevatron  collider and, to a
greater extent,  the LHC are  the almost ideal  places to explore  more
efficiently the parameter space of the CP-violating MSSM.

Another consequence   of such a minimal SUSY   scenario of explicit CP
violation is that  the mass splitting  between the two heaviest  Higgs
bosons $H_2$ and  $H_3$  may be  of   order 20$\%$ for heavy   charged
Higgs-boson  masses,  $M_{H^\pm}  \approx 300$  GeV   (see  also Fig.\ 
\ref{fig:scp1}(b)).  This  is   in   agreement with   earlier  results
reported in \cite{APLB}.  Furthermore,  we find that the strong mixing
of the Higgs bosons may induce large CP violation in the vertices $H_1
H^\pm W^\mp$, $H_1\bar{d}d$, $H_2\bar{d}d$,  etc., that could  even be
of order  unity.  Most interestingly,  CP violation  can be resonantly
enhanced  in high-energy  reactions mediated  by  scalar--pseudoscalar
transitions that involve the nearly degenerate $H_2$ and $H_3$ states,
especially when the mass  difference $M_{H_2} - M_{H_3}$ is comparable
to the decay   widths $\Gamma_{H_2}$ and $\Gamma_{H_3}$  \cite{ANPB}.  
Such  resonant  effects of   CP   violation may   be tested  at future
high-energy  colliders  such   as   the     LHC, NLC and    the    FMC
\cite{ANPB,PN,Osland,LHC,APRL,CD}.  Finally, the analysis presented in
this  paper   clearly demonstrates that   the   MSSM with  explicit CP
violation in the Higgs sector  constitutes an interesting  theoretical
framework, which  will have a significant  impact on $B$-meson decays,
dark-matter searches and electroweak baryogenesis.

\subsection*{Acknowledgements} 
We wish to thank Jisuke Kubo for useful discussions, and Patrick Janot
for providing us with experimental plots concerning Table 2.

\subsection*{Note added}
While finalizing our  paper, we became aware of  two very recent works
\cite{Demir,GGK}  that treat some of the  topics we have been studying
here.  In  \cite{Demir}, the author  mainly concentrated on the effect
of CP-violating Higgs  mixing on  the Higgs--fermion--fermion coupling
for relatively low values of $\mu$, i.e.\ $\mu  = 250$ GeV, and $M_a =
200$ GeV.  In this regime, CP violation in the  sector of the lightest
Higgs boson $H_1$ is  small.  Instead, we effectively consider smaller
values of $M_a$, i.e.\ $120 \simlt M_{H^+} \simlt 180$ GeV, and higher
values of $\mu$, which can  lead to significantly larger  CP-violating
effects in the $H_1$ sector.  In this  respect, our conclusions differ
from  those presented  in  \cite{Demir}; otherwise,  we find agreement
with the results regarding the $H_2\bar{f}f$ coupling.  The authors of
Ref.\ \cite{GGK} discuss  Higgs-boson production cross sections  for a
future $e^+e^-$ NLC in a general CP-violating two-Higgs-doublet model.
Here, instead,  we are mainly interested  in possible effects  at LEP2
and the upgraded Tevatron collider,  within the MSSM with  radiatively
induced      CP  violation in   the  Higgs    sector.   In addition to
\cite{Demir,GGK}, we present  analytic expressions for the Higgs-boson
masses  and mixing  angles,  and pay  particular  attention to the EDM
constraints.

\newpage

\def\theequation{\Alph{section}.\arabic{equation}}
\begin{appendix}
\setcounter{equation}{0}
\section{Analytic expressions of quartic couplings}

The dominant   contribution of    radiative interactions to    quartic
couplings  comes from  enhanced      Yukawa couplings of   the   third
generation. The relevant Lagrangians \cite{GH}, including CP-violating
sources, are given by
\begin{eqnarray}
  \label{Lsoft}
-{\cal L}_{\rm soft} & = & \widetilde{M}^2_Q \widetilde{Q}^\dagger 
\widetilde{Q}\, +\, \widetilde{M}^2_U \widetilde{U}^* \widetilde{U}\, +\,    
\widetilde{M}^2_D \widetilde{D}^* \widetilde{D}\, \nonumber\\
&&+\, \Big(\, h_bA_b\,\Phi_1^\dagger \widetilde{Q} \widetilde{D}\, -\,  
h_tA_t\,\Phi_2^T i\tau_2 \widetilde{Q} \widetilde{U}\ +\ {\rm H.c.}\, \Big),\\
  \label{LF}
-{\cal L}_F & = & h^2_b\, |\Phi^+_1 \widetilde{Q}|^2\, +\, 
h^2_t\, |\Phi^T_2 i\tau_2 \widetilde{Q}|^2\, \nonumber\\
&&-\, \Big(\, \mu h_b\, \widetilde{Q}^\dagger \Phi_2 \widetilde{D}^*\, +\,
\mu h_t \widetilde{Q}^\dagger i\tau_2 \Phi_1^* \widetilde{U}^*\ +\
{\rm H.c.}\, \Big) \nonumber\\
&& -\, \Big(\, h_b\, \widetilde{D}^* \Phi_1^T i\tau_2\, +\,
h_t \widetilde{U}^* \Phi_2^\dagger \Big)\, \Big(\, h_b\, i\tau_2 \Phi_1^*
\widetilde{D}\, - \, h_t\, \Phi_2 \widetilde{U}\, \Big) ,\\
  \label{LD}
-{\cal L}_D & = & \frac{g_w^2}{4}\, \Big[\, 2|\Phi^T_1 i\tau_2
\widetilde{Q}|^2\, +\, 2|\Phi^\dagger_2 \widetilde{Q}|^2\,
-\, \widetilde{Q}^\dagger \widetilde{Q}\, (\Phi^\dagger_1 \Phi_1\,
+\, \Phi_2^\dagger \Phi_2)\, \Big]\nonumber\\
&&+\, \frac{g'^2}{4}\, ( \Phi^\dagger_2 \Phi_2\, -\, \Phi_1^\dagger \Phi_1)\,
\Big[\, \frac{1}{3}\, (\widetilde{Q}^\dagger \widetilde{Q})\, -\,
\frac{4}{3}\, (\widetilde{U}^* \widetilde{U})\, +\,
\frac{2}{3}\, (\widetilde{D}^* \widetilde{D})\, \Big]\, ,\\
  \label{Lfermions} 
-{\cal L}_{\rm fermions} & = & h_b\, \Big[\, \bar{b}_R 
(t_L,b_L) \Phi^*_1\, +\, {\rm H.c.}\, \Big]\, +\, 
h_t\, \Big[\, \bar{t}_R (t_L,b_L) i\tau_2 \Phi_2\ +\ {\rm H.c.}\, \Big]\, ,
\end{eqnarray}
with $\widetilde{Q}^T = (\tilde{t}_L,\tilde{b}_L)$, $\widetilde{U}^*
= \tilde{t}_R$, $\widetilde{D}^* = \tilde{b}_R$.  In addition, we also
consider next-to-leading  order   QCD   quantum   corrections.   These
corrections have been computed  in full detail in \cite{KYS,CEQR}, and
may easily be implemented in the analysis, as they form a CP-invariant
subset of graphs by themselves.

Employing        the                interaction            Lagrangians
(\ref{Lsoft})--(\ref{Lfermions}),  it is straightforward to extend the
two-loop analytic results in \cite{CEQW} to  the case of CP violation. 
In this way, we find
\begin{eqnarray}
  \label{lambda1}
\lambda_1 \!\!&=&\!\!
   -\, \frac{g_w^2+g'^2}{8}\  \Big( 1\, -\, \frac{3}{8\pi^2}\
          h^2_b\, t \Big)\nonumber\\
   &&-\,\frac{3}{16\pi^2}\ h^4_b\, \Big[\, t\, +\, \frac{1}{2}\, X_b\, +\,
   \frac{1}{16\pi^2}\, \Big(\, \frac{3}{2}\, h^2_b\, +\, \frac{1}{2}\,
   h^2_t\, -\, 8g^2_s\, \Big)\, (X_b t\, +\, t^2 )\, \Big]\nonumber\\
   &&+\, \frac{3}{192\pi^2}\, h^4_t\, \frac{|\mu|^4}{M^4_{\rm SUSY}}\, 
   \Big[\, 1\, +\, \frac{1}{16\pi^2}\, (9h^2_t\, -\, 5h^2_b\, -\, 
   16g^2_s)t\, \Big] ,\\
  \label{lambda2}
\lambda_2 \!\!&=&\!\!
   -\, \frac{g_w^2+g'^2}{8}\  \Big( 1\, -\, \frac{3}{8\pi^2}\
          h^2_t\, t \Big)\nonumber\\
   &&-\,\frac{3}{16\pi^2}\ h^4_t\, \Big[\, t\, +\, \frac{1}{2}\, X_t\, +\,
   \frac{1}{16\pi^2}\, \Big(\, \frac{3}{2}\, h^2_t\, +\, \frac{1}{2}\,
   h^2_b\, -\, 8g^2_s\, \Big)\, (X_t t\, +\, t^2 )\, \Big]\nonumber\\
   &&+\, \frac{3}{192\pi^2}\, h^4_b\, \frac{|\mu|^4}{M^4_{\rm SUSY}}\, 
   \Big[\, 1\, +\, \frac{1}{16\pi^2}\, (9h^2_b\, -\, 5h^2_t\, -\, 
   16g^2_s)t\, \Big] ,\\
  \label{lambda3}
\lambda_3 \!\!&=&\!\!
   -\, \frac{g_w^2-g'^2}{4}\  \Big[\, 1\, -\, \frac{3}{16\pi^2}\
          (h^2_t\, +\, h^2_b) \, t\, \Big]\nonumber\\
   &&-\,\frac{3}{8\pi^2}\ h^2_th^2_b\, \Big[\, t\, +\, \frac{1}{2}\,
          X_{tb}\, +\, \frac{1}{16\pi^2}\, 
   ( h^2_t\, +\, h^2_b\, -\, 8g^2_s)\, (X_{tb} t\, +\, t^2 )\, \Big]\nonumber\\
   &&-\, \frac{3}{96\pi^2}\, h^4_t\, \Big(\, \frac{3|\mu|^2}{M^2_{\rm
   SUSY}}\, -\, \frac{|\mu|^2 |A_t|^2}{M^4_{\rm SUSY}}\, \Big)\,
   \Big[\, 1\, +\, \frac{1}{16\pi^2}\, (6h^2_t\, -\, 2h^2_b\, -\, 
   16g^2_s)t\, \Big]\nonumber\\
   &&-\, \frac{3}{96\pi^2}\, h^4_b\, \Big(\, \frac{3|\mu|^2}{M^2_{\rm
   SUSY}}\, -\, \frac{|\mu|^2 |A_b|^2}{M^4_{\rm SUSY}}\, \Big)\,
   \Big[\, 1\, +\, \frac{1}{16\pi^2}\, (6h^2_b\, -\, 2h^2_t\, -\, 
   16g^2_s)t\, \Big] ,\\
  \label{lambda4}
\lambda_4 \!\!&=&\!\! \frac{g_w^2}{2}\  \Big[\, 1\, -\, \frac{3}{16\pi^2}\
          (h^2_t\, +\, h^2_b) \, t\, \Big]\nonumber\\
   &&+\,\frac{3}{8\pi^2}\ h^2_th^2_b\, \Big[\, t\, +\, \frac{1}{2}\,
          X_{tb}\, +\, \frac{1}{16\pi^2}\, 
   ( h^2_t\, +\, h^2_b\, -\, 8g^2_s)\, (X_{tb} t\, +\, t^2 )\, \Big]\nonumber\\
   &&-\, \frac{3}{96\pi^2}\, h^4_t\, \Big(\, \frac{3|\mu|^2}{M^2_{\rm
   SUSY}}\, -\, \frac{|\mu|^2 |A_t|^2}{M^4_{\rm SUSY}}\, \Big)\,
   \Big[\, 1\, +\, \frac{1}{16\pi^2}\, (6h^2_t\, -\, 2h^2_b\, -\, 
   16g^2_s)t\, \Big]\nonumber\\
   &&-\, \frac{3}{96\pi^2}\, h^4_b\, \Big(\, \frac{3|\mu|^2}{M^2_{\rm
   SUSY}}\, -\, \frac{|\mu|^2 |A_b|^2}{M^4_{\rm SUSY}}\, \Big)\,
   \Big[\, 1\, +\, \frac{1}{16\pi^2}\, (6h^2_b\, -\, 2h^2_t\, -\, 
   16g^2_s)t\, \Big] ,\\
  \label{lambda5}
\lambda_5 \!\!&=&\!\! \frac{3}{192\pi^2}\, h^4_t\,
   \frac{\mu^2 A_t^2}{M^4_{\rm SUSY}}\,
   \Big[\, 1\, -\, \frac{1}{16\pi^2}\, (2h^2_b\, -\, 6h^2_t\, +\, 
   16g^2_s)t\, \Big]\nonumber\\
   &&+\, \frac{3}{192\pi^2}\, h^4_b\, 
   \frac{\mu^2 A_b^2}{M^4_{\rm SUSY}}\, 
   \Big[\, 1\, -\, \frac{1}{16\pi^2}\, (2h^2_t\, -\, 6h^2_b\, +\, 
   16g^2_s)t\, \Big] ,\\
  \label{lambda6}
\lambda_6 \!\!&=&\!\! -\, \frac{3}{96\pi^2}\, h^4_t\,
   \frac{|\mu|^2 \mu A_t}{M^4_{\rm SUSY}}\,
   \Big[\, 1\, -\, \frac{1}{16\pi^2}\, \Big(\, \frac{7}{2}\, h^2_b\, 
   -\, \frac{15}{2}\, h^2_t\, +\, 16g^2_s\, \Big)\, t\, \Big]\nonumber\\
   &&\!\!\!+\, \frac{3}{96\pi^2}\, h^4_b\, \frac{\mu}{M_{\rm SUSY}}\, 
   \Big(\, \frac{6A_b}{M_{\rm SUSY}}\, -\, \frac{|A_b|^2 A_b}{M^3_{\rm
   SUSY}}\, \Big)
  \Big[\, 1\, -\, \frac{1}{16\pi^2}\, \Big(\, \frac{1}{2}\,
   h^2_t\, -\, \frac{9}{2}\, h^2_b\, +\, 16g^2_s\, \Big)\, t\, \Big], \qquad\\
  \label{lambda7}
\lambda_7 \!\!&=&\!\! -\, \frac{3}{96\pi^2}\, h^4_b\,
   \frac{|\mu|^2 \mu A_b}{M^4_{\rm SUSY}}\,
   \Big[\, 1\, -\, \frac{1}{16\pi^2}\, \Big(\, \frac{7}{2}\, h^2_t\, 
   -\, \frac{15}{2}\, h^2_b\, +\, 16g^2_s\, \Big)\, t\, \Big]\nonumber\\
   &&\!\!\!+\, \frac{3}{96\pi^2}\, h^4_t\, \frac{\mu}{M_{\rm SUSY}}\, 
   \Big(\, \frac{6A_t}{M_{\rm SUSY}}\, -\, 
   \frac{|A_t|^2 A_t}{M^3_{\rm SUSY}}\,\Big) 
   \Big[\, 1\, -\, \frac{1}{16\pi^2}\, \Big(\, \frac{1}{2}\,
   h^2_b\, -\, \frac{9}{2}\, h^2_t\, +\, 16g^2_s\, \Big)\, t\, \Big] ,
\end{eqnarray}
where $t = \ln ( M^2_{\rm SUSY}/\overline{m}^2_t )$ and
\begin{eqnarray}
  \label{hthb}
h_t &=& \frac{\sqrt{2}\, m_t (\overline{m}_t)}{v \sin\beta }\ ,\, \qquad
h_b \ \, =\ \, \frac{\sqrt{2}\, m_b (\overline{m}_t)}{v \cos\beta }\ ,\\
  \label{Xtbtb}
X_t &=& \frac{2|A_t|^2}{M^2_{\rm SUSY}}\ \Big(\, 1\, -\, 
     \frac{|A_t|^2}{12M^2_{\rm SUSY}}\, \Big)\, ,\nonumber\\
X_b &=& \frac{2|A_b|^2}{M^2_{\rm SUSY}}\ \Big(\, 1\, -\, 
     \frac{|A_b|^2}{12M^2_{\rm SUSY}}\, \Big)\, , \nonumber\\
X_{tb} &=& \frac{ |A_t|^2+|A_b|^2 + 2{\rm Re}(A^*_bA_t) }
     {2\,M^2_{\rm SUSY}}\, -\, \frac{|\mu|^2}{M^2_{\rm SUSY}}\,
     -\, \frac{|\,|\mu|^2 -A^*_bA_t\,|^2}{6\,M^4_{\rm SUSY}}\ .\quad
\end{eqnarray}
In Eq. (\ref{hthb}), $\overline{m}_t$ is the top-quark pole mass, which
is related to the on-shell running mass $m_t$ through
\begin{equation}
  \label{mt}
m_t (\overline{m}_t)\ =\ \frac{\overline{m}_t}{1 + \frac{4}{3\pi}\, 
 \alpha_s (\overline{m}_t )}\ .
\end{equation}
It  is  important to   remark  that the  above   expressions  are  not
equivalent   to the  ones   that would   be   obtained by   taking the
expressions  given in Ref.\ \cite{CEQW},   and considering all  mixing
parameters to be complex.  If this were  done, incorrect results would
be obtained.

The  RG analysis under consideration  assumes a single step decoupling
of the scalar-quark fields. This assumption is only  valid if the mass
splitting among the scalar-quark  mass eigenstates is relatively small
\cite{CEQW}.  Specifically, the expansion becomes more trustworthy if
\begin{equation}
  \label{split}
\frac{m_{\tilde{t}_1}^2 - m_{\tilde{t}_2}^2}
{m_{\tilde{t}_1}^2 + m_{\tilde{t}_2}^2}\ \simlt\ 0.5,
\end{equation}
where $m_{\tilde{t}_1}^2$ and $m_{\tilde{t}_2}^2$ are the squared mass
eigenvalues of the scalar-top  quarks.    This last restriction   also
applies to  other Higgs-mass analyses  that have been performed at the
next-to-leading  order   \cite{Mh,HH,CEQW}.  If  one  assumes that the
approximate inequality  (\ref{split}) holds   true, the scale  $M_{\rm
  SUSY}^2$ may then be safely defined as the arithmetic average of the
scalar-top mass eigenvalues squared,
\begin{equation}
  \label{Msusy}
M_{\rm SUSY}^2\ =\ \frac{1}{2}\, \Big(\, m_{\tilde{t}_1}^2\, +\,
 m_{\tilde{t}_2}^2\, \Big)\,.
\end{equation}
One should bear  in  mind that  our RG  analysis   also relies on   an
expansion of  the   effective  Higgs potential   up  to operators   of
dimension 4.    The contribution  of higher-dimensional  operators may
only be neglected if $2  |m_t A_t| \simlt M_{\rm  SUSY}^2$ and $2 |m_t
\mu| \simlt   \tan\beta\, M_{\rm  SUSY}^2$.   Moreover, in   the  high
$\tan\beta$  regime, in   which the bottom-Yukawa  interactions become
more relevant, we  have assumed that  the scalar-bottom masses are  of
the order of  the  scalar-top ones and that   similar bounds on  their
respective mixing  mass parameters  are  fulfilled.  Observe  that, in
order  to evaluate  Higgs-boson pole masses, Higgs-vacuum-polarization
contributions should be included in  the calculation.  In general, for
our  choice   of the  renormalization   scale, the vacuum-polarization
contributions are small.   These contributions would only be necessary
in   a   calculation  of  Higgs-boson  masses   that  goes  beyond the
approximation presented here.  Therefore, the present approach is very
analogous  to that studied  in Ref.\ \cite{CQW}  in the MSSM framework
with CP-invariant Higgs potential.

The computation   of the Higgs-boson  mass-matrix  elements considered
here and  in   \cite{CEQW}  is     still affected  by      theoretical
uncertainties, most noticeably, those associated with two-loop, finite
threshold corrections to the effective  quartic couplings of the Higgs
potential.  Recently, a partial, diagrammatic, two-loop calculation of
the mass  of the  lightest CP-even  Higgs  boson has been  carried out
\cite{HEHOWE}.  In the  limit  of   large $M_{H^+}$, the    additional
two-loop threshold corrections  lead to a  slight modification of  the
dependence of the lightest CP-even  Higgs-boson mass on the scalar-top
mixing  parameters.  For  instance, although the   upper bound  on the
lightest CP-even Higgs-boson    mass for TeV scalar-quark  masses   is
approximately equal to   that obtained through  next-to-leading order,
which is also the approach we followed  in the present work, the upper
bound on the Higgs-boson mass is reached for slightly different values
of $|\tilde{A}_t|$, i.e.\ $|\tilde{A}_t| \simeq 2M_{\rm SUSY}$ instead
of $|\tilde{A}_t| = \sqrt{6}\, M_{\rm  SUSY}$, with a weak  dependence
on the sign  of $\tilde{A}_t$. Similar results  were obtained by means
of  a  two-loop calculation  of  the effective potential \cite{Zhang}. 
Nevertheless,   a   complete  diagrammatic analysis of   the  two-loop
corrections induced by the $t$-Yukawa coupling,  which are included at
the leading-logarithmic level in our computation, is still lacking.

\end{appendix}


\newpage

\begin{thebibliography}{99}
  
\bibitem{Higgs1} LEP Committee, 12 November, 1998, CERN/LEPC 98-9.
  The experimental collaborations at CERN report the following lower
  mass bounds on the SM Higgs boson at 95$\%$ CL: 95.5 GeV (L3)
  [http://hpl3sn02.cern.ch/analysis/latestresults.html], 94 GeV (OPAL)
  [http://www1.cern.ch/Opal/plots/plane/lepc98.html], 94.1 GeV
  (DELPHI) [http://delphiwww.cern.ch/delfigs/figures/figures.html];
  90.4 GeV (ALEPH) [http://alephwww.cern.ch/ALPUB/oldconf/oldconf99.html].
 
\bibitem{Higgs2} For recent  analyses, see for  instance J.  Erler and
  P.  Langacker,  to appear in Proceedings  of the ``5th International
  Wein Symposium:  A Conference on Physics  Beyond the  Standard Model
  (WEIN 98),'' Santa Fe,  NM,  14--21 June  1998, hep-ph/9809352; G.   
  Degrassi, P.  Gambino, M. Passera and A.  Sirlin, Phys.\ Lett.\ {\bf
    B418} (1998) 209; G.  D'Agostini and G.  Degrassi, hep-ph/9902226
  
\bibitem{SUSY} For a review, see, J.F. Gunion, H.E. Haber, G. Kane
  and S. Dawson, ``The Higgs Hunters Guide,'' (Addison-Wesley,
  Reading, MA, 1990).
  
\bibitem{Mh}  M.S.  Berger,  Phys.\ Rev.\ {\bf  D41}  (1990)  225;  Y. 
  Okada, M.  Yamaguchi and T. Yanagida, Prog.\ Theor.\ Phys.\ {\bf 85}
  (1991) 1; Phys.\ Lett.\ {\bf B262}  (1991)  54; H.E.   Haber and R.  
  Hempfling, Phys.\ Rev.\  Lett.\ {\bf 66} (1991)  1815; J.  Ellis, G. 
  Ridolfi and F.    Zwirner, Phys.\ Lett.\ {\bf  B257}  (1991) 83;  R. 
  Barbieri, M.  Frigeni and  F.  Caravaglios, Phys.\ Lett.\ {\bf B258}
  (1991) 167;    P.H.    Chankowski,  Warsaw   preprint 1991, IFT-7-91
  (unpublished);  J.R.  Espinosa and M.   Quir\'os, Phys.\ Lett.\ {\bf
    B266} (1991) 389; J.L.  Lopez and  D.V.  Nanopoulos, Phys.\ Lett.\ 
  {\bf B266}  (1991) 397; M.    Carena, K.  Sasaki and  C.E.M. Wagner,
  Nucl.\ Phys.\ {\bf  B381} (1992) 66; P.H.   Chankowski, S.  Pokorski
  and J.   Rosiek, Phys.\ Lett.\ {\bf  B281} (1992) 100; Nucl.\ Phys.\ 
  {\bf  B423}  (1994) 437;  D.M.   Pierce, A.    Papadopoulos and S.B. 
  Johnson, Phys.\   Rev.\ Lett.\ {\bf 68}  (1992)  3678; A.  Brignole,
  Phys.\ Lett.\ {\bf B281}  (1992) 284; V. Barger, M.S.  Berger and P. 
  Ohmann, Phys.\ Rev.\ {\bf D49} (1994) 4908; G.L.  Kane, C. Kolda, L.
  Roszkowski and J.D.   Wells, Phys.\ Rev.\ {\bf D49}  (1994) 6173; R. 
  Hempfling and A.H.  Hoang, Phys.\ Lett.\ {\bf B331} (1994) 99;
  P. Langacker and N. Polonsky, Phys.\ Rev.\ {\bf D50} (1994) 2199.
  
\bibitem{HH} H.E. Haber and R. Hempfling, Phys.\ Rev.\ {\bf D48}
  (1993) 4280.

\bibitem{CEQW}  M.   Carena, J.R. Espinosa, M.    Quir\'os  and C.E.M. 
  Wagner, Phys.\ Lett.\ {\bf B355} (1995) 209.
  
\bibitem{CHE}   J. Ellis, T. Falk,   K.  Olive and  M. Schmitt, Phys.\ 
  Lett.\ {\bf  B388} (1996) 97;  Phys.\ Lett.\ {\bf B413}  (1997) 355;
  S.A. Abel and  B.C. Allanach, Phys.\ Lett.\  {\bf  B431} (1998) 339;
  J.A. Casas, J.R. Espinosa  and H.E. Haber,  Nucl.\ Phys.\ {\bf B526}
  (1998) 3.
  
\bibitem{CCPW} M. Carena, P. Chankowski, S. Pokorski and C.E.M.
  Wagner, Phys.\ Lett.\ {\bf B441} (1998) 205.

\bibitem{APLB} A. Pilaftsis, Phys.\ Lett.\ {\bf B435} (1998) 88;
  Phys.\ Rev.\ {\bf D58} (1998) 096010.

\bibitem{ANPB} A. Pilaftsis, Nucl.\ Phys.\ {\bf B504} (1997) 61.
  
\bibitem{PN} A. Pilaftsis  and  M. Nowakowski, Int.\ J.   Mod.\ Phys.\ 
  {\bf  A9} (1994) 1097; G.   Cveti$\check{\rm c}$, Phys.\ Rev.\  {\bf
    D48} (1993) 5280; B.  Grzadkowski, Phys.\ Lett.\ {\bf B338} (1994)
  71.
  
\bibitem{Osland} C.A. Boe,   O.M. Ogreid, P.  Osland  and J.-Z. Zhang,
  hep-ph/9811505; see also A.  Skjold  and P. Osland,  Nucl.\  Phys.\
  {\bf B453} (1995) 3.
  
\bibitem{LHC} W. Bernreuther, A.  Brandenburg and M. Flesch, Phys.\ 
  Rev.\ {\bf D56} (1997) 90; hep-ph/9812387.

\bibitem{APRL} A. Pilaftsis, Phys.\ Rev.\ Lett.\ {\bf 77} (1996) 4996.

\bibitem{CD} S.-Y. Choi and M. Drees, Phys.\ Rev.\ Lett.\ {\bf 81} (1998)
  5509.
  
\bibitem{EWbau} M.  Carena,   M.  Quir\'os and  C.E.M.  Wagner, Phys.\ 
  Lett.\ {\bf B380} (1996) 81; Nucl.\ Phys.\ {\bf B524} (1998) 3; J.R.
  Espinosa,  Nucl.\ Phys.\ {\bf  B475} (1996) 273;  D.  Delepine, J.M. 
  Gerard, R.  Gonzalez-Felipe and J.  Weyers, Phys.\ Lett.\ {\bf B386}
  (1996) 183;  A.  Riotto, Phys.\ Rev.\  {\bf  D53} (1996)  5834; J.R. 
  Espinosa and B.  De  Carlos, Nucl.\ Phys.\ {\bf B503}  (1997) 24; D. 
  B\"odeker, P.  John, M.  Laine and M.G.  Schmidt, Nucl.\ Phys.\ {\bf
    B497} (1997) 387; M. Carena, M.  Quiros,  A. Riotto, I.  Vilja and
  C.E.M.  Wagner, Nucl.\ Phys.\ {\bf B503}  (1997) 387; J.M. Cline, M. 
  Joyce and M.  Kainulainen,  Phys.\ Lett.\ {\bf  B417} (1998) 79; M.  
  Laine and K. Rummukainen, Phys.\  Rev.\ Lett.\ {\bf 80} (1998) 5259;
  Nucl.\  Phys.\ {\bf B535}  (1998) 423; J.M.    Cline and G.D. Moore,
  Phys.\ Rev.\ Lett.\ {\bf 81}  (1998) 3317; M.  Losada, Nucl.\ Phys.\ 
  {\bf B537}  (1999) 3; K. Funakubo, hep-ph/9809517;  J.  Grant and M. 
  Hindmarsh,    hep-ph/9811289;    M.  Laine   and     K. Rummukainen,
  hep-ph/9811369;  A.B. Lahanas,     V.C.  Spanos  and  V.    Zarikas,
  hep-ph/9812535.
  
\bibitem{KYS} J. Kodaira, Y. Yasui and K. Sasaki, Phys.\ Rev.\ {\bf
    D50} (1994) 7035.
  
\bibitem{CEQR} J.A. Casas, J.R. Espinosa, M. Quir\'os and A. Riotto,
  Nucl.\ Phys.\ {\bf B436} (1995) 3; (E) {\bf B439} (1995) 466.

\bibitem{EDM} J. Ellis, S.  Ferrara and D.V. Nanopoulos, Phys.\ Lett.\ 
  {\bf B114} (1982) 231; W.  Buchm\"uller and D. Wyler, Phys.\  Lett.\ 
  {\bf B121} (1983) 321; J. Polchinski and M. Wise, Phys.\ Lett.\ {\bf
    B125} (1983) 393; F.   del Aguila, M.  Gavela,  J.  Grifols and A. 
  Mendez, Phys.\ Lett.\ {\bf B126} (1983) 71;  D.V.  Nanopoulos and M. 
  Srednicki, Phys.\ Lett.\ {\bf B128} (1983) 61.
  
\bibitem{DGH} M. Dugan, B.  Grinstein and L.  Hall, Nucl.\ Phys.\ {\bf
    B255} (1985) 413.

\bibitem{FOS} T. Falk, K.A. Olive and M. Srednicki, Phys.\ Lett.\ {\bf
    B354} (1995) 99; T. Falk and K.A. Olive, hep-ph/9806236.
  
\bibitem{KO} Y. Kizukuri and N. Oshimo, Phys.\  Rev.\ {\bf D46} (1992)
  3025.

\bibitem{IN} T. Ibrahim   and P. Nath,   Phys.\ Lett.\ {\bf  B418}
  (1998) 98; M. Brhlik, G.J. Good and G.L. Kane, hep-ph/9810457.
  
\bibitem{AF} See, for example,   S.A. Abel and J.-M. Fr\`ere,   Phys.\ 
  Rev.\ {\bf D55} (1997) 1623, and references therein.

\bibitem{CKP} D. Chang, W.Y. Keung and A. Pilaftsis, Phys.\ Rev.\ 
  Lett.\ {\bf 82} (1999) 900.
  
\bibitem{BRLW} G.C. Branco and M.N. Rebelo, Phys.\ Lett.\ {\bf B160}
  (1985) 117; J. Liu and L. Wolfenstein, Nucl.\ Phys.\ {\bf B289}
  (1987) 1.

\bibitem{NM} N. Maekawa, Phys.\ Lett.\ {\bf B282} (1992) 387.

\bibitem{APNH} A. Pomarol, Phys.\ Lett.\ {\bf  B287} (1992) 331; 
 N. Haba, Phys.\ Lett.\ {\bf B398} (1997) 305.
 
\bibitem{KL}  O.C.W. Kong and F.-L.    Lin,  Phys.\ Lett.\ {\bf  B419}
  (1998) 217.
  
\bibitem{SW} S. Weinberg, Phys.\ Rev.\ Lett.\ {\bf 63} (1989) 2333; E.
  Braaten,  C.S.  Li and  T.C.  Yuan,  Phys.\   Rev.\ Lett.\ {\bf  64}
  (1990) 1709.
  
\bibitem{BZ} S.M.  Barr and A.  Zee, Phys.\ Rev.\ Lett.\ {\bf 65}
  (1990) 21.

\bibitem{PDG} C. Caso et al.  (Particle Data Group), Eur.\ Phys.\
  J. {\bf C3} (1998) 1.
  
\bibitem{abdullah} K. Abdullah, C. Carlberg, E.D. Commins, H. Gould
  and S.B. Ross, Phys.\ Rev.\ Lett.\ {\bf 65} (1990) 2347.

\bibitem{commins} E.D. Commins, S.B. Ross, D. DeMille and B.C. Regan,
  Phys.\ Rev.\ {\bf A50} (1994) 2960.

\bibitem{altarev} I.S. Altarev et al., Phys.\ Atom.\ Nucl.\ {\bf 59}
  (1996) 1152. 
  
\bibitem{FPT}   W. Fischler, S.   Paban and  S.  Thomas, Phys.\ Lett.\ 
  {\bf B289} (1992) 373, and references therein.
  
\bibitem{LEP2}  M.  Carena, P.  Zerwas,   and convenors of  the  Higgs
  Physics Working  Group, in Physics at  LEP2, eds.\ G.  Altarelli, T. 
  Sj\"ostrand and F. Zwirner, (Report  CERN 96-01, Geneva 1996), Vol.\ 
  1.
  
\bibitem{Workshop} M. Carena and J. Lykken, ``Report of the Physics at
  Run  II  Supersymmetry/Higgs  Workshop,''  Fermilab, 1999,  eds., in
  preparation.
  
\bibitem{Alex} A. M\'endez  and A. Pomarol,  Phys.\  Lett.\ {\bf B272}
  (1991) 313; J.F.   Gunion, B.   Grzadkowski,   H.E.  Haber and  J.   
  Kalinowski, Phys.\ Rev.\ Lett.\ {\bf 79} (1997) 982.
  
\bibitem{Howie} H. Baer, B.W. Harris and X. Tata, Phys. Rev.  {\bf
    D59} (1999) 015003.

\bibitem{CMW} M. Carena, S. Mrenna and C.E.M. Wagner, hep-ph/9808312,
  to be published in Phys. Rev. D.
  
\bibitem{LL} L. Lavoura, Phys.\ Rev.\ {\bf D51} (1995) 5256. 

\bibitem{DM}  N.G. Deshpande and E. Ma,  Phys.\ Rev.\ {\bf D16} (1977)
  1583; Phys.\ Rev.\ {\bf D18} (1978) 2574.
    
\bibitem{darwin/yee} D.  Chang and   W.-Y. Keung, Phys.\   Lett.\ {\bf
    B305}  (1993) 261; D. Chang, W.-Y.  Keung  and I. Phillips, Phys.\ 
  Rev.\ {\bf D48} (1993) 3225.
  
\bibitem{BKW} K.S. Babu, C. Kolda, J. March-Russell and F. Wilczek,
  Phys.\ Rev.\ {\bf D59} (1999) 016004.

\bibitem{EMa} E. Ma, Phys.\ Rev.\ {\bf D39} (1989) 1922.
  
\bibitem{dmb} R. Hempfling, Phys.\ Rev.\ {\bf D49} (1994) 6168; L.
  Hall, R. Rattazzi and U. Sarid, Phys.\ Rev.\ {\bf D50} (1994) 7048;
  M. Carena, M. Olechowski, S. Pokorski and C.E.M. Wagner, Nucl.\ 
  Phys.\ {\bf B426} (1994) 269; D. Pierce, J. Bagger, K.  Matchev and
  R. Zhang, Nucl.\ Phys.\ {\bf B491} (1997) 3.
  
\bibitem{sola} J.A. Coarasa, R.A.  Jimenez and J. Sola, Phys.\ Lett.\ 
  {\bf B389} (1996) 312; R.A. Jimenez and J.  Sola, Phys.\ Lett.\ {\bf
    B389} (1996) 53; K.T.  Matchev and D.M. Pierce, Phys.\ Lett.\ {\bf
    B445} (1999) 331; P.H. Chankowski, J. Ellis, M. Olechowski and S.
  Pokorski, hep-ph/9808275; K.S. Babu and C. Kolda, hep-ph/9811308.

\bibitem{Demir} D.A. Demir, hep-ph/9901389.
  
\bibitem{GGK} B. Grzadkowski, J.F. Gunion and J. Kalinowski,
  hep-ph/9902308.

\bibitem{GH} J.F. Gunion and H.E. Haber, Nucl.\ Phys.\ {\bf B272}
  (1986) 1; (E) {\bf B402} (1993) 567.
  
\bibitem{CQW} M. Carena, M.  Quiros and C.E.M. Wagner, Nucl.\ Phys.\ 
  {\bf B461} (1996) 407; H.E. Haber, R. Hempfling and A.H. Hoang, Z.
  Phys.\ {\bf C75} (1997) 539.

\bibitem{HEHOWE} S. Heinemeyer, W. Hollik and G. Weiglein, Phys.\ 
  Rev.\ {\bf D58} (1998) 091701; Phys. Lett. {\bf B440} (1998) 96.

\bibitem{Zhang} R.-J. Zhang, Phys.\ Lett.\ {\bf B447} (1999) 89.

\end{thebibliography}
\end{document}